\def\RELEASE{1}  %
\def\ANON{0}     %
\def\SQUEEZE{0}  %

\documentclass[sigplan,10pt,screen]{acmart}
\renewcommand\footnotetextcopyrightpermission[1]{}
\PassOptionsToPackage{hyphens}{url}\usepackage{hyperref}

\usepackage[noend]{algpseudocode}
\usepackage{algorithm}
\usepackage{algorithmicx}
\usepackage{booktabs}
\usepackage{caption}
\usepackage{comment}
\usepackage[inline]{enumitem}
\usepackage{graphicx}
\usepackage{listings}
\usepackage{multirow}
\usepackage{xcolor}
\usepackage{xspace}
\usepackage{libertine}
\usepackage{microtype}
\usepackage{tikz}
\usepackage{paralist}
\usepackage{gnuplot-lua-tikz}

\microtypecontext{spacing=nonfrench}

\hypersetup{
  pdflang={en},
  pdfdisplaydoctitle}

\definecolor[named]{OurPurple}{cmyk}{0.55,1,0,0.15}
\definecolor[named]{OurDarkBlue}{cmyk}{1,0.58,0,0.21}
\hypersetup{
  colorlinks,
  linkcolor=OurPurple,
  citecolor=OurPurple,
  urlcolor=OurDarkBlue,
  filecolor=OurDarkBlue}

\newcommand{\algorithmautorefname}{Algorithm}
\newcommand{\algorithmname}{Algorithm}

\if \SQUEEZE 1
\setlist[itemize]{
  leftmargin=*,
  itemsep=2pt,
  topsep=2pt}

\renewcommand\figurename{Fig.}
\renewcommand*\figureautorefname{Fig.}
\renewcommand\tablename{Tab.}
\renewcommand*\tableautorefname{Tab.}
\renewcommand\algorithmname{Alg.}
\renewcommand*\algorithmautorefname{Alg.}

\fi
\def\Snospace~{\S{}}
\def\sectionautorefname{\Snospace}
\def\subsectionautorefname{\Snospace}
\def\subsubsectionautorefname{\Snospace}

\if \RELEASE 1
  \def\NOTES{0}
\else
  \def\NOTES{1}
\fi
\if \NOTES 1
  \newcommand{\XXX}[1]{{\color{red}{XXX {#1}}}}
  \newcommand{\matheus}[1]{{\color{violet}{[\textbf{MS:} {#1}]}}}
  \newcommand{\liam}[1]{{\color{orange}{[\textbf{LA:} {#1}]}}}
  \newcommand{\simon}[1]{{\color{olive}{[\textbf{SP:} {#1}]}}}
  \newcommand{\antoine}[1]{{\color{teal}{[\textbf{AK:} {#1}]}}}
  \newcommand{\todo}[1]{{\color{blue}{TODO: {#1}}}}
\else
  \newcommand{\XXX}[1]{}
  \newcommand{\matheus}[1]{}
  \newcommand{\liam}[1]{}
  \newcommand{\simon}[1]{}
  \newcommand{\antoine}[1]{}
  \newcommand{\todo}[1]{}
\fi

\newcommand{\eg}{e.g.\xspace}
\newcommand{\ie}{i.e.\xspace}
\newcommand{\cf}{cf.\xspace}

  \newcommand{\sys}{Virtuoso\xspace}
  \newcommand{\sysplot}{virtuoso\xspace}

\lstdefinelanguage{C}{
  morekeywords={
    auto,break,case,char,const,continue,default,do,double,else,enum,extern,
    float,for,goto,if,inline,int,long,register,restrict,return,short,signed,
    sizeof,static,struct,switch,typedef,union,unsigned,void,volatile,while,
    _Alignas,_Alignof,_Atomic,_Bool,_Complex,_Generic,_Imaginary,_Noreturn,
    _Static_assert,_Thread_local
  },
  sensitive=true,
  morecomment=[l]{//},
  morecomment=[s]{/*}{*/},
  morestring=[b]",
}

\lstMakeShortInline[columns=fixed]|

\begin{document}
\date{}
\title{Tail Contagion: Sub-$\mu$s Time Protection in Shared Software Network Datapaths}

    \author{Matheus Stolet}%
    \affiliation{%
      \institution{Max Planck Institute for Software Systems}%
      \city{Saarbrücken}%
      \country{Germany}%
    }%
    \email{mstolet@mpi-sws.org}%

    \author{Liam Arzola}%
    \affiliation{%
      \institution{UC San Diego}%
      \city{San Diego}%
      \country{USA}%
    }
    \email{larzola@ucsd.edu}%

    \author{Simon Peter}%
    \affiliation{%
      \institution{University of Washington}%
      \city{Seattle}%
      \country{USA}%
    }%
    \email{simpeter@cs.washington.edu}%

    \author{Antoine Kaufmann}%
    \affiliation{%
      \institution{Max Planck Institute for Software Systems}%
      \city{Saarbrücken}%
      \country{Germany}%
    }%
    \email{antoinek@mpi-sws.org}%
\begin{abstract}
Shared software datapaths underpin modern datacentre networking.
They implement mechanisms such as virtual switching, network
virtualisation tunneling, or reliable transport, and enforce policies, such as tenant
rate limits, virtual network isolation, or congestion control.
However, because multiple applications, containers, or VMs share them, often
across tenants, they pose a tail latency isolation challenge.
Current isolation approaches either sacrifice efficiency via
coarse-grained core partitioning or provide weak tail latency 
isolation when sharing cores with basic rate limits.

This paper presents \sys, a time protection mechanism for shared software
datapaths
that provides strong cross-tenant tail latency isolation while
preserving low overhead and $\mu$s-scale latency.
Our key insight is that tail latency is fundamentally a time metric, so
byte or packet throughput is the wrong metric for controlling interference
when packet processing costs vary.
Our design instead enforces isolation through per-tenant
CPU-time budgets at datapath
intervention points within run-to-completion loops, without relying on
preemption.
In a case study, we instantiate \sys in
the TAS TCP stack and demonstrate a
7.8$\times$ reduction in victim tail latency under adversarial interference
while keeping throughput within 5\% of unmodified TAS. We also observe a
3$\times$ per-core efficiency improvement compared to siloed datapaths
under bursty workloads.
\end{abstract}
 
\settopmatter{printfolios=true, printacmref=false}
\maketitle
\pagestyle{plain}

\section{Introduction}

Shared software datapaths are a core component of modern datacentre
networking.
They process packets and implement functions such as virtual
switching~\cite{software:ovs}, tunneling for network
virtualisation~\cite{dalton:andromeda}, reliable
transport~\cite{kaufmann:tas,marty:snap,niu:netkernel}, and policy
enforcement~\cite{software:ovs,dalton:andromeda}.
These datapaths are shared across applications, containers, or VMs and
multiplex access to the shared physical network.
Operators implement these functions in software, on host
CPUs~\cite{kaufmann:tas,marty:snap} or on DPUs~\cite{product:amazon:nitro},
to preserve the programmability needed to evolve infrastructure protocols,
virtualisation mechanisms, and policies.
Shared software datapaths are therefore a necessary part of many modern network
stacks.

That same sharing creates a cross-tenant tail latency interference problem.
Shared datapath cores couple tenants in time: work done for one tenant
directly delays service to another.
On shared network datapath cores, common-case packet processing often spans
only hundreds of CPU cycles~\cite{farshin:packetmill,kaufmann:tas}, while
end-to-end latency budgets are only a few to a few tens of microseconds.
To sustain high performance, datapaths rely on
batching~\cite{miao:smart_batching}
and run-to-completion execution~\cite{panda:netbricks},
so interference occurs in non-preemptive windows.
At these timescales, even a small burst or unusually expensive packets from a
co-located tenant can materially inflate tail
latency~\cite{kaffes:shinjuku,fried:caladan,iyer:concord}.
The resulting challenge is strong in-datapath tail latency isolation:
the ability of a system to control cross-tenant tail latency interference.

Existing approaches force an efficiency/protection trade-off at these
timescales.
Coarse-grained designs partition tenants onto dedicated
cores~\cite{grant:fairnic,marty:snap}, improving isolation but stranding
resources and forfeiting efficient fine-grained sharing.
Fine-grained shared datapaths~\cite{niu:netkernel,kaufmann:tas} retain the
efficiency benefits of pooling, but provide weak tail
latency isolation when tenants contend
on the same core.
Volume-based controls for fine-grained
shared datapaths~\cite{shieh:seawall,ballani:oktopus,spec:tctbf,rfc:sr_three_color_marker,rfc:tr_three_color_marker,rfc:diff_three_color_marker}
try to recover tail latency isolation while preserving sharing,
but still account in traffic volume rather than execution time and
are thus not able to bound cross-tenant latency interference.
Related schedulers rely on mechanisms such as preemption, centralised dispatch,
work stealing, or additional
queueing~\cite{prekas:zygos,demoulin:persephone,iyer:concord,ousterhout:shenango,kaffes:shinjuku,fried:caladan},
whose overheads are incompatible with sub-$\mu$s packet processing in a shared
datapath.
Current choices therefore offer either efficient sharing without strong
in-datapath tail latency isolation, or stronger isolation without efficient sharing.

In a shared datapath, tail latency isolation means bounding the increase in a
tenant's tail latency caused by co-located tenants' contention for shared
datapath CPU time.
That increase is driven by short-term contention over CPU
time on shared cores, not simply by traffic volume.
Yet, in-datapath protection mechanisms typically account for packets or bytes per
second.
Packet processing cost varies with cache locality, rare branches, flow-state
footprint, coherence traffic, and batch composition, so the same traffic volume
can still consume different amounts of CPU time (\autoref{fig:iso-tas}).
Tail latency isolation must therefore be expressed in CPU time, not traffic volume.

Leveraging this insight, we present \sys, a \emph{time protection mechanism for shared
software datapaths}.
While prior time protection work has largely been motivated by timing side
channels~\cite{ge:time_protection,heiser:time_protection_proof}, we apply it to
a finer-grained setting where run-to-completion datapath work leaves little
slack for enforcement overhead.
\sys assigns every tenant a CPU-time budget in each datapath core, charges that
budget as packets are processed, and enforces budgets only at intervention points
where a run-to-completion datapath can act cheaply.
Budgeted execution time bounds the delay for a task executed in
the datapath and can therefore control the additional tail delay
that a co-located tenant can impose through shared CPU time
contention.
To preserve efficiency, \sys retains batching and repays temporary overruns with
bounded deficit instead of relying on preemption or breaking batches.
For receive processing, \sys
charges the work to tenants once attribution becomes known and enforces budget state at the
next eligible intervention point.

We instantiate \sys in TAS, a state-of-the-art kernel-bypass TCP
stack~\cite{kaufmann:tas}, and evaluate it against both shared and siloed
baselines.
Our results show that \sys protects victim tail latency under adversarial
interference, reducing victim tail latency by 7.8$\times$ relative to
unmodified shared TAS while keeping throughput within 5\% of the unprotected
shared datapath.
It also preserves much of the efficiency advantage that unprotected sharing
retains over partitioned deployments, improving the per-core efficiency
by 3$\times$ compared to siloed datapaths under bursty workloads.
These results show that strong in-datapath tail latency isolation
does not require giving
up the efficiency benefits that motivated shared software datapaths in the
first place.

This paper makes the following contributions:

\smallskip
\begin{compactitem}[\labelitemi]
  \item We identify and characterise tail latency interference in shared software
    network datapaths as a time protection problem.

  \item We design \sys, a time protection mechanism that enforces per-tenant
    CPU-time budgets in run-to-completion shared software datapaths.

  \item We instantiate \sys in TAS and show strong tail latency isolation
    under interference while retaining efficiency close
    to an unprotected shared
    datapath and incurring low overhead.
\end{compactitem}
\smallskip

\noindent
We will release \sys as open-source on publication.
\section{Background}%
\label{sec:bg}

We now define the shared software datapaths we study and use OvS and
TAS as concrete examples.
We then characterise the fast-path execution model that makes tail latency
isolation difficult, and summarise the main in-datapath isolation approaches
used today.

\subsection{Shared Software Network Datapaths}
Shared software datapaths are packet-processing pipelines
that execute in software and serve multiple tenants,
such as applications, containers, or VMs,
within one implementation instance.
They run as independent services or as part of operating systems or hypervisors,
either on the host processor or on SmartNIC/DPU
cores~\cite{product:amazon:nitro,shashidhara:flextoe}.
They implement a range of different functions.
For example,
software switches multiplex VM traffic and implement network
virtualisation~\cite{software:ovs,product:amazon:nitro,dalton:andromeda},
and optimised network stacks centralise host networking up to the transport
layer in optimised
datapaths~\cite{marty:snap,kaufmann:tas,shashidhara:flextoe,moon:acceltcp}.

We use OpenvSwitch virtual switch and the TAS TCP stack as illustrative
examples.
Both systems rely on a control and datapath split, with an optimised datapath
running on typically dedicated cores, augmented with an out-of-band control path
for more complex but infrequent operations.
The performance-optimised OvS-DPDK datapath polls NIC RX queues (\texttt{RX})
and VM TX queues (\texttt{TX}), and then classifies and transforms packets, all
on behalf of multiple tenant VMs.
The TAS TCP stack datapath handles common-case packet processing for receiving
and sending TCP data by polling NIC RX queues (\texttt{RX}), application TX
queues (\texttt{TX}), and the internal flow transmission scheduler
(\texttt{SCHED}), implementing all necessary per-packet TCP protocol
processing steps for \texttt{RX} and \texttt{TX}.
Internally, both datapaths employ run-to-completion processing and batching when
processing packets, accesses to shared state data structures, and en-/dequeueing
packets from NIC queues multiplexed across tenants.

\subsection{Performance Characteristics and Interference}
Shared datapaths are resource efficient because of resource multiplexing and
microarchitectural benefits.
They reduce resource stranding by pooling resources such as cores or buffer
memory and thereby replacing per-tenant peak provisioning.
Micro-architecturally, shared network datapaths centralise packet processing
onto a small set of cores and thus reap the locality benefits of only
executing the same small set of operations accessing the same data structures
on these cores.
Individual per-packet operations are typically very small. Both OvS and
TAS spend only 488--862 CPU cycles total per packet across the 2 or 3
execution tasks (\autoref{tab:cycles}).
Common optimisations such as batched packet
processing~\cite{miao:smart_batching} and group
prefetching~\cite{kalia:gpupproc}
further boost execution efficiency when consolidating execution of these small
packet handlers.

\begin{table}[t]%
  \centering%
  \captionsetup{skip=1pt}%
  \small%
  \begin{tabular}{@{}lllll@{}}%
    & \texttt{RX} & \texttt{TX} & \texttt{SCHED} & Total \\
    \midrule
    OvS & 341 & 147 & - &  488 \\
    TAS & 317 & 164 & 381 & 862 \\
  \end{tabular}%
  \vspace{1em}%
  \caption{CPU cycles spent per-packet on the datapath of TAS and OvS.}%
  \label{tab:cycles}%
\end{table}%

Shared worker cores couple tenants in time.
Work done for one tenant directly delays work for another, and this dependence
is amplified by shared microarchitectural state such as the cache and TLB.
At this scale, and even with these optimisations, packet processing times can
vary substantially due to poor locality or rare but expensive code paths.

Batching further increases this coupling.
High-performance datapaths batch packets to amortise fixed overheads in polling,
descriptor handling, and per-iteration bookkeeping.
Batching also typically improves locality and provides software prefetches for
state accesses likely to miss with sufficient time to complete~\cite{kalia:gpupproc}.
Since batches mix tenants, however, they become the unit of execution and
dispatch.
The same mechanism that improves efficiency also enlarges the atomic
delay one tenant can impose on its peers.

Receive processing is also asymmetric.
On shared NIC RX queues, the datapath must often fetch a packet and parse its
headers before it can determine which tenant should be charged for the work.
As a result, a substantial fraction of the reception cost may be incurred before
the datapath knows which tenant owns the packet.
In some cases, NIC hardware can steer packets into per-tenant RX queues and
avoid this ambiguity.
However, that requires protocol-aware steering for the full stack, including
mechanisms such as network virtualisation, and more NIC queues generally reduce
efficiency in the driver~\cite{sanaee:backdraft} and on
PCIe~\cite{stephens:loom}.

\begin{figure}%
\centering
\includegraphics[width=0.48\textwidth]{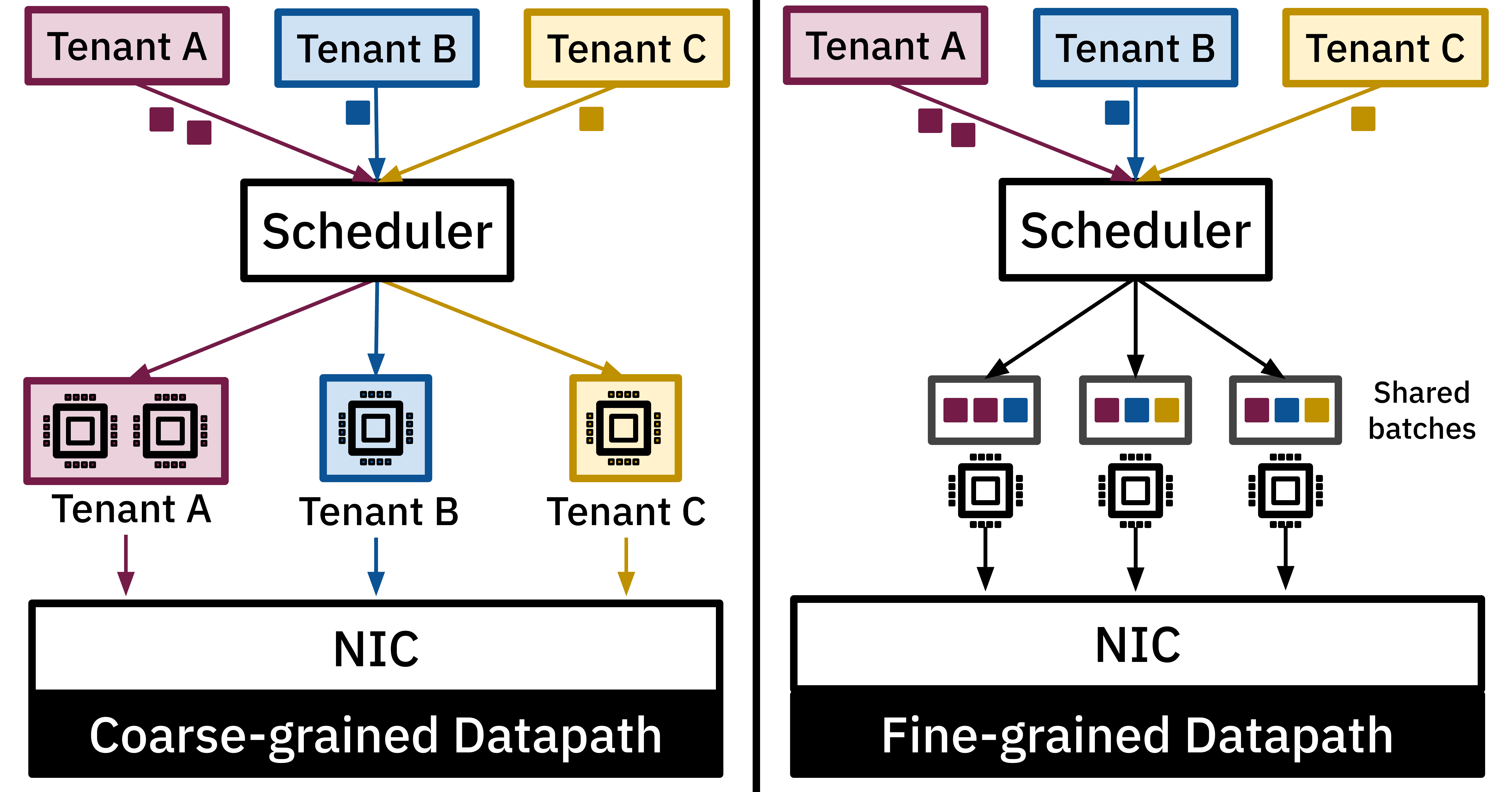}%
\caption{Coarse-grained shared datapaths partition tenants to cores
and have better performance isolation.
Fine-grained shared datapaths batch the processing from
multiple tenants and have better efficiency.}%
\label{fig:shared_datapath}%
\Description[Coarse grained and fine grained datapath visualisation]
  {Fully described in the text.}%
\end{figure}

\subsection{Existing Isolation Approaches Are Insufficient}
Within shared datapaths, the main approaches to tail latency isolation fall in
two categories (\autoref{fig:shared_datapath}):
coarse-grained sharing with core partitioning,
and fine-grained sharing with rate limits.
Coarse-grained sharing partitions tenants to exclusive cores, and then
reallocates cores out-of-band between tenants based on
utilisation~\cite{marty:snap,grant:fairnic}.
Fine-grained sharing instead processes packets from multiple tenants on the
same cores, and then relies on volume-based controls for tenants, such as
packet or byte rate limits~\cite{software:ovs,niu:netkernel,kaufmann:tas}.

\begin{figure}%
\centering%
\begin{tikzpicture}[gnuplot]
\tikzset{every node/.append style={font={\fontsize{8.0pt}{9.6pt}\selectfont}}}
\path (0.000,0.000) rectangle (8.458,3.048);
\gpfill{rgb color={0.392,0.561,1.000}} (1.717,2.800)--(2.084,2.800)--(2.084,2.940)--(1.717,2.940)--cycle;
\gpcolor{color=gp lt color border}
\gpsetlinetype{gp lt border}
\gpsetdashtype{gp dt solid}
\gpsetlinewidth{1.00}
\draw[gp path] (1.717,2.800)--(1.717,2.940)--(2.084,2.940)--(2.084,2.800)--cycle;
\gpfill{rgb color={0.471,0.369,0.941}} (3.925,2.800)--(4.292,2.800)--(4.292,2.940)--(3.925,2.940)--cycle;
\draw[gp path] (3.925,2.800)--(3.925,2.940)--(4.292,2.940)--(4.292,2.800)--cycle;
\gpfill{rgb color={0.863,0.149,0.498}} (6.132,2.800)--(6.499,2.800)--(6.499,2.940)--(6.132,2.940)--cycle;
\draw[gp path] (6.132,2.800)--(6.132,2.940)--(6.499,2.940)--(6.499,2.800)--cycle;
\node[gp node left] at (2.232,2.834) {1 core};
\node[gp node left] at (4.440,2.834) {2 cores};
\node[gp node left] at (6.647,2.834) {3 cores};
\gpdefrectangularnode{gp plot 1}{\pgfpoint{0.761cm}{2.621cm}}{\pgfpoint{8.118cm}{3.047cm}}
\gpcolor{color=gp lt color axes}
\gpsetlinetype{gp lt axes}
\gpsetdashtype{gp dt axes}
\gpsetlinewidth{0.50}
\draw[gp path] (0.761,0.396)--(3.213,0.396);
\gpcolor{color=gp lt color border}
\gpsetlinetype{gp lt border}
\gpsetdashtype{gp dt solid}
\gpsetlinewidth{1.00}
\draw[gp path] (0.761,0.396)--(0.851,0.396);
\node[gp node right] at (0.614,0.396) {0};
\gpcolor{color=gp lt color axes}
\gpsetlinetype{gp lt axes}
\gpsetdashtype{gp dt axes}
\gpsetlinewidth{0.50}
\draw[gp path] (0.761,0.975)--(3.213,0.975);
\gpcolor{color=gp lt color border}
\gpsetlinetype{gp lt border}
\gpsetdashtype{gp dt solid}
\gpsetlinewidth{1.00}
\draw[gp path] (0.761,0.975)--(0.851,0.975);
\node[gp node right] at (0.614,0.975) {1};
\gpcolor{color=gp lt color axes}
\gpsetlinetype{gp lt axes}
\gpsetdashtype{gp dt axes}
\gpsetlinewidth{0.50}
\draw[gp path] (0.761,1.553)--(3.213,1.553);
\gpcolor{color=gp lt color border}
\gpsetlinetype{gp lt border}
\gpsetdashtype{gp dt solid}
\gpsetlinewidth{1.00}
\draw[gp path] (0.761,1.553)--(0.851,1.553);
\node[gp node right] at (0.614,1.553) {2};
\gpcolor{color=gp lt color axes}
\gpsetlinetype{gp lt axes}
\gpsetdashtype{gp dt axes}
\gpsetlinewidth{0.50}
\draw[gp path] (0.761,2.132)--(3.213,2.132);
\gpcolor{color=gp lt color border}
\gpsetlinetype{gp lt border}
\gpsetdashtype{gp dt solid}
\gpsetlinewidth{1.00}
\draw[gp path] (0.761,2.132)--(0.851,2.132);
\node[gp node right] at (0.614,2.132) {3};
\draw[gp path] (1.394,0.396)--(1.394,0.486);
\node[gp node center] at (1.394,0.150) {total};
\draw[gp path] (2.580,0.396)--(2.580,0.486);
\node[gp node center] at (2.580,0.150) {per-core};
\draw[gp path] (3.213,0.396)--(3.123,0.396);
\draw[gp path] (3.213,0.975)--(3.123,0.975);
\draw[gp path] (3.213,1.553)--(3.123,1.553);
\draw[gp path] (3.213,2.132)--(3.123,2.132);
\draw[gp path] (0.761,2.132)--(0.761,0.396)--(3.213,0.396)--(3.213,2.132)--cycle;
\gpfill{rgb color={0.392,0.561,1.000}} (0.919,0.396)--(1.237,0.396)--(1.237,0.601)--(0.919,0.601)--cycle;
\draw[gp path] (0.919,0.396)--(0.919,0.600)--(1.236,0.600)--(1.236,0.396)--cycle;
\gpfill{rgb color={0.392,0.561,1.000}} (2.106,0.396)--(2.423,0.396)--(2.423,1.009)--(2.106,1.009)--cycle;
\draw[gp path] (2.106,0.396)--(2.106,1.008)--(2.422,1.008)--(2.422,0.396)--cycle;
\gpfill{rgb color={0.471,0.369,0.941}} (1.236,0.396)--(1.553,0.396)--(1.553,0.797)--(1.236,0.797)--cycle;
\draw[gp path] (1.236,0.396)--(1.236,0.796)--(1.552,0.796)--(1.552,0.396)--cycle;
\gpfill{rgb color={0.471,0.369,0.941}} (2.422,0.396)--(2.739,0.396)--(2.739,0.997)--(2.422,0.997)--cycle;
\draw[gp path] (2.422,0.396)--(2.422,0.996)--(2.738,0.996)--(2.738,0.396)--cycle;
\gpfill{rgb color={0.863,0.149,0.498}} (1.552,0.396)--(1.869,0.396)--(1.869,0.988)--(1.552,0.988)--cycle;
\draw[gp path] (1.552,0.396)--(1.552,0.987)--(1.868,0.987)--(1.868,0.396)--cycle;
\gpfill{rgb color={0.863,0.149,0.498}} (2.738,0.396)--(3.056,0.396)--(3.056,0.988)--(2.738,0.988)--cycle;
\draw[gp path] (2.738,0.396)--(2.738,0.987)--(3.055,0.987)--(3.055,0.396)--cycle;
\draw[gp path] (0.761,2.132)--(0.761,0.396)--(3.213,0.396)--(3.213,2.132)--cycle;
\node[gp node center,rotate=-270.0] at (0.160,1.264) {Normalised Tput};
\node[gp node center] at (1.987,2.378) {(a) 0 ms};
\gpdefrectangularnode{gp plot 2}{\pgfpoint{0.761cm}{0.396cm}}{\pgfpoint{3.213cm}{2.132cm}}
\gpcolor{color=gp lt color axes}
\gpsetlinetype{gp lt axes}
\gpsetdashtype{gp dt axes}
\gpsetlinewidth{0.50}
\draw[gp path] (3.214,0.396)--(5.666,0.396);
\gpcolor{color=gp lt color border}
\gpsetlinetype{gp lt border}
\gpsetdashtype{gp dt solid}
\gpsetlinewidth{1.00}
\draw[gp path] (3.214,0.396)--(3.304,0.396);
\gpcolor{color=gp lt color axes}
\gpsetlinetype{gp lt axes}
\gpsetdashtype{gp dt axes}
\gpsetlinewidth{0.50}
\draw[gp path] (3.214,0.975)--(5.666,0.975);
\gpcolor{color=gp lt color border}
\gpsetlinetype{gp lt border}
\gpsetdashtype{gp dt solid}
\gpsetlinewidth{1.00}
\draw[gp path] (3.214,0.975)--(3.304,0.975);
\gpcolor{color=gp lt color axes}
\gpsetlinetype{gp lt axes}
\gpsetdashtype{gp dt axes}
\gpsetlinewidth{0.50}
\draw[gp path] (3.214,1.553)--(5.666,1.553);
\gpcolor{color=gp lt color border}
\gpsetlinetype{gp lt border}
\gpsetdashtype{gp dt solid}
\gpsetlinewidth{1.00}
\draw[gp path] (3.214,1.553)--(3.304,1.553);
\gpcolor{color=gp lt color axes}
\gpsetlinetype{gp lt axes}
\gpsetdashtype{gp dt axes}
\gpsetlinewidth{0.50}
\draw[gp path] (3.214,2.132)--(5.666,2.132);
\gpcolor{color=gp lt color border}
\gpsetlinetype{gp lt border}
\gpsetdashtype{gp dt solid}
\gpsetlinewidth{1.00}
\draw[gp path] (3.214,2.132)--(3.304,2.132);
\draw[gp path] (3.847,0.396)--(3.847,0.486);
\node[gp node center] at (3.847,0.150) {total};
\draw[gp path] (5.033,0.396)--(5.033,0.486);
\node[gp node center] at (5.033,0.150) {per-core};
\draw[gp path] (5.666,0.396)--(5.576,0.396);
\draw[gp path] (5.666,0.975)--(5.576,0.975);
\draw[gp path] (5.666,1.553)--(5.576,1.553);
\draw[gp path] (5.666,2.132)--(5.576,2.132);
\draw[gp path] (3.214,2.132)--(3.214,0.396)--(5.666,0.396)--(5.666,2.132)--cycle;
\gpfill{rgb color={0.392,0.561,1.000}} (3.372,0.396)--(3.690,0.396)--(3.690,0.827)--(3.372,0.827)--cycle;
\draw[gp path] (3.372,0.396)--(3.372,0.826)--(3.689,0.826)--(3.689,0.396)--cycle;
\gpfill{rgb color={0.392,0.561,1.000}} (4.559,0.396)--(4.876,0.396)--(4.876,1.686)--(4.559,1.686)--cycle;
\draw[gp path] (4.559,0.396)--(4.559,1.685)--(4.875,1.685)--(4.875,0.396)--cycle;
\gpfill{rgb color={0.471,0.369,0.941}} (3.689,0.396)--(4.006,0.396)--(4.006,0.982)--(3.689,0.982)--cycle;
\draw[gp path] (3.689,0.396)--(3.689,0.981)--(4.005,0.981)--(4.005,0.396)--cycle;
\gpfill{rgb color={0.471,0.369,0.941}} (4.875,0.396)--(5.192,0.396)--(5.192,1.275)--(4.875,1.275)--cycle;
\draw[gp path] (4.875,0.396)--(4.875,1.274)--(5.191,1.274)--(5.191,0.396)--cycle;
\gpfill{rgb color={0.863,0.149,0.498}} (4.005,0.396)--(4.322,0.396)--(4.322,1.007)--(4.005,1.007)--cycle;
\draw[gp path] (4.005,0.396)--(4.005,1.006)--(4.321,1.006)--(4.321,0.396)--cycle;
\gpfill{rgb color={0.863,0.149,0.498}} (5.191,0.396)--(5.509,0.396)--(5.509,1.007)--(5.191,1.007)--cycle;
\draw[gp path] (5.191,0.396)--(5.191,1.006)--(5.508,1.006)--(5.508,0.396)--cycle;
\draw[gp path] (3.214,2.132)--(3.214,0.396)--(5.666,0.396)--(5.666,2.132)--cycle;
\node[gp node center] at (4.440,2.378) {(b) 1 ms};
\gpdefrectangularnode{gp plot 3}{\pgfpoint{3.214cm}{0.396cm}}{\pgfpoint{5.666cm}{2.132cm}}
\gpcolor{color=gp lt color axes}
\gpsetlinetype{gp lt axes}
\gpsetdashtype{gp dt axes}
\gpsetlinewidth{0.50}
\draw[gp path] (5.666,0.396)--(8.118,0.396);
\gpcolor{color=gp lt color border}
\gpsetlinetype{gp lt border}
\gpsetdashtype{gp dt solid}
\gpsetlinewidth{1.00}
\draw[gp path] (5.666,0.396)--(5.756,0.396);
\gpcolor{color=gp lt color axes}
\gpsetlinetype{gp lt axes}
\gpsetdashtype{gp dt axes}
\gpsetlinewidth{0.50}
\draw[gp path] (5.666,0.975)--(8.118,0.975);
\gpcolor{color=gp lt color border}
\gpsetlinetype{gp lt border}
\gpsetdashtype{gp dt solid}
\gpsetlinewidth{1.00}
\draw[gp path] (5.666,0.975)--(5.756,0.975);
\gpcolor{color=gp lt color axes}
\gpsetlinetype{gp lt axes}
\gpsetdashtype{gp dt axes}
\gpsetlinewidth{0.50}
\draw[gp path] (5.666,1.553)--(8.118,1.553);
\gpcolor{color=gp lt color border}
\gpsetlinetype{gp lt border}
\gpsetdashtype{gp dt solid}
\gpsetlinewidth{1.00}
\draw[gp path] (5.666,1.553)--(5.756,1.553);
\gpcolor{color=gp lt color axes}
\gpsetlinetype{gp lt axes}
\gpsetdashtype{gp dt axes}
\gpsetlinewidth{0.50}
\draw[gp path] (5.666,2.132)--(8.118,2.132);
\gpcolor{color=gp lt color border}
\gpsetlinetype{gp lt border}
\gpsetdashtype{gp dt solid}
\gpsetlinewidth{1.00}
\draw[gp path] (5.666,2.132)--(5.756,2.132);
\draw[gp path] (6.299,0.396)--(6.299,0.486);
\node[gp node center] at (6.299,0.150) {total};
\draw[gp path] (7.485,0.396)--(7.485,0.486);
\node[gp node center] at (7.485,0.150) {per-core};
\draw[gp path] (8.118,0.396)--(8.028,0.396);
\draw[gp path] (8.118,0.975)--(8.028,0.975);
\draw[gp path] (8.118,1.553)--(8.028,1.553);
\draw[gp path] (8.118,2.132)--(8.028,2.132);
\draw[gp path] (5.666,2.132)--(5.666,0.396)--(8.118,0.396)--(8.118,2.132)--cycle;
\gpfill{rgb color={0.392,0.561,1.000}} (5.824,0.396)--(6.142,0.396)--(6.142,0.964)--(5.824,0.964)--cycle;
\draw[gp path] (5.824,0.396)--(5.824,0.963)--(6.141,0.963)--(6.141,0.396)--cycle;
\gpfill{rgb color={0.392,0.561,1.000}} (7.011,0.396)--(7.328,0.396)--(7.328,2.098)--(7.011,2.098)--cycle;
\draw[gp path] (7.011,0.396)--(7.011,2.097)--(7.327,2.097)--(7.327,0.396)--cycle;
\gpfill{rgb color={0.471,0.369,0.941}} (6.141,0.396)--(6.458,0.396)--(6.458,0.970)--(6.141,0.970)--cycle;
\draw[gp path] (6.141,0.396)--(6.141,0.969)--(6.457,0.969)--(6.457,0.396)--cycle;
\gpfill{rgb color={0.471,0.369,0.941}} (7.327,0.396)--(7.644,0.396)--(7.644,1.257)--(7.327,1.257)--cycle;
\draw[gp path] (7.327,0.396)--(7.327,1.256)--(7.643,1.256)--(7.643,0.396)--cycle;
\gpfill{rgb color={0.863,0.149,0.498}} (6.457,0.396)--(6.774,0.396)--(6.774,0.963)--(6.457,0.963)--cycle;
\draw[gp path] (6.457,0.396)--(6.457,0.962)--(6.773,0.962)--(6.773,0.396)--cycle;
\gpfill{rgb color={0.863,0.149,0.498}} (7.643,0.396)--(7.961,0.396)--(7.961,0.963)--(7.643,0.963)--cycle;
\draw[gp path] (7.643,0.396)--(7.643,0.962)--(7.960,0.962)--(7.960,0.396)--cycle;
\draw[gp path] (5.666,2.132)--(5.666,0.396)--(8.118,0.396)--(8.118,2.132)--cycle;
\node[gp node center] at (6.892,2.378) {(c) 5 ms};
\gpdefrectangularnode{gp plot 4}{\pgfpoint{5.666cm}{0.396cm}}{\pgfpoint{8.118cm}{2.132cm}}
\end{tikzpicture}
\caption{Fine-grained sharing improves datapath CPU resource efficiency for
bursty workload compared to coarse-grained core partitioning. Compares
aggregate and per-core throughput for fine-grained sharing with different
numbers of cores, normalised to 3 cores with coarse-grained sharing.}%
\Description[Shared dapath efficiency]{Shared datapath with one core does not maintain
throughput with non-bursty and 1ms burst interval workloads. Shared datapath with one and
two cores is more efficient than siloed datapath under
3 milisecond and 5 milisecond bursty workload.}%
\label{fig:efficiency-burst}%
\end{figure}

This resource siloing with coarse-grained sharing results in lower resource
efficiency compared to fine-grained sharing, as unused datapath core cycles
cannot be used by other tenants.
For example, with fine-grained sharing, one tenant's packet burst can be
processed in another tenant's unused cycles.
We quantify this in \autoref{fig:efficiency-burst}, with three
memcached~\cite{software:memcached} servers running on a machine comparing two
datapath isolation configurations, under workloads with different burstiness.
For coarse-grained sharing, we run three separate TAS instances, each
with a dedicated datapath core and a distinct SR-IOV NIC virtual function.
For fine-grained sharing, we run one instance with three shared cores.
With 5\,ms bursts, fine-grained sharing achieves 201\% higher per-core
throughput, while it matches the throughput of coarse-grained sharing without
bursts.

While fine-grained sharing achieves higher resource efficiency, current isolation
approaches provide substantially weaker tail latency isolation under
cross-tenant interference
compared to the partitioned cores with coarse-grained sharing.
We quantify this in \autoref{fig:iso-conns}, where we show tenant tail latency
with an adversarial neighbour, in both OvS and TAS.
Both OvS and TAS run with a single shared datapath core, and we again use a
memcached server as the victim and adversary applications (both pinned to
separate cores).
In this experiment, the adversary increases the number of connections to create
contention in the shared datapath thereby increasing the victim tenant's tail
latency by up to 27$\times$.

\begin{figure}%
\centering%
\begin{tikzpicture}[gnuplot]
\tikzset{every node/.append style={font={\fontsize{8.0pt}{9.6pt}\selectfont}}}
\path (0.000,0.000) rectangle (8.458,3.048);
\gpcolor{color=gp lt color axes}
\gpsetlinetype{gp lt axes}
\gpsetdashtype{gp dt axes}
\gpsetlinewidth{0.50}
\draw[gp path] (1.201,0.787)--(7.722,0.787);
\gpcolor{color=gp lt color border}
\gpsetlinetype{gp lt border}
\gpsetdashtype{gp dt solid}
\gpsetlinewidth{1.00}
\draw[gp path] (1.201,0.787)--(1.291,0.787);
\node[gp node right] at (1.054,0.787) {$0$};
\gpcolor{color=gp lt color axes}
\gpsetlinetype{gp lt axes}
\gpsetdashtype{gp dt axes}
\gpsetlinewidth{0.50}
\draw[gp path] (1.201,1.142)--(7.722,1.142);
\gpcolor{color=gp lt color border}
\gpsetlinetype{gp lt border}
\gpsetdashtype{gp dt solid}
\gpsetlinewidth{1.00}
\draw[gp path] (1.201,1.142)--(1.291,1.142);
\node[gp node right] at (1.054,1.142) {$300$};
\gpcolor{color=gp lt color axes}
\gpsetlinetype{gp lt axes}
\gpsetdashtype{gp dt axes}
\gpsetlinewidth{0.50}
\draw[gp path] (1.201,1.498)--(7.722,1.498);
\gpcolor{color=gp lt color border}
\gpsetlinetype{gp lt border}
\gpsetdashtype{gp dt solid}
\gpsetlinewidth{1.00}
\draw[gp path] (1.201,1.498)--(1.291,1.498);
\node[gp node right] at (1.054,1.498) {$600$};
\gpcolor{color=gp lt color axes}
\gpsetlinetype{gp lt axes}
\gpsetdashtype{gp dt axes}
\gpsetlinewidth{0.50}
\draw[gp path] (1.201,1.853)--(7.722,1.853);
\gpcolor{color=gp lt color border}
\gpsetlinetype{gp lt border}
\gpsetdashtype{gp dt solid}
\gpsetlinewidth{1.00}
\draw[gp path] (1.201,1.853)--(1.291,1.853);
\node[gp node right] at (1.054,1.853) {$900$};
\gpcolor{color=gp lt color axes}
\gpsetlinetype{gp lt axes}
\gpsetdashtype{gp dt axes}
\gpsetlinewidth{0.50}
\draw[gp path] (1.201,2.209)--(1.348,2.209);
\draw[gp path] (2.851,2.209)--(7.722,2.209);
\gpcolor{color=gp lt color border}
\gpsetlinetype{gp lt border}
\gpsetdashtype{gp dt solid}
\gpsetlinewidth{1.00}
\draw[gp path] (1.201,2.209)--(1.291,2.209);
\node[gp node right] at (1.054,2.209) {$1200$};
\gpcolor{color=gp lt color axes}
\gpsetlinetype{gp lt axes}
\gpsetdashtype{gp dt axes}
\gpsetlinewidth{0.50}
\draw[gp path] (1.201,2.564)--(1.348,2.564);
\draw[gp path] (2.851,2.564)--(7.722,2.564);
\gpcolor{color=gp lt color border}
\gpsetlinetype{gp lt border}
\gpsetdashtype{gp dt solid}
\gpsetlinewidth{1.00}
\draw[gp path] (1.201,2.564)--(1.291,2.564);
\node[gp node right] at (1.054,2.564) {$1500$};
\draw[gp path] (1.201,0.787)--(1.201,0.877);
\node[gp node center] at (1.201,0.541) {$0$};
\draw[gp path] (2.387,0.787)--(2.387,0.877);
\node[gp node center] at (2.387,0.541) {$20$};
\draw[gp path] (3.572,0.787)--(3.572,0.877);
\node[gp node center] at (3.572,0.541) {$40$};
\draw[gp path] (4.758,0.787)--(4.758,0.877);
\node[gp node center] at (4.758,0.541) {$60$};
\draw[gp path] (5.944,0.787)--(5.944,0.877);
\node[gp node center] at (5.944,0.541) {$80$};
\draw[gp path] (7.129,0.787)--(7.129,0.877);
\node[gp node center] at (7.129,0.541) {$100$};
\draw[gp path] (7.722,0.787)--(7.632,0.787);
\draw[gp path] (7.722,1.142)--(7.632,1.142);
\draw[gp path] (7.722,1.498)--(7.632,1.498);
\draw[gp path] (7.722,1.853)--(7.632,1.853);
\draw[gp path] (7.722,2.209)--(7.632,2.209);
\draw[gp path] (7.722,2.564)--(7.632,2.564);
\draw[gp path] (1.201,2.801)--(1.201,0.787)--(7.722,0.787)--(7.722,2.801)--cycle;
\node[gp node right] at (1.789,2.498) {ovs};
\gpcolor{rgb color={0.471,0.369,0.941}}
\gpsetlinewidth{3.00}
\draw[gp path] (1.936,2.498)--(2.704,2.498);
\draw[gp path] (1.260,0.852)--(2.387,1.203)--(3.572,1.620)--(4.758,2.008)--(5.944,2.444)%
  --(7.129,2.615);
\gpsetpointsize{6.00}
\gp3point{gp mark 5}{}{(1.260,0.852)}
\gp3point{gp mark 5}{}{(2.387,1.203)}
\gp3point{gp mark 5}{}{(3.572,1.620)}
\gp3point{gp mark 5}{}{(4.758,2.008)}
\gp3point{gp mark 5}{}{(5.944,2.444)}
\gp3point{gp mark 5}{}{(7.129,2.615)}
\gp3point{gp mark 5}{}{(2.320,2.498)}
\gpcolor{color=gp lt color border}
\node[gp node right] at (1.789,2.252) {tas};
\gpcolor{rgb color={0.392,0.561,1.000}}
\draw[gp path] (1.936,2.252)--(2.704,2.252);
\draw[gp path] (1.260,0.814)--(2.387,0.843)--(3.572,0.871)--(4.758,1.582)--(5.944,2.151)%
  --(7.129,2.473);
\gp3point{gp mark 7}{}{(1.260,0.814)}
\gp3point{gp mark 7}{}{(2.387,0.843)}
\gp3point{gp mark 7}{}{(3.572,0.871)}
\gp3point{gp mark 7}{}{(4.758,1.582)}
\gp3point{gp mark 7}{}{(5.944,2.151)}
\gp3point{gp mark 7}{}{(7.129,2.473)}
\gp3point{gp mark 7}{}{(2.320,2.252)}
\gpcolor{color=gp lt color border}
\gpsetlinewidth{1.00}
\draw[gp path] (1.201,2.801)--(1.201,0.787)--(7.722,0.787)--(7.722,2.801)--cycle;
\node[gp node center,rotate=-270.0] at (0.232,1.671) {Victim 99p Lat [µs]};
\node[gp node center] at (4.461,0.172) {Adversary Connections};
\gpdefrectangularnode{gp plot 1}{\pgfpoint{1.201cm}{0.787cm}}{\pgfpoint{7.722cm}{2.801cm}}
\end{tikzpicture}
\caption{An adversary tenant can compromise the tail latency of a co-located
  victim tenant by creating contention in shared cores via connection multiplicity.}%
\label{fig:iso-conns}%
\Description[Shared datapath isolation]{TAS and OvS adversary increases number of connections
  and victim tail latency increases with connection count}%
\end{figure}

Overall, existing isolation approaches force a trade-off between the strong
tail latency isolation of coarse-grained sharing and the higher
efficiency of fine-grained sharing.
\section{Time Protection Challenges}
\label{sec:time_protection}
Foundational work framed time protection as the 
temporal analogue of memory protection:
if resources are shared, the system should prevent one protection 
domain from perturbing another in time~\cite{ge:time_protection,heiser:time_protection_proof}.
The existing literature investigated this abstraction in the context of security, 
where such perturbations enable side/covert channel attacks. 
Shared network datapaths exhibit the same fundamental coupling, 
but with the consequence of performance interference.
As presented earlier, tenants in a shared datapath inflate one another’s 
tail latency by competing for CPU time. 
This section frames time protection in the context of
shared network datapaths and identifies the challenges of
applying it to a network datapath with sub-$\mu$s tasks.
We identify four challenges: variable packet costs, cross-tenant batching,
tight overhead budgets, and delayed attribution on \texttt{RX}.

\subsection{C1: Packet Processing Times are Variable}
Packet processing cost varies from packet to packet, even for traffic of the
same volume.
Shared microarchitectural state, such as the cache, TLB, and branch predictor,
means that packets can have different costs depending on whether their flow
state is hot or cold, whether they take extra branches, or whether they incur
additional table probes or cache misses (\autoref{fig:iso-tas} and
\autoref{fig:iso-ovs}).
At microsecond timescales, these effects represent a large fraction of the
latency budget, making traffic volume a poor proxy for datapath CPU time.

\textbf{\textit{Equal traffic volume does not imply equal CPU time.}}
We first show this in OvS. \autoref{fig:iso-ovs} shows that equal traffic
volume does not imply equal datapath CPU time.
Victim and adversary run in separate VMs on separate cores, while sharing a
single OvS poll-mode driver (PMD) core.
Each VM runs memcached, and we rate limit the clients to equalise throughput
across the two tenants.
We install flow rules that recirculate adversary 
packets through OvS while victim packets follow the normal path.
We compare victim--victim, adversary--adversary, and victim--adversary
pairings.
The extra datapath work increases the average per-request cost from 1,858 cycles
for victim--victim to 4,670 cycles for victim--adversary when one victim is
replaced by the adversary. The more expensive
per-request costs inflate the victim's tail latency because the extra datapath
work creates contention that packet- and byte-based controls do not
capture. 

\begin{figure}%
\centering%
\begin{tikzpicture}[gnuplot]
\tikzset{every node/.append style={font={\fontsize{8.0pt}{9.6pt}\selectfont}}}
\path (0.000,0.000) rectangle (8.458,2.540);
\gpfill{rgb color={0.392,0.561,1.000}} (0.846,2.319)--(1.268,2.319)--(1.268,2.444)--(0.846,2.444)--cycle;
\gpcolor{color=gp lt color border}
\gpsetlinetype{gp lt border}
\gpsetdashtype{gp dt solid}
\gpsetlinewidth{1.00}
\draw[gp path] (0.846,2.319)--(0.846,2.444)--(1.268,2.444)--(1.268,2.319)--cycle;
\gpfill{rgb color={0.471,0.369,0.941}} (3.383,2.319)--(3.805,2.319)--(3.805,2.444)--(3.383,2.444)--cycle;
\draw[gp path] (3.383,2.319)--(3.383,2.444)--(3.805,2.444)--(3.805,2.319)--cycle;
\gpfill{rgb color={0.863,0.149,0.498}} (6.089,2.319)--(6.511,2.319)--(6.511,2.444)--(6.089,2.444)--cycle;
\draw[gp path] (6.089,2.319)--(6.089,2.444)--(6.511,2.444)--(6.511,2.319)--cycle;
\node[gp node left] at (1.438,2.349) {victim [v]};
\node[gp node left] at (3.975,2.349) {adversary [a]};
\node[gp node left] at (6.681,2.349) {aggregate};
\gpdefrectangularnode{gp plot 1}{\pgfpoint{0.000cm}{2.159cm}}{\pgfpoint{8.457cm}{2.539cm}}
\gpcolor{color=gp lt color axes}
\gpsetlinetype{gp lt axes}
\gpsetdashtype{gp dt axes}
\gpsetlinewidth{0.50}
\draw[gp path] (1.099,0.254)--(2.706,0.254);
\gpcolor{color=gp lt color border}
\gpsetlinetype{gp lt border}
\gpsetdashtype{gp dt solid}
\gpsetlinewidth{1.00}
\draw[gp path] (1.099,0.254)--(1.189,0.254);
\node[gp node right] at (0.952,0.254) {$0$};
\gpcolor{color=gp lt color axes}
\gpsetlinetype{gp lt axes}
\gpsetdashtype{gp dt axes}
\gpsetlinewidth{0.50}
\draw[gp path] (1.099,0.677)--(2.706,0.677);
\gpcolor{color=gp lt color border}
\gpsetlinetype{gp lt border}
\gpsetdashtype{gp dt solid}
\gpsetlinewidth{1.00}
\draw[gp path] (1.099,0.677)--(1.189,0.677);
\node[gp node right] at (0.952,0.677) {$20$};
\gpcolor{color=gp lt color axes}
\gpsetlinetype{gp lt axes}
\gpsetdashtype{gp dt axes}
\gpsetlinewidth{0.50}
\draw[gp path] (1.099,1.100)--(2.706,1.100);
\gpcolor{color=gp lt color border}
\gpsetlinetype{gp lt border}
\gpsetdashtype{gp dt solid}
\gpsetlinewidth{1.00}
\draw[gp path] (1.099,1.100)--(1.189,1.100);
\node[gp node right] at (0.952,1.100) {$40$};
\gpcolor{color=gp lt color axes}
\gpsetlinetype{gp lt axes}
\gpsetdashtype{gp dt axes}
\gpsetlinewidth{0.50}
\draw[gp path] (1.099,1.523)--(2.706,1.523);
\gpcolor{color=gp lt color border}
\gpsetlinetype{gp lt border}
\gpsetdashtype{gp dt solid}
\gpsetlinewidth{1.00}
\draw[gp path] (1.099,1.523)--(1.189,1.523);
\node[gp node right] at (0.952,1.523) {$60$};
\draw[gp path] (1.367,0.254)--(1.367,0.344);
\node[gp node center] at (1.367,0.008) {v-v};
\draw[gp path] (1.903,0.254)--(1.903,0.344);
\node[gp node center] at (1.903,0.008) {a-a};
\draw[gp path] (2.438,0.254)--(2.438,0.344);
\node[gp node center] at (2.438,0.008) {v-a};
\draw[gp path] (2.706,0.254)--(2.616,0.254);
\draw[gp path] (2.706,0.677)--(2.616,0.677);
\draw[gp path] (2.706,1.100)--(2.616,1.100);
\draw[gp path] (2.706,1.523)--(2.616,1.523);
\draw[gp path] (1.099,1.523)--(1.099,0.254)--(2.706,0.254)--(2.706,1.523)--cycle;
\gpfill{rgb color={0.392,0.561,1.000}} (1.238,0.254)--(1.368,0.254)--(1.368,1.312)--(1.238,1.312)--cycle;
\draw[gp path] (1.238,0.254)--(1.238,1.311)--(1.367,1.311)--(1.367,0.254)--cycle;
\gpfill{rgb color={0.392,0.561,1.000}} (1.367,0.254)--(1.496,0.254)--(1.496,1.313)--(1.367,1.313)--cycle;
\draw[gp path] (1.367,0.254)--(1.367,1.312)--(1.495,1.312)--(1.495,0.254)--cycle;
\gpfill{rgb color={0.471,0.369,0.941}} (1.774,0.254)--(1.904,0.254)--(1.904,1.308)--(1.774,1.308)--cycle;
\draw[gp path] (1.774,0.254)--(1.774,1.307)--(1.903,1.307)--(1.903,0.254)--cycle;
\gpfill{rgb color={0.471,0.369,0.941}} (1.903,0.254)--(2.032,0.254)--(2.032,1.310)--(1.903,1.310)--cycle;
\draw[gp path] (1.903,0.254)--(1.903,1.309)--(2.031,1.309)--(2.031,0.254)--cycle;
\gpfill{rgb color={0.392,0.561,1.000}} (2.310,0.254)--(2.439,0.254)--(2.439,1.313)--(2.310,1.313)--cycle;
\draw[gp path] (2.310,0.254)--(2.310,1.312)--(2.438,1.312)--(2.438,0.254)--cycle;
\gpfill{rgb color={0.471,0.369,0.941}} (2.438,0.254)--(2.568,0.254)--(2.568,1.312)--(2.438,1.312)--cycle;
\draw[gp path] (2.438,0.254)--(2.438,1.311)--(2.567,1.311)--(2.567,0.254)--cycle;
\draw[gp path] (1.099,1.523)--(1.099,0.254)--(2.706,0.254)--(2.706,1.523)--cycle;
\node[gp node center,rotate=-270.0] at (0.351,0.888) {Tput [KReq/s]};
\node[gp node center] at (1.902,1.892) {(a) Throughput};
\gpdefrectangularnode{gp plot 2}{\pgfpoint{1.099cm}{0.254cm}}{\pgfpoint{2.706cm}{1.523cm}}
\gpcolor{color=gp lt color axes}
\gpsetlinetype{gp lt axes}
\gpsetdashtype{gp dt axes}
\gpsetlinewidth{0.50}
\draw[gp path] (3.806,0.254)--(5.412,0.254);
\gpcolor{color=gp lt color border}
\gpsetlinetype{gp lt border}
\gpsetdashtype{gp dt solid}
\gpsetlinewidth{1.00}
\draw[gp path] (3.806,0.254)--(3.896,0.254);
\node[gp node right] at (3.659,0.254) {$0$};
\gpcolor{color=gp lt color axes}
\gpsetlinetype{gp lt axes}
\gpsetdashtype{gp dt axes}
\gpsetlinewidth{0.50}
\draw[gp path] (3.806,0.677)--(5.412,0.677);
\gpcolor{color=gp lt color border}
\gpsetlinetype{gp lt border}
\gpsetdashtype{gp dt solid}
\gpsetlinewidth{1.00}
\draw[gp path] (3.806,0.677)--(3.896,0.677);
\node[gp node right] at (3.659,0.677) {$30$};
\gpcolor{color=gp lt color axes}
\gpsetlinetype{gp lt axes}
\gpsetdashtype{gp dt axes}
\gpsetlinewidth{0.50}
\draw[gp path] (3.806,1.100)--(5.412,1.100);
\gpcolor{color=gp lt color border}
\gpsetlinetype{gp lt border}
\gpsetdashtype{gp dt solid}
\gpsetlinewidth{1.00}
\draw[gp path] (3.806,1.100)--(3.896,1.100);
\node[gp node right] at (3.659,1.100) {$60$};
\gpcolor{color=gp lt color axes}
\gpsetlinetype{gp lt axes}
\gpsetdashtype{gp dt axes}
\gpsetlinewidth{0.50}
\draw[gp path] (3.806,1.523)--(5.412,1.523);
\gpcolor{color=gp lt color border}
\gpsetlinetype{gp lt border}
\gpsetdashtype{gp dt solid}
\gpsetlinewidth{1.00}
\draw[gp path] (3.806,1.523)--(3.896,1.523);
\node[gp node right] at (3.659,1.523) {$90$};
\draw[gp path] (4.341,0.254)--(4.341,0.344);
\node[gp node center] at (4.341,0.008) {v-v};
\draw[gp path] (4.877,0.254)--(4.877,0.344);
\node[gp node center] at (4.877,0.008) {v-a};
\draw[gp path] (5.412,0.254)--(5.322,0.254);
\draw[gp path] (5.412,0.677)--(5.322,0.677);
\draw[gp path] (5.412,1.100)--(5.322,1.100);
\draw[gp path] (5.412,1.523)--(5.322,1.523);
\draw[gp path] (3.806,1.523)--(3.806,0.254)--(5.412,0.254)--(5.412,1.523)--cycle;
\gpfill{rgb color={0.392,0.561,1.000}} (4.221,0.254)--(4.463,0.254)--(4.463,0.692)--(4.221,0.692)--cycle;
\draw[gp path] (4.221,0.254)--(4.221,0.691)--(4.462,0.691)--(4.462,0.254)--cycle;
\gpfill{rgb color={0.392,0.561,1.000}} (4.756,0.254)--(4.998,0.254)--(4.998,1.369)--(4.756,1.369)--cycle;
\draw[gp path] (4.756,0.254)--(4.756,1.368)--(4.997,1.368)--(4.997,0.254)--cycle;
\draw[gp path] (3.806,1.523)--(3.806,0.254)--(5.412,0.254)--(5.412,1.523)--cycle;
\node[gp node center,rotate=-270.0] at (3.058,0.888) {Lat [us]};
\node[gp node center] at (4.609,1.892) {(b) 99p Latency};
\gpdefrectangularnode{gp plot 3}{\pgfpoint{3.806cm}{0.254cm}}{\pgfpoint{5.412cm}{1.523cm}}
\gpcolor{color=gp lt color axes}
\gpsetlinetype{gp lt axes}
\gpsetdashtype{gp dt axes}
\gpsetlinewidth{0.50}
\draw[gp path] (6.512,0.254)--(8.118,0.254);
\gpcolor{color=gp lt color border}
\gpsetlinetype{gp lt border}
\gpsetdashtype{gp dt solid}
\gpsetlinewidth{1.00}
\draw[gp path] (6.512,0.254)--(6.602,0.254);
\node[gp node right] at (6.365,0.254) {$0$};
\gpcolor{color=gp lt color axes}
\gpsetlinetype{gp lt axes}
\gpsetdashtype{gp dt axes}
\gpsetlinewidth{0.50}
\draw[gp path] (6.512,0.571)--(8.118,0.571);
\gpcolor{color=gp lt color border}
\gpsetlinetype{gp lt border}
\gpsetdashtype{gp dt solid}
\gpsetlinewidth{1.00}
\draw[gp path] (6.512,0.571)--(6.602,0.571);
\node[gp node right] at (6.365,0.571) {$2$};
\gpcolor{color=gp lt color axes}
\gpsetlinetype{gp lt axes}
\gpsetdashtype{gp dt axes}
\gpsetlinewidth{0.50}
\draw[gp path] (6.512,0.889)--(8.118,0.889);
\gpcolor{color=gp lt color border}
\gpsetlinetype{gp lt border}
\gpsetdashtype{gp dt solid}
\gpsetlinewidth{1.00}
\draw[gp path] (6.512,0.889)--(6.602,0.889);
\node[gp node right] at (6.365,0.889) {$4$};
\gpcolor{color=gp lt color axes}
\gpsetlinetype{gp lt axes}
\gpsetdashtype{gp dt axes}
\gpsetlinewidth{0.50}
\draw[gp path] (6.512,1.206)--(8.118,1.206);
\gpcolor{color=gp lt color border}
\gpsetlinetype{gp lt border}
\gpsetdashtype{gp dt solid}
\gpsetlinewidth{1.00}
\draw[gp path] (6.512,1.206)--(6.602,1.206);
\node[gp node right] at (6.365,1.206) {$6$};
\gpcolor{color=gp lt color axes}
\gpsetlinetype{gp lt axes}
\gpsetdashtype{gp dt axes}
\gpsetlinewidth{0.50}
\draw[gp path] (6.512,1.523)--(8.118,1.523);
\gpcolor{color=gp lt color border}
\gpsetlinetype{gp lt border}
\gpsetdashtype{gp dt solid}
\gpsetlinewidth{1.00}
\draw[gp path] (6.512,1.523)--(6.602,1.523);
\node[gp node right] at (6.365,1.523) {$8$};
\draw[gp path] (6.780,0.254)--(6.780,0.344);
\node[gp node center] at (6.780,0.008) {v-v};
\draw[gp path] (7.315,0.254)--(7.315,0.344);
\node[gp node center] at (7.315,0.008) {a-a};
\draw[gp path] (7.850,0.254)--(7.850,0.344);
\node[gp node center] at (7.850,0.008) {v-a};
\draw[gp path] (8.118,0.254)--(8.028,0.254);
\draw[gp path] (8.118,0.571)--(8.028,0.571);
\draw[gp path] (8.118,0.889)--(8.028,0.889);
\draw[gp path] (8.118,1.206)--(8.028,1.206);
\draw[gp path] (8.118,1.523)--(8.028,1.523);
\draw[gp path] (6.512,1.523)--(6.512,0.254)--(8.118,0.254)--(8.118,1.523)--cycle;
\gpfill{rgb color={0.863,0.149,0.498}} (6.646,0.254)--(6.915,0.254)--(6.915,0.493)--(6.646,0.493)--cycle;
\draw[gp path] (6.646,0.254)--(6.646,0.492)--(6.914,0.492)--(6.914,0.254)--cycle;
\gpfill{rgb color={0.863,0.149,0.498}} (7.181,0.254)--(7.450,0.254)--(7.450,1.286)--(7.181,1.286)--cycle;
\draw[gp path] (7.181,0.254)--(7.181,1.285)--(7.449,1.285)--(7.449,0.254)--cycle;
\gpfill{rgb color={0.863,0.149,0.498}} (7.717,0.254)--(7.985,0.254)--(7.985,0.890)--(7.717,0.890)--cycle;
\draw[gp path] (7.717,0.254)--(7.717,0.889)--(7.984,0.889)--(7.984,0.254)--cycle;
\draw[gp path] (6.512,1.523)--(6.512,0.254)--(8.118,0.254)--(8.118,1.523)--cycle;
\node[gp node center,rotate=-270.0] at (5.911,0.888) {Passes/Req};
\node[gp node center] at (7.315,1.892) {(c) Datapath Passes};
\gpdefrectangularnode{gp plot 4}{\pgfpoint{6.512cm}{0.254cm}}{\pgfpoint{8.118cm}{1.523cm}}
\end{tikzpicture}
\caption{Throughput, latency and packet datapath passes for different
  application pairings. An adversary forces its packets to recirculate
  the OvS datapath,
  increasing processing cost and victim tail latency despite both victim and
  adversary being rate limited to the same number of requests per second.}%
\Description[OvS datapath isolation]{Victim and adversary have the same
  throughput. Victim 99p latency increases. Adversary increases number of
  datapath passes per request. Adversary increases number of cycles per
  request.}%
\label{fig:iso-ovs}%
\end{figure}

The same problem appears even in TAS, where packet processing time varies with
working-set size and locality rather than an explicit change in the datapath.
In \autoref{fig:iso-tas}, we modify TAS to provide round-robin transmit
fairness and pin memcached servers to isolated cores.
The victim maintains five client connections, whereas the adversary opens 5000.
As in OvS, both tenants are rate limited on the client to equalise aggregate
throughput and isolate the effect of connection count.
Two victim applications have low tail latency and L1 misses.
Replacing one victim with the adversary increases tail latency despite identical
throughput: the adversary's larger connection working set increases
cache pressure and per-packet cost rises from 628 cycles per request with victim--victim
to 943 cycles with victim--adversary.

\begin{figure}%
\centering%
\begin{tikzpicture}[gnuplot]
\tikzset{every node/.append style={font={\fontsize{8.0pt}{9.6pt}\selectfont}}}
\path (0.000,0.000) rectangle (8.458,2.540);
\gpfill{rgb color={0.392,0.561,1.000}} (0.846,2.319)--(1.268,2.319)--(1.268,2.444)--(0.846,2.444)--cycle;
\gpcolor{color=gp lt color border}
\gpsetlinetype{gp lt border}
\gpsetdashtype{gp dt solid}
\gpsetlinewidth{1.00}
\draw[gp path] (0.846,2.319)--(0.846,2.444)--(1.268,2.444)--(1.268,2.319)--cycle;
\gpfill{rgb color={0.471,0.369,0.941}} (3.383,2.319)--(3.805,2.319)--(3.805,2.444)--(3.383,2.444)--cycle;
\draw[gp path] (3.383,2.319)--(3.383,2.444)--(3.805,2.444)--(3.805,2.319)--cycle;
\gpfill{rgb color={0.863,0.149,0.498}} (6.089,2.319)--(6.511,2.319)--(6.511,2.444)--(6.089,2.444)--cycle;
\draw[gp path] (6.089,2.319)--(6.089,2.444)--(6.511,2.444)--(6.511,2.319)--cycle;
\node[gp node left] at (1.438,2.349) {victim [v]};
\node[gp node left] at (3.975,2.349) {adversary [a]};
\node[gp node left] at (6.681,2.349) {aggregate};
\gpdefrectangularnode{gp plot 1}{\pgfpoint{0.000cm}{2.159cm}}{\pgfpoint{8.457cm}{2.539cm}}
\gpcolor{color=gp lt color axes}
\gpsetlinetype{gp lt axes}
\gpsetdashtype{gp dt axes}
\gpsetlinewidth{0.50}
\draw[gp path] (1.099,0.254)--(2.706,0.254);
\gpcolor{color=gp lt color border}
\gpsetlinetype{gp lt border}
\gpsetdashtype{gp dt solid}
\gpsetlinewidth{1.00}
\draw[gp path] (1.099,0.254)--(1.189,0.254);
\node[gp node right] at (0.952,0.254) {$0$};
\gpcolor{color=gp lt color axes}
\gpsetlinetype{gp lt axes}
\gpsetdashtype{gp dt axes}
\gpsetlinewidth{0.50}
\draw[gp path] (1.099,0.571)--(2.706,0.571);
\gpcolor{color=gp lt color border}
\gpsetlinetype{gp lt border}
\gpsetdashtype{gp dt solid}
\gpsetlinewidth{1.00}
\draw[gp path] (1.099,0.571)--(1.189,0.571);
\node[gp node right] at (0.952,0.571) {$150$};
\gpcolor{color=gp lt color axes}
\gpsetlinetype{gp lt axes}
\gpsetdashtype{gp dt axes}
\gpsetlinewidth{0.50}
\draw[gp path] (1.099,0.889)--(2.706,0.889);
\gpcolor{color=gp lt color border}
\gpsetlinetype{gp lt border}
\gpsetdashtype{gp dt solid}
\gpsetlinewidth{1.00}
\draw[gp path] (1.099,0.889)--(1.189,0.889);
\node[gp node right] at (0.952,0.889) {$300$};
\gpcolor{color=gp lt color axes}
\gpsetlinetype{gp lt axes}
\gpsetdashtype{gp dt axes}
\gpsetlinewidth{0.50}
\draw[gp path] (1.099,1.206)--(2.706,1.206);
\gpcolor{color=gp lt color border}
\gpsetlinetype{gp lt border}
\gpsetdashtype{gp dt solid}
\gpsetlinewidth{1.00}
\draw[gp path] (1.099,1.206)--(1.189,1.206);
\node[gp node right] at (0.952,1.206) {$450$};
\gpcolor{color=gp lt color axes}
\gpsetlinetype{gp lt axes}
\gpsetdashtype{gp dt axes}
\gpsetlinewidth{0.50}
\draw[gp path] (1.099,1.523)--(2.706,1.523);
\gpcolor{color=gp lt color border}
\gpsetlinetype{gp lt border}
\gpsetdashtype{gp dt solid}
\gpsetlinewidth{1.00}
\draw[gp path] (1.099,1.523)--(1.189,1.523);
\node[gp node right] at (0.952,1.523) {$600$};
\draw[gp path] (1.367,0.254)--(1.367,0.344);
\node[gp node center] at (1.367,0.008) {v-v};
\draw[gp path] (1.903,0.254)--(1.903,0.344);
\node[gp node center] at (1.903,0.008) {a-a};
\draw[gp path] (2.438,0.254)--(2.438,0.344);
\node[gp node center] at (2.438,0.008) {v-a};
\draw[gp path] (2.706,0.254)--(2.616,0.254);
\draw[gp path] (2.706,0.571)--(2.616,0.571);
\draw[gp path] (2.706,0.889)--(2.616,0.889);
\draw[gp path] (2.706,1.206)--(2.616,1.206);
\draw[gp path] (2.706,1.523)--(2.616,1.523);
\draw[gp path] (1.099,1.523)--(1.099,0.254)--(2.706,0.254)--(2.706,1.523)--cycle;
\gpfill{rgb color={0.392,0.561,1.000}} (1.238,0.254)--(1.368,0.254)--(1.368,1.313)--(1.238,1.313)--cycle;
\draw[gp path] (1.238,0.254)--(1.238,1.312)--(1.367,1.312)--(1.367,0.254)--cycle;
\gpfill{rgb color={0.392,0.561,1.000}} (1.367,0.254)--(1.496,0.254)--(1.496,1.312)--(1.367,1.312)--cycle;
\draw[gp path] (1.367,0.254)--(1.367,1.311)--(1.495,1.311)--(1.495,0.254)--cycle;
\gpfill{rgb color={0.471,0.369,0.941}} (1.774,0.254)--(1.904,0.254)--(1.904,1.259)--(1.774,1.259)--cycle;
\draw[gp path] (1.774,0.254)--(1.774,1.258)--(1.903,1.258)--(1.903,0.254)--cycle;
\gpfill{rgb color={0.471,0.369,0.941}} (1.903,0.254)--(2.032,0.254)--(2.032,1.260)--(1.903,1.260)--cycle;
\draw[gp path] (1.903,0.254)--(1.903,1.259)--(2.031,1.259)--(2.031,0.254)--cycle;
\gpfill{rgb color={0.392,0.561,1.000}} (2.310,0.254)--(2.439,0.254)--(2.439,1.313)--(2.310,1.313)--cycle;
\draw[gp path] (2.310,0.254)--(2.310,1.312)--(2.438,1.312)--(2.438,0.254)--cycle;
\gpfill{rgb color={0.471,0.369,0.941}} (2.438,0.254)--(2.568,0.254)--(2.568,1.255)--(2.438,1.255)--cycle;
\draw[gp path] (2.438,0.254)--(2.438,1.254)--(2.567,1.254)--(2.567,0.254)--cycle;
\draw[gp path] (1.099,1.523)--(1.099,0.254)--(2.706,0.254)--(2.706,1.523)--cycle;
\node[gp node center,rotate=-270.0] at (0.204,0.888) {Tput [KReq/s]};
\node[gp node center] at (1.902,1.892) {(a) Throughput};
\gpdefrectangularnode{gp plot 2}{\pgfpoint{1.099cm}{0.254cm}}{\pgfpoint{2.706cm}{1.523cm}}
\gpcolor{color=gp lt color axes}
\gpsetlinetype{gp lt axes}
\gpsetdashtype{gp dt axes}
\gpsetlinewidth{0.50}
\draw[gp path] (3.806,0.254)--(5.412,0.254);
\gpcolor{color=gp lt color border}
\gpsetlinetype{gp lt border}
\gpsetdashtype{gp dt solid}
\gpsetlinewidth{1.00}
\draw[gp path] (3.806,0.254)--(3.896,0.254);
\node[gp node right] at (3.659,0.254) {$0$};
\gpcolor{color=gp lt color axes}
\gpsetlinetype{gp lt axes}
\gpsetdashtype{gp dt axes}
\gpsetlinewidth{0.50}
\draw[gp path] (3.806,0.508)--(5.412,0.508);
\gpcolor{color=gp lt color border}
\gpsetlinetype{gp lt border}
\gpsetdashtype{gp dt solid}
\gpsetlinewidth{1.00}
\draw[gp path] (3.806,0.508)--(3.896,0.508);
\node[gp node right] at (3.659,0.508) {$30$};
\gpcolor{color=gp lt color axes}
\gpsetlinetype{gp lt axes}
\gpsetdashtype{gp dt axes}
\gpsetlinewidth{0.50}
\draw[gp path] (3.806,0.762)--(5.412,0.762);
\gpcolor{color=gp lt color border}
\gpsetlinetype{gp lt border}
\gpsetdashtype{gp dt solid}
\gpsetlinewidth{1.00}
\draw[gp path] (3.806,0.762)--(3.896,0.762);
\node[gp node right] at (3.659,0.762) {$60$};
\gpcolor{color=gp lt color axes}
\gpsetlinetype{gp lt axes}
\gpsetdashtype{gp dt axes}
\gpsetlinewidth{0.50}
\draw[gp path] (3.806,1.015)--(5.412,1.015);
\gpcolor{color=gp lt color border}
\gpsetlinetype{gp lt border}
\gpsetdashtype{gp dt solid}
\gpsetlinewidth{1.00}
\draw[gp path] (3.806,1.015)--(3.896,1.015);
\node[gp node right] at (3.659,1.015) {$90$};
\gpcolor{color=gp lt color axes}
\gpsetlinetype{gp lt axes}
\gpsetdashtype{gp dt axes}
\gpsetlinewidth{0.50}
\draw[gp path] (3.806,1.269)--(5.412,1.269);
\gpcolor{color=gp lt color border}
\gpsetlinetype{gp lt border}
\gpsetdashtype{gp dt solid}
\gpsetlinewidth{1.00}
\draw[gp path] (3.806,1.269)--(3.896,1.269);
\node[gp node right] at (3.659,1.269) {$120$};
\gpcolor{color=gp lt color axes}
\gpsetlinetype{gp lt axes}
\gpsetdashtype{gp dt axes}
\gpsetlinewidth{0.50}
\draw[gp path] (3.806,1.523)--(5.412,1.523);
\gpcolor{color=gp lt color border}
\gpsetlinetype{gp lt border}
\gpsetdashtype{gp dt solid}
\gpsetlinewidth{1.00}
\draw[gp path] (3.806,1.523)--(3.896,1.523);
\node[gp node right] at (3.659,1.523) {$150$};
\draw[gp path] (4.341,0.254)--(4.341,0.344);
\node[gp node center] at (4.341,0.008) {v-v};
\draw[gp path] (4.877,0.254)--(4.877,0.344);
\node[gp node center] at (4.877,0.008) {v-a};
\draw[gp path] (5.412,0.254)--(5.322,0.254);
\draw[gp path] (5.412,0.508)--(5.322,0.508);
\draw[gp path] (5.412,0.762)--(5.322,0.762);
\draw[gp path] (5.412,1.015)--(5.322,1.015);
\draw[gp path] (5.412,1.269)--(5.322,1.269);
\draw[gp path] (5.412,1.523)--(5.322,1.523);
\draw[gp path] (3.806,1.523)--(3.806,0.254)--(5.412,0.254)--(5.412,1.523)--cycle;
\gpfill{rgb color={0.392,0.561,1.000}} (4.221,0.254)--(4.463,0.254)--(4.463,0.788)--(4.221,0.788)--cycle;
\draw[gp path] (4.221,0.254)--(4.221,0.787)--(4.462,0.787)--(4.462,0.254)--cycle;
\gpfill{rgb color={0.392,0.561,1.000}} (4.756,0.254)--(4.998,0.254)--(4.998,1.284)--(4.756,1.284)--cycle;
\draw[gp path] (4.756,0.254)--(4.756,1.283)--(4.997,1.283)--(4.997,0.254)--cycle;
\draw[gp path] (3.806,1.523)--(3.806,0.254)--(5.412,0.254)--(5.412,1.523)--cycle;
\node[gp node center,rotate=-270.0] at (2.955,0.888) {Lat [µs]};
\node[gp node center] at (4.609,1.892) {(b) 99p Latency};
\gpdefrectangularnode{gp plot 3}{\pgfpoint{3.806cm}{0.254cm}}{\pgfpoint{5.412cm}{1.523cm}}
\gpcolor{color=gp lt color axes}
\gpsetlinetype{gp lt axes}
\gpsetdashtype{gp dt axes}
\gpsetlinewidth{0.50}
\draw[gp path] (6.512,0.254)--(8.118,0.254);
\gpcolor{color=gp lt color border}
\gpsetlinetype{gp lt border}
\gpsetdashtype{gp dt solid}
\gpsetlinewidth{1.00}
\draw[gp path] (6.512,0.254)--(6.602,0.254);
\node[gp node right] at (6.365,0.254) {$0$};
\gpcolor{color=gp lt color axes}
\gpsetlinetype{gp lt axes}
\gpsetdashtype{gp dt axes}
\gpsetlinewidth{0.50}
\draw[gp path] (6.512,0.508)--(8.118,0.508);
\gpcolor{color=gp lt color border}
\gpsetlinetype{gp lt border}
\gpsetdashtype{gp dt solid}
\gpsetlinewidth{1.00}
\draw[gp path] (6.512,0.508)--(6.602,0.508);
\node[gp node right] at (6.365,0.508) {$5$};
\gpcolor{color=gp lt color axes}
\gpsetlinetype{gp lt axes}
\gpsetdashtype{gp dt axes}
\gpsetlinewidth{0.50}
\draw[gp path] (6.512,0.762)--(8.118,0.762);
\gpcolor{color=gp lt color border}
\gpsetlinetype{gp lt border}
\gpsetdashtype{gp dt solid}
\gpsetlinewidth{1.00}
\draw[gp path] (6.512,0.762)--(6.602,0.762);
\node[gp node right] at (6.365,0.762) {$10$};
\gpcolor{color=gp lt color axes}
\gpsetlinetype{gp lt axes}
\gpsetdashtype{gp dt axes}
\gpsetlinewidth{0.50}
\draw[gp path] (6.512,1.015)--(8.118,1.015);
\gpcolor{color=gp lt color border}
\gpsetlinetype{gp lt border}
\gpsetdashtype{gp dt solid}
\gpsetlinewidth{1.00}
\draw[gp path] (6.512,1.015)--(6.602,1.015);
\node[gp node right] at (6.365,1.015) {$15$};
\gpcolor{color=gp lt color axes}
\gpsetlinetype{gp lt axes}
\gpsetdashtype{gp dt axes}
\gpsetlinewidth{0.50}
\draw[gp path] (6.512,1.269)--(8.118,1.269);
\gpcolor{color=gp lt color border}
\gpsetlinetype{gp lt border}
\gpsetdashtype{gp dt solid}
\gpsetlinewidth{1.00}
\draw[gp path] (6.512,1.269)--(6.602,1.269);
\node[gp node right] at (6.365,1.269) {$20$};
\gpcolor{color=gp lt color axes}
\gpsetlinetype{gp lt axes}
\gpsetdashtype{gp dt axes}
\gpsetlinewidth{0.50}
\draw[gp path] (6.512,1.523)--(8.118,1.523);
\gpcolor{color=gp lt color border}
\gpsetlinetype{gp lt border}
\gpsetdashtype{gp dt solid}
\gpsetlinewidth{1.00}
\draw[gp path] (6.512,1.523)--(6.602,1.523);
\node[gp node right] at (6.365,1.523) {$25$};
\draw[gp path] (6.780,0.254)--(6.780,0.344);
\node[gp node center] at (6.780,0.008) {v-v};
\draw[gp path] (7.315,0.254)--(7.315,0.344);
\node[gp node center] at (7.315,0.008) {a-a};
\draw[gp path] (7.850,0.254)--(7.850,0.344);
\node[gp node center] at (7.850,0.008) {v-a};
\draw[gp path] (8.118,0.254)--(8.028,0.254);
\draw[gp path] (8.118,0.508)--(8.028,0.508);
\draw[gp path] (8.118,0.762)--(8.028,0.762);
\draw[gp path] (8.118,1.015)--(8.028,1.015);
\draw[gp path] (8.118,1.269)--(8.028,1.269);
\draw[gp path] (8.118,1.523)--(8.028,1.523);
\draw[gp path] (6.512,1.523)--(6.512,0.254)--(8.118,0.254)--(8.118,1.523)--cycle;
\gpfill{rgb color={0.863,0.149,0.498}} (6.646,0.254)--(6.915,0.254)--(6.915,0.585)--(6.646,0.585)--cycle;
\draw[gp path] (6.646,0.254)--(6.646,0.584)--(6.914,0.584)--(6.914,0.254)--cycle;
\gpfill{rgb color={0.863,0.149,0.498}} (7.181,0.254)--(7.450,0.254)--(7.450,1.398)--(7.181,1.398)--cycle;
\draw[gp path] (7.181,0.254)--(7.181,1.397)--(7.449,1.397)--(7.449,0.254)--cycle;
\gpfill{rgb color={0.863,0.149,0.498}} (7.717,0.254)--(7.985,0.254)--(7.985,1.007)--(7.717,1.007)--cycle;
\draw[gp path] (7.717,0.254)--(7.717,1.006)--(7.984,1.006)--(7.984,0.254)--cycle;
\draw[gp path] (6.512,1.523)--(6.512,0.254)--(8.118,0.254)--(8.118,1.523)--cycle;
\node[gp node center,rotate=-270.0] at (5.764,0.888) {Misses/Req};
\node[gp node center] at (7.315,1.892) {(c) L1 Misses};
\gpdefrectangularnode{gp plot 4}{\pgfpoint{6.512cm}{0.254cm}}{\pgfpoint{8.118cm}{1.523cm}}
\end{tikzpicture}
\caption{Throughput, latency and L1 cache misses for different
  application pairings. An adversary intensifies its cache footprint in TAS with many
  connections and increases L1 misses and contention. Volume-based fairness
  mechanisms fail to detect this contention, and the latency of a victim
  sharing the datapath core increases.}%
\Description[TAS datapath isolation]{Victim and adversary have the same
  throughput. Victim 99p latency increases. Adversary increases number of L1
  misses per request. Adversary increases number of cycles per request.}%
\label{fig:iso-tas}%
\end{figure}

Across both systems, equal traffic volume can mask substantially different
datapath CPU costs.
Time protection must therefore account and enforce CPU time directly.

\subsection{C2: Cross-tenant Batching is Required for Efficiency
but Complicates Protection}
High-performance datapaths batch packets from multiple tenants
to amortise fixed overheads in polling, 
descriptor handling, and per-iteration 
bookkeeping~\cite{borisov:batching_performance_estimation,rizzo:netmap,nagle:tcp_cc,belay:ix,miao:smart_batching}. 
This improves throughput; however,
the batch becomes the unit of execution and the core 
stops interleaving tenants at fine granularity.
Enforcing protection per-packet forces the datapath
either to break batches and lose efficiency or to allow
batches to overrun budgets and weaken protection;
the same mechanism that improves efficiency 
enlarges the atomic delay a tenant can impose on its peers.
Time protection has to balance the efficiency extracted from batching
and bound tail latency interference.

\subsection{C3: The Datapath Can Only Tolerate Low Overhead}
Microsecond scale datapaths leave almost no slack for control
logic~\cite{liu:ceio,cai:host_stack_overheads,kaufmann:tas,preeti:toasty}.
However, the overheads of mechanisms that enforce fair sharing and protection on the datapath, 
such as preemption~\cite{humphries:against_ctx_switches,belay:ix,peter:arrakis,ousterhout:shenango}, 
queueing~\cite{fried:junction}, and synchronisation~\cite{kaufmann:tas,jeong:mtcp,han:megapipe},
rarely remain constant or have low overhead because 
they induce cross-core communication, cache-line bouncing, 
and scheduler work under load, which in turn increases tail 
latency.
Therefore, time protection mechanisms must add
constant and bounded overhead.

\subsection{C4: Delayed Attribution Prevents Early Scheduling at \texttt{RX}}
At receive, the datapath must parse packet headers to 
determine which tenants to charge.
In contrast, during transmission, a packet is already
associated with a queue, flow, or context linked
to a tenant, allowing immediate attribution.
This creates an asymmetry between transmission and
reception, where enforcing protection for receive
typically requires additional indirections or re-queueing,
a source of cache misses and latency variance~\cite{ghasemirahni:reframer}.
For example, the datapath could insert packets from tenants
that violate protection into a queue for later processing,
but at the microsecond scale the additional cache miss
from delayed processing becomes expensive. 
Time protection must thus handle this asymmetry with minimal
overhead for protected tenants.

\section{\sys Time Protection}
\label{sec:design}
\begin{figure}%
\centering
\includegraphics[width=0.48\textwidth]{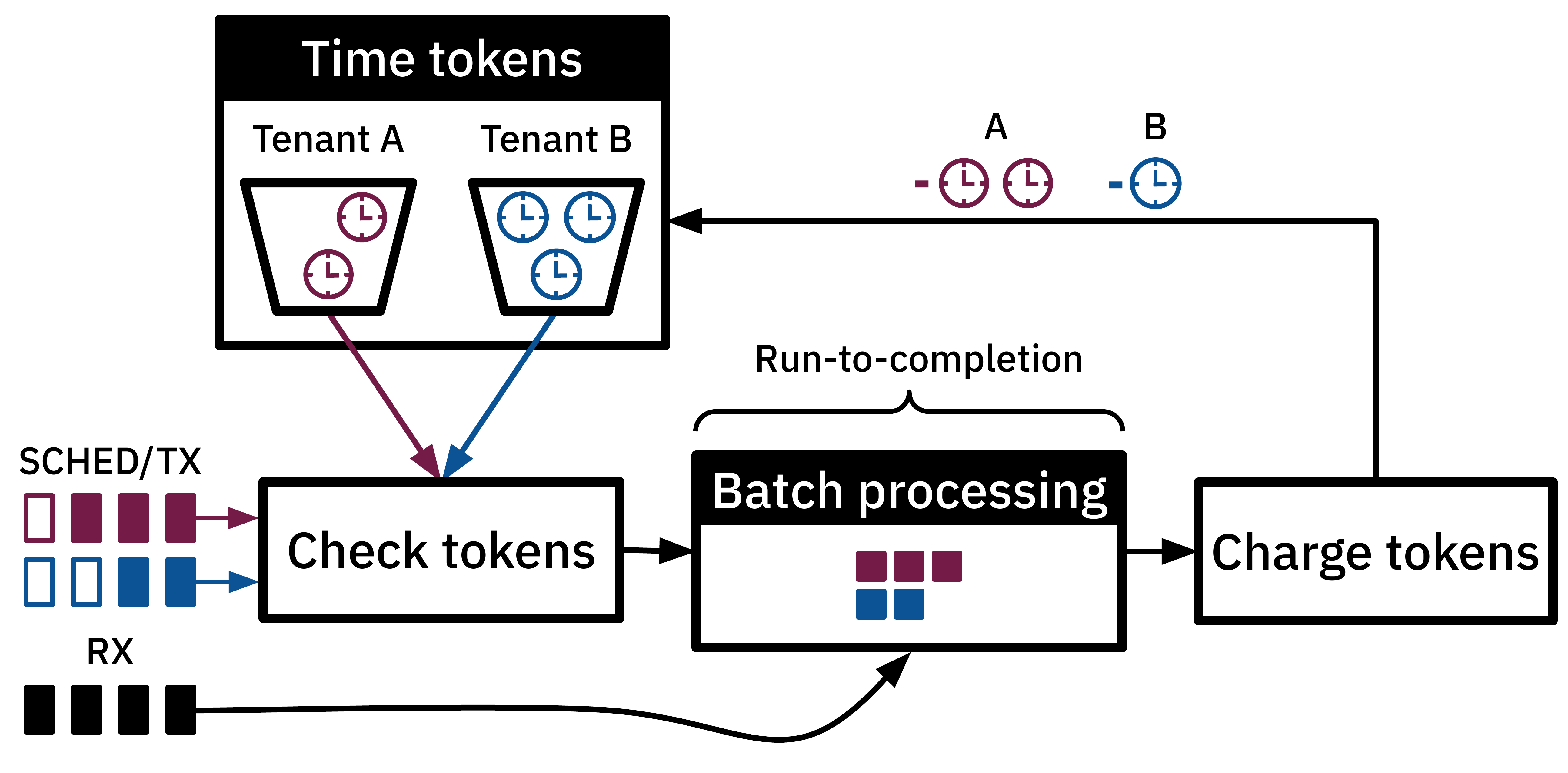}%
\caption{Intervention points enforce token availability
at \texttt{SCHED} and \texttt{TX}. 
The datapath executes
each batch run-to-completion and charges tenants for the
time spent on their packets.
The \texttt{RX} path accounts CPU time and enforces
protection at the next intervention point because
it only identifies the destination tenant for a packet
after processing headers.}%
\Description[\sys time protection visualisation.]{Fully described in the text.}%
\label{fig:virtuoso_overview}%
\end{figure}

In this section, we present the design of \sys, our time protection mechanism
for shared network datapaths.
\autoref{fig:virtuoso_overview} summarises its accounting and enforcement
mechanisms, and \autoref{tab:mechanisms} maps them to the design principles and
challenges.
Later, we instantiate \sys in TAS (\autoref{sec:impl}) and evaluate it in
\autoref{sec:eval}.

\begin{table*}[t]%
  \centering%
  \captionsetup{skip=1pt}%
  \small%
  \begin{tabular}{@{}llll@{}}%
    \toprule
    \textbf{Mechanism} & \textbf{Challenges} & \textbf{Principles} & \textbf{Role} \\
    \midrule
    Accounting          & C1, C4             & P1, P3              & Charge CPU time in cycles \\
    Scheduling          & C2, C3             & P1                  & Batch scheduling with bounded deficit \\
    Enforcement         & C2, C3, C4         & P2, P3              & Gate work at \texttt{TX}/\texttt{SCHED} and defer it for \texttt{RX} \\
    Budget Refill       & C3                 & P1, P2              & Calculate and refill budgets with low overhead \\
    \bottomrule
  \end{tabular}%
  \vspace{1em}%
  \caption{Mapping of mechanisms to challenges and principles.}%
  \label{tab:mechanisms}%
\end{table*}%

\subsection{Design Principles}
\label{subsec:principles}
Given C1--C4, our design adopts three principles for enforcing time protection
in shared datapaths.
They preserve batching, accommodate the asymmetry between the \texttt{RX} and \texttt{TX} paths,
and keep datapath overhead low.

\textbf{\textit{P1: Use a token bucket with time as tokens together
    with deficit round-robin scheduling.}}
Following C1, our design uses a token bucket, represented as a budget,
to rate limit tenants.
Tokens in a budget represent CPU cycles that a tenant can spend in the
datapath.
After each batch, the datapath measures the CPU time spent on behalf of each
tenant and deducts the corresponding number of tokens from that tenant's
budget.
To schedule tenants fairly in time while preserving run-to-completion batching,
we combine these per-tenant budgets with deficit round-robin scheduling, carrying forward
bounded time debt across rounds (C2).

\textbf{\textit{P2: Enforce protection at intervention points 
with bounded deficit and avoid preemption.}}
Intervention points are datapath control points at which
the system can attribute work to a tenant and enforce
protection before consuming more CPU time.
They use the token budgets assigned
to each tenant to make scheduling decisions and
enforce protection.
Intervention points execute whole batches, 
so a tenant may temporarily exceed its budget. We limit this 
overshoot with bounded deficit, meaning that tenants can
accrue small debt that is bounded to the completion of \textit{one}
batched task and subsequent intervention points gate a tenant
until it has recovered from the deficit. Modern network datapaths have tasks
on the order of a couple of hundred
CPU cycles (\autoref{tab:cycles}), so the accrued deficit
remains bounded (C2).
Furthermore, bounded deficits allow the
datapath to maintain protection without introducing new queueing,
synchronisation, or preemption into the datapath (C3).

\textbf{\textit{P3: Handle RX with delayed attribution and enforcement.}}
On reception, the datapath can only 
identify tenants after
fetching the packet into a cache line and parsing its headers (C4).
Because of the delayed attribution, 
the design delays enforcement to after the \texttt{RX} batch is completely processed.
This avoids the expensive costs of enqueueing packets for
processing at a later time~\cite{ghasemirahni:reframer}, 
preserves the efficiency of 
mixed-tenant batching, and bounds the CPU-time consumption
of out-of-budget tenants (C2); a tenant's debt is bounded
by the cost of \textit{one} \texttt{RX} batch. Once the datapath resolves
attribution, it charges the tenant and enforces
the budget at the next eligible intervention point.
For open-loop traffic, this delayed-enforcement path can be paired with
post-attribution drops or explicit feedback once a tenant exhausts its budget.

\subsection{Architecture}
\label{subsec:architecture}
Our approach presents a split architecture to enforce
time protection. This split separates latency critical packet processing 
from global coordination and control. 
The fast path executes common-case datapath work under tight 
timing constraints, while a separate slow path manages 
cross-core coordination and other variable cost functions. 
This decomposition isolates unpredictable control work from the datapath 
and allows fine-grained enforcement without introducing heavyweight 
synchronisation into the critical path.

\textbf{\textit{Bounded fast path.}}
The fast path is restricted to small, predictable operations 
on core-local state. It executes common-case datapath 
tasks only at well-defined intervention points and 
avoids blocking, cross-core coordination, and variable-cost 
control logic on the critical path. As a result, accounting 
and scheduling can be applied at fine granularity 
without preemption. The worst-case overshoot is 
therefore one bounded task for a tenant, and the fast path
penalises excesses with budget deficits that must
be repaid before servicing that tenant again.

\textbf{\textit{Coordinating slow path.}}
The slow path handles coordination that requires a 
global view of tenants and cores. It periodically 
replenishes core-local budgets and applies 
global weight and cap policies to prevent idle-credit buildup
outside the packet-processing path. This lets our approach 
combine lightweight local enforcement on the fast path with 
coordinated system-wide control at longer timescales.

\subsection{Accounting}
\label{subsec:accountant}
To enforce time protection, the datapath must first attribute
CPU time precisely to the tenant on whose behalf work is performed. 
Our approach therefore builds its protection mechanism on lightweight, 
fine-grained accounting in the fast path, using CPU cycles as 
the common currency of resource use. The accountant
feeds these numbers to the scheduler, described 
in \autoref{subsec:scheduler}, so it can enforce protection.

\textbf{\textit{Time tokens via TSC accounting.}}
The first step towards time protection is to accurately account
for resource use.
The fast path measures CPU consumption with the CPU timestamp counter (TSC)
before and after servicing a tenant in a batch. 
Reading the TSC is lightweight and precise. 
Using the TSC gives cycle-level resolution with a 
constant cost pair of reads that can be issued inline on 
the fast path, 
so the accounting overhead is negligible compared to per-packet processing work.
The fast path then subtracts the
cycles consumed from the respective tenant's resource budget
at the end of a batch.

\textbf{\textit{Core-local resource accounting.}}
Each fast path core tracks resources available to
and used on behalf of individual tenants through a local token budget table, storing
each tenant's available budget on that core. The slow path periodically
updates the fast path value and
replenishes tokens by performing an atomic add to the tenant's
table entry. Periodic replenishments minimise synchronisation on the
fast path and improve overall throughput, adding minimal
overheads as the datapath scales the number of cores (\autoref{fig:rpc-throughput-scaling}).

\textbf{\textit{Delayed RX accounting.}}
The datapath dequeues a batch of packets from the NIC
and it has to decode header information before attributing
the packet to a tenant. This delayed attribution complicates
enforcement. The fast path could insert packets from
out-of-budget tenants into queues, for processing
when the tenant has available budget, 
but at the microsecond scale the cost of fetching a packet into the
cache, adding the packet to a queue, and then incurring another
cache miss when the packet is dequeued for processing at a
later time is prohibitive~\cite{ghasemirahni:reframer}.
Our design instead admits received packets into a
processing batch, accounts for the time spent by each tenant,
lets them accrue bounded deficits, and has them repay for
deficits during transmission. Deficits are bounded
by the time to process a batch 
and are enforced immediately at the next intervention point.

\subsection{Scheduling and Enforcement}
\label{subsec:scheduler}

The fast path enforces time protection with 
fine-grained scheduling based on per-tenant budgets that
use TSC cycles as tokens.
The fast path first chooses tenants and, when 
sending packets, determines which
tenant gets to transmit next.
Before the fast path starts a task on behalf of a tenant, 
the core consults the tenant's budget, and
if it is zero or negative moves on to a different tenant. 
Out-of-token tenants are serviced only if no funded tenants have
work in a batch.

\textbf{\textit{Intervention points in the datapath.}}
We identify three main intervention points where network datapaths
perform CPU intensive tasks for tenants and enforce
protection: receiving packets (\texttt{RX}), 
transmitting packets (\texttt{TX}), and scheduling packets 
for transmission (\texttt{SCHED}). 
\texttt{RX} dequeues incoming packets from the NIC, parses the packets, and
implements the necessary processing before forwarding the payload
to the tenant.
\texttt{SCHED} checks outgoing queues from tenant applications to the fast path
for new transmission requests and schedules them.
\texttt{TX} assembles packets and enqueues them in the NIC.

\textbf{\textit{Deficit-based scheduling.}}
In kernel bypass stacks such as TAS, fast path tasks at
intervention points have sub-$\mu$s latencies (\autoref{tab:cycles}).
This provides our approach with two key
advantages. First, preemption is not necessary, as individual
packet processing tasks complete very
quickly.
Second, fine-grained batch scheduling and accurate accounting
enable strong tail latency isolation, even without knowing
concrete task lengths. Tasks normally range between 200-500 cycles, 
and after they complete the next scheduling decision can compensate based on
the updated budget. Even if a task overruns the budget, it
will only be by a small amount of cycles and the system still precisely
accounts for this with negative budgets, akin to deficit round-robin
scheduling~\cite{shreedhar:drr}.

\textbf{\textit{Preserving batching without weakening protection.}}
The fast path retains batching as an efficiency mechanism.
At each intervention point, 
it may process a small, bounded batch for the selected 
tenant to amortise NIC access, queue management, and other 
fixed per-iteration costs. The CPU time spent by each tenant
in the batch 
is then charged to the tenant's budget, and any overrun 
is carried forward as deficit in subsequent scheduling decisions. 
In this way, batching improves throughput and locality without 
introducing long non-preemptive windows.

\textbf{\textit{Work-conserving fast path scheduler.}}
The fast path scheduler improves efficiency and throughput
by being work-conserving. The scheduler admits out-of-budget
tenants at an intervention point if there are no tenants
with a positive budget and available work. This allows the system
to service more packets, even when tenants are throttled,
without weakening protection. Work-conserving batches run 
only when no other tenant on that core is 
eligible for service under the protected scheduler. 
They therefore consume only slack time and do 
not delay any funded tenant beyond the bounded
task already in service.

\textbf{\textit{Enforcing protection at RX.}}
\texttt{RX} cannot enforce protection at admission before attribution, so 
it accounts preliminary work immediately, carries the charge
forward, and enforces the deficit at the next \texttt{SCHED}/\texttt{TX} decision.
For feedback-paced bidirectional protocols, the system is self-correcting when
both endpoints enforce similar budgets.
The accounting mechanism keeps track of the cycle deficit accrued when
replenishing the budget, so tenants that deplete their budget on \texttt{RX} tasks as a
result have fewer cycles available for \texttt{SCHED} and \texttt{TX} tasks.
Consequently, senders that do not receive replies will stop sending.
For open-loop traffic or tighter receiver-side protection, the datapath can
apply post-attribution drops once it resolves tenant ownership, or pair budget
exhaustion with explicit feedback.
In \autoref{sec:impl} we show how the budget mechanism can integrate 
with congestion control to reduce drops.

\subsection{Budget Refill}

While scheduling and enforcement occur locally on 
each fast path core, a slow path coordinator complements decentralised
fast path decisions. 
The goal of the coordinator is to replenish each tenant's core-local
budget off the fast path, so that over time, each tenant receives
a configured weighted share of CPU time in each fast path core, while
preventing idle tenants from banking unbounded credit and later causing
large bursts. Scheduling and deficit repayment remain local to each
core, so the interference bound still comes from the local enforcement
at intervention points; the refill mechanism only determines how quickly
budget replenishes and how much idle credit a tenant may accumulate.

\textbf{\textit{Periodic budget refill.}}
A separate slow path core periodically replenishes the per-core
budgets on the fast path.
Separation into a parallel decentralised fast path and a central
slow path enables scalable and efficient coordination of the
frequently accessed per-core budgets.
The slow path refills the total budget in periodic 1 $\mu$s
intervals and
distributes the new budget to each tenant.
The distribution among tenants is controlled by a tenant weight $w_t$,
configured by the operator. By default, each tenant has the same weight.

Update tokens represent elapsed cycles in a fast path 
core between budget refills and the slow path distributes these tokens
to tenants to control their share of fast path CPU time.
We compute tokens $e_{tc}$ for tenant $t$ at core $c$ by recording the
timestamp $\tau^\prime$ for the current update and the timestamp $\tau$ for
the previous update.  The allocator scales the elapsed time
$\tau^\prime - \tau$ by a constant boost $B$.
$B$ compensates for any fast path CPU cycles not explicitly accounted
to any tenant to avoid over-committing processor cycles.
We multiply the product of the boost and elapsed cycles by the
tenant's $w_t$, divided by the sum of the weights of all $n$ tenants.
\begin{equation}
  e_{tc} = \frac{B (\tau^\prime - \tau) w_t}{\sum_{i=1}^{n} w_i}
  \label{eq:budget_increment}
\end{equation}

\textbf{\textit{Preventing tenants from accumulating budget.}}
The operator also configures a budget cap $C$ for each tenant. This cap
prevents tenants from accumulating arbitrary budgets during
low utilisation periods and starving other tenants with bursty
activity. $C$ restricts the number of CPU cycles the datapath can
spend on behalf of a tenant per fast path core between update periods.
The slow path calculates the updated per-core
budget $b^\prime_{tc} = \min{\{C, b_{tc} + e_{tc}\}}$ for tenant $t$ on core $c$.

\subsection{Time Protection Bound}
Our approach provides a per-core bound on the direct
scheduling interference. 
Let $D_{task}$ denote the maximum duration 
of one \texttt{SCHED} or \texttt{TX} task, and let $D_{rx}$ denote the 
maximum duration of one \texttt{RX} batch before attribution.
On a given core, if a funded tenant has pending eligible work,
an out-of-budget tenant can delay that tenant's next task
by at most $D_{task}+D_{rx}$ before enforcement.
The out-of-budget tenant takes at max $D_{task}$ to complete
one already started bounded task, and $D_{rx}$ for resolving
a received packet's ownership, and the tenant must repay any
deficits before it receives further service.

\textbf{\textit{Refill Recovery and Funded Interference.}}
\sys exposes two additional controls beyond the direct 
interference bound from out-of-budget work. First, if 
tenant $t$ on core $c$ needs $\delta_{tc}$ cycles to become 
eligible again and receives $e_{tc}$ cycles per refill interval, 
then its refill-recovery delay is at
most $\lceil \delta_{tc}/e_{tc} \rceil P$, where $P$ is the 
refill period.
$P$ thus encodes a trade-off: a small value of $P$ decreases the 
refill-recovery delay, but increases overhead because of more
frequent synchronisation.
Second, among funded tenants, 
interference is dictated by two bounded quantities: the budget cap $C$, 
which bounds how much service an idle tenant can 
bring between refill periods, and the intervention-point quantum,
which deficit round-robin uses to interleave funded tenants at fine granularity.

\textbf{\textit{Scope.}}
We do not claim a universal bound on all residual microarchitectural
interference because shared cache and TLB effects can persist across
tasks.
Instead, we show in \autoref{fig:iso-latency} 
that datapath time protection substantially reduces interference in practice.
Under open-loop traffic, receiver protection relies on
post-attribution drops or 
explicit congestion notification (ECN) marking (\autoref{fig:ecn-drops}). 
In their absence, our approach bounds unattributed work per \texttt{RX} batch,
but does not guarantee zero cross-tenant perturbation if both
endpoints do not enforce similar budgets.
\section{Instantiating \sys in TAS}%
\label{sec:impl}
We instantiate \sys, our time protection mechanism,
in TAS to show how it maps onto a real kernel-bypass TCP stack
that exposes the challenges of time protection at the $\mu$s-scale.
TAS is a particularly demanding case study because it centralises
transport-layer processing in a shared datapath, including per-flow state,
flow scheduling, and bidirectional RX/TX handling, rather than simpler packet
forwarding alone.
At the same time, TAS separates fast-path packet processing from slow-path
control, making it a good vehicle for instantiating \sys.
This section describes how the design of \sys maps onto TAS and the concrete
implementation choices required to embed it into the datapath.

\sys maps directly onto TAS's execution structure.
In TAS, a tenant is an application that uses TAS as a separate TCP
acceleration service.
The fast-path \texttt{RX}, \texttt{SCHED}, and \texttt{TX} loops serve as the intervention points from
\autoref{sec:design}.
TAS's slow path provides the coordination locus for budget refill and global
policy.
Tenant identity is already available on \texttt{SCHED} and \texttt{TX} through application and
flow context, and on \texttt{RX} is resolved after header parsing and flow lookup.

\textbf{\textit{Implementing \sys in TAS.}}
We integrate \sys into TAS by adding inline accounting and budget checks to its
fast-path \texttt{RX}, \texttt{SCHED}, and \texttt{TX} loops.
The fast path reads the TSC before and after servicing a tenant in a batch,
and
uses the elapsed cycles to charge that tenant's budget.
Each fast-path context maintains a core-local budget table, and intervention
points consult this table before starting more work on behalf of a tenant.
TAS's slow path periodically replenishes per-tenant budgets by atomically
adding tokens to the corresponding table entries.
These changes embed \sys into TAS through local modifications to the existing
fast-path loops and slow-path coordination logic.

TAS also instantiates the abstract protection bound from
\autoref{sec:design} with concrete short execution windows.
As shown in \autoref{tab:cycles}, the \texttt{RX}, \texttt{TX}, and \texttt{SCHED} phases in TAS consume
317, 381, and 164 cycles per packet at saturation.
In our evaluated configuration, the individually scheduled TAS tasks are
typically on the order of 200 to 500 cycles, depending on packet size.
In TAS, $D_{task}$ therefore corresponds to one short \texttt{SCHED} or \texttt{TX} task, while
$D_{rx}$ is limited to the portion of \texttt{RX} that executes before tenant ownership
is resolved.
Even in this more demanding transport datapath, the enforceable units remain
short enough for bounded-deficit scheduling without preemption.

\textbf{\textit{Receiver-Side Feedback in TAS.}}
In TAS, receiver-side protection under open-loop traffic benefits from an
explicit feedback path once a tenant exhausts its budget.
Because TAS already implements transport congestion control, we realize
this path by marking ECN bits on packets from tenants whose budget falls below
a threshold, so senders receive explicit notifications that a receiver is
running out-of-budget.
This extension is specific to TAS's transport-layer setting.

\textbf{\textit{Lessons from TAS.}}
Our case study shows that \sys can be added to a demanding shared transport
datapath with modest structural change when the datapath already separates
fast-path processing from slow-path control.
TAS provides natural intervention points and a convenient coordination locus
for global budget refill, allowing \sys to be integrated through local
modifications instead of redesign.
\section{Evaluation}%
\label{sec:eval}
In this section we evaluate how well time
protection addresses performance interference
in a shared datapath with tight latency budgets.
We evaluate the TAS implementation of \sys primarily against
the default shared TAS datapath, which preserves the efficiency of
fine-grained sharing but provides no protection against shared-core
interference, and a siloed TAS datapath, which provides
dedicated-core isolation but compromises sharing.
We answer the following questions:

\begin{compactitem}[\labelitemi]
\item Can datapath time protection provide tail latency isolation with shared resources?

\item Does time protection enable efficient sharing?

\item What is the performance cost of enforcing time protection? 

\item How sensitive is \sys to its budget parameters?

\item Is receiver-side feedback effective for open-loop traffic?
\end{compactitem}

\textbf{\textit{Testbed.}}
We configure two identical machines as client and server.
They are directly connected with a pair of 100\,Gbps Mellanox
ConnectX-5 Ethernet adapters.
Both machines have two Intel Xeon Gold 6152 processors at 2.1\,GHz,
each with 22 cores for a total of 44 cores and 187\,GB of RAM per
machine.
We run Linux kernel 6.1 with Debian 11.

\textbf{\textit{Baselines.}}
As baselines for comparison we use
shared TAS and siloed TAS using SR-IOV.
In the overhead microbenchmarks, we additionally compare against the Linux
network stack.
In shared TAS,
a single TAS instance multiplexes traffic
from multiple applications; this is the standard
configuration described in the TAS paper~\cite{kaufmann:tas}.
In the siloed baseline each
datapath instance is attached to a distinct
SR-IOV virtual function, datapath cores are
pinned to exclusive cores, and applications are
partitioned to exclusive instances.

\subsection{\sys Provides Tail Latency Isolation}
We evaluate \sys's ability
to protect a victim tenant's tail latency from co-located
interference by measuring latency and throughput for a
victim tenant, and throughput for an
adversary tenant attempting to introduce performance
interference in an RPC echo server application.
We examine interference caused by packet and byte load
by separately varying the number of adversary cores and
the size of the adversary messages. For the former,
the adversary uses a fixed message size of 64 bytes and
a total of 10000 connections. For the latter, the adversary
uses one core and 10000 total connections. 
The victim application opens one connection with 64 B messages
and is assigned one core. 

We compare this workload across different system configurations:
\sys with two fast path cores, shared TAS with two fast path cores,
and siloed TAS with two instances each with one fast path core. In all
cases, processes and fast path cores are pinned to dedicated cores
and \sys assigns the same CPU-time weight $w_t$ to each tenant.
We use the following budget parameters for the evaluations
in this section: boost = 0.85, cap = 15,000, and an update period of 1$\mu$s.

\textbf{\textit{Latency.}}
The results in \autoref{fig:iso-latency} show 
that \sys's fine-grained time protection preserves tail latency isolation
under attempted packet and byte volume interference. 
In contrast, shared TAS 
exhibits substantial tail latency inflation as the adversary 
increases either message size or the number of cores.
At five adversarial cores
the 99p tail latency of the different systems is 30$\mu$s for \sys, 
234$\mu$s for shared TAS, and 9$\mu$s for siloed TAS.
Compared to siloed TAS, \sys inevitably shows a controlled increase 
in victim tail latency because both tenants 
share a core and therefore split its CPU time according to
equal weights. However, performance 
stabilizes after this initial sharing and does not degrade 
further as the adversary increases load. A similar pattern
is exhibited as the adversary increases the message size.
Siloed TAS and \sys keep stable tail latencies, while
shared TAS observes a steep degradation.

\textbf{\textit{Throughput.}}
Siloed TAS avoids latency interference by placing tenants on separate cores, 
but this isolation comes at the cost of reduced aggregate throughput
as shown in \autoref{fig:iso-throughput}. 
For example, when the system is saturated with five adversary cores, 
siloed TAS experiences a 25\% drop in combined victim and adversary 
throughput relative to shared TAS. In contrast, \sys reduces throughput 
only by 4\% because fine-grained sharing enables 
it to utilise idle victim cycles for adversary traffic 
without sacrificing tail latency isolation.
A similar trend appears as message size increases.
The throughput delta observed between \sys and siloed TAS
could be further reduced by optimising budget parameters adaptively
to the workloads, so that slack used for protection dynamically
shifts with workload characteristics. We leave this as future work.

\begin{figure}%
  \centering%
\begin{tikzpicture}[gnuplot]
\tikzset{every node/.append style={font={\fontsize{8.0pt}{9.6pt}\selectfont}}}
\path (0.000,0.000) rectangle (8.458,3.048);
\gpcolor{color=gp lt color border}
\gpcolor{rgb color={0.996,0.380,0.000}}
\gpsetlinewidth{1.00}
\gpsetpointsize{4.00}
\gp3point{gp mark 11}{}{(1.944,3.038)}
\gpcolor{color=gp lt color border}
\gpcolor{rgb color={0.471,0.369,0.941}}
\gp3point{gp mark 5}{}{(4.989,3.038)}
\gpcolor{color=gp lt color border}
\gpcolor{rgb color={0.392,0.561,1.000}}
\gp3point{gp mark 7}{}{(1.944,2.783)}
\gpcolor{color=gp lt color border}
\gpcolor{rgb color={0.863,0.149,0.498}}
\gp3point{gp mark 9}{}{(4.989,2.783)}
\gpcolor{color=gp lt color border}
\node[gp node left] at (2.477,3.038) {linux};
\node[gp node left] at (5.522,3.038) {shared-tas};
\node[gp node left] at (2.477,2.783) {\sysplot};
\node[gp node left] at (5.522,2.783) {siloed-tas};
\gpcolor{rgb color={0.996,0.380,0.000}}
\gpsetlinetype{gp lt border}
\gpsetdashtype{gp dt solid}
\gpsetlinewidth{3.00}
\draw[gp path](1.640,3.039)--(2.249,3.039);
\gpcolor{rgb color={0.471,0.369,0.941}}
\draw[gp path](4.685,3.039)--(5.294,3.039);
\gpcolor{rgb color={0.392,0.561,1.000}}
\draw[gp path](1.640,2.783)--(2.249,2.783);
\gpcolor{rgb color={0.863,0.149,0.498}}
\draw[gp path](4.685,2.783)--(5.294,2.783);
\gpdefrectangularnode{gp plot 1}{\pgfpoint{0.422cm}{2.621cm}}{\pgfpoint{8.034cm}{3.047cm}}
\gpcolor{color=gp lt color axes}
\gpsetlinetype{gp lt axes}
\gpsetdashtype{gp dt axes}
\gpsetlinewidth{0.50}
\draw[gp path] (0.676,0.365)--(3.551,0.365);
\gpcolor{color=gp lt color border}
\gpsetlinetype{gp lt border}
\gpsetdashtype{gp dt solid}
\gpsetlinewidth{1.00}
\draw[gp path] (0.676,0.365)--(0.766,0.365);
\node[gp node right] at (0.529,0.365) {$0$};
\gpcolor{color=gp lt color axes}
\gpsetlinetype{gp lt axes}
\gpsetdashtype{gp dt axes}
\gpsetlinewidth{0.50}
\draw[gp path] (0.676,0.833)--(3.551,0.833);
\gpcolor{color=gp lt color border}
\gpsetlinetype{gp lt border}
\gpsetdashtype{gp dt solid}
\gpsetlinewidth{1.00}
\draw[gp path] (0.676,0.833)--(0.766,0.833);
\node[gp node right] at (0.529,0.833) {$100$};
\gpcolor{color=gp lt color axes}
\gpsetlinetype{gp lt axes}
\gpsetdashtype{gp dt axes}
\gpsetlinewidth{0.50}
\draw[gp path] (0.676,1.300)--(3.551,1.300);
\gpcolor{color=gp lt color border}
\gpsetlinetype{gp lt border}
\gpsetdashtype{gp dt solid}
\gpsetlinewidth{1.00}
\draw[gp path] (0.676,1.300)--(0.766,1.300);
\node[gp node right] at (0.529,1.300) {$200$};
\gpcolor{color=gp lt color axes}
\gpsetlinetype{gp lt axes}
\gpsetdashtype{gp dt axes}
\gpsetlinewidth{0.50}
\draw[gp path] (0.676,1.768)--(3.551,1.768);
\gpcolor{color=gp lt color border}
\gpsetlinetype{gp lt border}
\gpsetdashtype{gp dt solid}
\gpsetlinewidth{1.00}
\draw[gp path] (0.676,1.768)--(0.766,1.768);
\node[gp node right] at (0.529,1.768) {$300$};
\gpcolor{color=gp lt color axes}
\gpsetlinetype{gp lt axes}
\gpsetdashtype{gp dt axes}
\gpsetlinewidth{0.50}
\draw[gp path] (0.676,2.236)--(3.551,2.236);
\gpcolor{color=gp lt color border}
\gpsetlinetype{gp lt border}
\gpsetdashtype{gp dt solid}
\gpsetlinewidth{1.00}
\draw[gp path] (0.676,2.236)--(0.766,2.236);
\node[gp node right] at (0.529,2.236) {$400$};
\draw[gp path] (0.676,0.365)--(0.676,0.455);
\node[gp node center] at (0.676,0.119) {$0$};
\draw[gp path] (1.155,0.365)--(1.155,0.455);
\node[gp node center] at (1.155,0.119) {$1$};
\draw[gp path] (1.634,0.365)--(1.634,0.455);
\node[gp node center] at (1.634,0.119) {$2$};
\draw[gp path] (2.114,0.365)--(2.114,0.455);
\node[gp node center] at (2.114,0.119) {$3$};
\draw[gp path] (2.593,0.365)--(2.593,0.455);
\node[gp node center] at (2.593,0.119) {$4$};
\draw[gp path] (3.072,0.365)--(3.072,0.455);
\node[gp node center] at (3.072,0.119) {$5$};
\draw[gp path] (3.551,0.365)--(3.551,0.455);
\node[gp node center] at (3.551,0.119) {$6$};
\draw[gp path] (3.551,0.365)--(3.461,0.365);
\draw[gp path] (3.551,0.833)--(3.461,0.833);
\draw[gp path] (3.551,1.300)--(3.461,1.300);
\draw[gp path] (3.551,1.768)--(3.461,1.768);
\draw[gp path] (3.551,2.236)--(3.461,2.236);
\draw[gp path] (0.676,2.376)--(0.676,0.365)--(3.551,0.365)--(3.551,2.376)--cycle;
\gpcolor{rgb color={0.996,0.380,0.000}}
\gpsetlinewidth{3.00}
\draw[gp path] (1.155,1.151)--(1.634,1.300)--(2.114,1.469)--(2.593,1.777)--(3.072,2.231);
\gp3point{gp mark 11}{}{(1.155,1.151)}
\gp3point{gp mark 11}{}{(1.634,1.300)}
\gp3point{gp mark 11}{}{(2.114,1.469)}
\gp3point{gp mark 11}{}{(2.593,1.777)}
\gp3point{gp mark 11}{}{(3.072,2.231)}
\gpcolor{rgb color={0.471,0.369,0.941}}
\draw[gp path] (1.155,0.538)--(1.634,0.697)--(2.114,1.043)--(2.593,1.211)--(3.072,1.459);
\gp3point{gp mark 5}{}{(1.155,0.538)}
\gp3point{gp mark 5}{}{(1.634,0.697)}
\gp3point{gp mark 5}{}{(2.114,1.043)}
\gp3point{gp mark 5}{}{(2.593,1.211)}
\gp3point{gp mark 5}{}{(3.072,1.459)}
\gpcolor{rgb color={0.863,0.149,0.498}}
\draw[gp path] (1.155,0.407)--(1.634,0.407)--(2.114,0.407)--(2.593,0.407)--(3.072,0.407);
\gp3point{gp mark 9}{}{(1.155,0.407)}
\gp3point{gp mark 9}{}{(1.634,0.407)}
\gp3point{gp mark 9}{}{(2.114,0.407)}
\gp3point{gp mark 9}{}{(2.593,0.407)}
\gp3point{gp mark 9}{}{(3.072,0.407)}
\gpcolor{rgb color={0.392,0.561,1.000}}
\draw[gp path] (1.155,0.477)--(1.634,0.477)--(2.114,0.487)--(2.593,0.491)--(3.072,0.505);
\gp3point{gp mark 7}{}{(1.155,0.477)}
\gp3point{gp mark 7}{}{(1.634,0.477)}
\gp3point{gp mark 7}{}{(2.114,0.487)}
\gp3point{gp mark 7}{}{(2.593,0.491)}
\gp3point{gp mark 7}{}{(3.072,0.505)}
\gpcolor{color=gp lt color border}
\gpsetlinewidth{1.00}
\draw[gp path] (0.676,2.376)--(0.676,0.365)--(3.551,0.365)--(3.551,2.376)--cycle;
\node[gp node center,rotate=-270.0] at (-0.218,1.370) {Vic 99p Lat [us]};
\node[gp node center] at (2.113,-0.249) {Adv Cores};
\gpdefrectangularnode{gp plot 2}{\pgfpoint{0.676cm}{0.365cm}}{\pgfpoint{3.551cm}{2.376cm}}
\gpcolor{color=gp lt color axes}
\gpsetlinetype{gp lt axes}
\gpsetdashtype{gp dt axes}
\gpsetlinewidth{0.50}
\draw[gp path] (4.736,0.365)--(7.611,0.365);
\gpcolor{color=gp lt color border}
\gpsetlinetype{gp lt border}
\gpsetdashtype{gp dt solid}
\gpsetlinewidth{1.00}
\draw[gp path] (4.736,0.365)--(4.826,0.365);
\node[gp node right] at (4.589,0.365) {$0$};
\gpcolor{color=gp lt color axes}
\gpsetlinetype{gp lt axes}
\gpsetdashtype{gp dt axes}
\gpsetlinewidth{0.50}
\draw[gp path] (4.736,0.833)--(7.611,0.833);
\gpcolor{color=gp lt color border}
\gpsetlinetype{gp lt border}
\gpsetdashtype{gp dt solid}
\gpsetlinewidth{1.00}
\draw[gp path] (4.736,0.833)--(4.826,0.833);
\node[gp node right] at (4.589,0.833) {$100$};
\gpcolor{color=gp lt color axes}
\gpsetlinetype{gp lt axes}
\gpsetdashtype{gp dt axes}
\gpsetlinewidth{0.50}
\draw[gp path] (4.736,1.300)--(7.611,1.300);
\gpcolor{color=gp lt color border}
\gpsetlinetype{gp lt border}
\gpsetdashtype{gp dt solid}
\gpsetlinewidth{1.00}
\draw[gp path] (4.736,1.300)--(4.826,1.300);
\node[gp node right] at (4.589,1.300) {$200$};
\gpcolor{color=gp lt color axes}
\gpsetlinetype{gp lt axes}
\gpsetdashtype{gp dt axes}
\gpsetlinewidth{0.50}
\draw[gp path] (4.736,1.768)--(7.611,1.768);
\gpcolor{color=gp lt color border}
\gpsetlinetype{gp lt border}
\gpsetdashtype{gp dt solid}
\gpsetlinewidth{1.00}
\draw[gp path] (4.736,1.768)--(4.826,1.768);
\node[gp node right] at (4.589,1.768) {$300$};
\gpcolor{color=gp lt color axes}
\gpsetlinetype{gp lt axes}
\gpsetdashtype{gp dt axes}
\gpsetlinewidth{0.50}
\draw[gp path] (4.736,2.236)--(7.611,2.236);
\gpcolor{color=gp lt color border}
\gpsetlinetype{gp lt border}
\gpsetdashtype{gp dt solid}
\gpsetlinewidth{1.00}
\draw[gp path] (4.736,2.236)--(4.826,2.236);
\node[gp node right] at (4.589,2.236) {$400$};
\draw[gp path] (5.215,0.365)--(5.215,0.455);
\node[gp node center] at (5.215,0.119) {128};
\draw[gp path] (5.694,0.365)--(5.694,0.455);
\draw[gp path] (6.174,0.365)--(6.174,0.455);
\node[gp node center] at (6.174,0.119) {512};
\draw[gp path] (6.653,0.365)--(6.653,0.455);
\draw[gp path] (7.132,0.365)--(7.132,0.455);
\node[gp node center] at (7.132,0.119) {2048};
\draw[gp path] (7.611,0.365)--(7.521,0.365);
\draw[gp path] (7.611,0.833)--(7.521,0.833);
\draw[gp path] (7.611,1.300)--(7.521,1.300);
\draw[gp path] (7.611,1.768)--(7.521,1.768);
\draw[gp path] (7.611,2.236)--(7.521,2.236);
\draw[gp path] (4.736,2.376)--(4.736,0.365)--(7.611,0.365)--(7.611,2.376)--cycle;
\gpcolor{rgb color={0.471,0.369,0.941}}
\gpsetlinewidth{3.00}
\draw[gp path] (5.215,0.524)--(5.694,0.547)--(6.174,1.595)--(6.653,2.198)--(7.132,1.768);
\gp3point{gp mark 5}{}{(5.215,0.524)}
\gp3point{gp mark 5}{}{(5.694,0.547)}
\gp3point{gp mark 5}{}{(6.174,1.595)}
\gp3point{gp mark 5}{}{(6.653,2.198)}
\gp3point{gp mark 5}{}{(7.132,1.768)}
\gpcolor{rgb color={0.863,0.149,0.498}}
\draw[gp path] (5.215,0.407)--(5.694,0.407)--(6.174,0.407)--(6.653,0.412)--(7.132,0.412);
\gp3point{gp mark 9}{}{(5.215,0.407)}
\gp3point{gp mark 9}{}{(5.694,0.407)}
\gp3point{gp mark 9}{}{(6.174,0.407)}
\gp3point{gp mark 9}{}{(6.653,0.412)}
\gp3point{gp mark 9}{}{(7.132,0.412)}
\gpcolor{rgb color={0.392,0.561,1.000}}
\draw[gp path] (5.215,0.459)--(5.694,0.449)--(6.174,0.454)--(6.653,0.487)--(7.132,0.463);
\gp3point{gp mark 7}{}{(5.215,0.459)}
\gp3point{gp mark 7}{}{(5.694,0.449)}
\gp3point{gp mark 7}{}{(6.174,0.454)}
\gp3point{gp mark 7}{}{(6.653,0.487)}
\gp3point{gp mark 7}{}{(7.132,0.463)}
\gpcolor{rgb color={0.996,0.380,0.000}}
\draw[gp path] (5.215,1.333)--(5.694,1.202)--(6.174,1.305)--(6.653,1.455)--(7.132,1.324);
\gp3point{gp mark 11}{}{(5.215,1.333)}
\gp3point{gp mark 11}{}{(5.694,1.202)}
\gp3point{gp mark 11}{}{(6.174,1.305)}
\gp3point{gp mark 11}{}{(6.653,1.455)}
\gp3point{gp mark 11}{}{(7.132,1.324)}
\gpcolor{color=gp lt color border}
\gpsetlinewidth{1.00}
\draw[gp path] (4.736,2.376)--(4.736,0.365)--(7.611,0.365)--(7.611,2.376)--cycle;
\node[gp node center,rotate=-270.0] at (3.841,1.370) {Vic 99p Lat [us]};
\node[gp node center] at (6.173,-0.249) {Adv Message Size [B]};
\gpdefrectangularnode{gp plot 3}{\pgfpoint{4.736cm}{0.365cm}}{\pgfpoint{7.611cm}{2.376cm}}
\end{tikzpicture}
  \caption{Victim tenant tail latency under increasing adversarial load.
    \sys bounds tail latency inflation as the adversary increases packet volume via additional core allocations and byte volume via larger message sizes.}%
  \label{fig:iso-latency}%
\end{figure}

\begin{figure}%
  \centering%
\begin{tikzpicture}[gnuplot]
\tikzset{every node/.append style={font={\fontsize{8.0pt}{9.6pt}\selectfont}}}
\path (0.000,0.000) rectangle (8.458,6.350);
\gpcolor{color=gp lt color border}
\gpcolor{rgb color={0.863,0.149,0.498}}
\gpsetlinewidth{1.00}
\gpsetpointsize{4.00}
\gp3point{gp mark 9}{}{(1.944,6.100)}
\gpcolor{color=gp lt color border}
\gpcolor{rgb color={0.392,0.561,1.000}}
\gp3point{gp mark 7}{}{(4.989,6.100)}
\gpcolor{color=gp lt color border}
\gpcolor{rgb color={0.471,0.369,0.941}}
\gp3point{gp mark 5}{}{(1.944,5.798)}
\gpcolor{color=gp lt color border}
\gpcolor{rgb color={0.996,0.380,0.000}}
\gp3point{gp mark 11}{}{(4.989,5.798)}
\gpcolor{color=gp lt color border}
\node[gp node left] at (2.477,6.100) {siloed-tas};
\node[gp node left] at (5.522,6.100) {\sysplot};
\node[gp node left] at (2.477,5.798) {shared-tas};
\node[gp node left] at (5.522,5.798) {linux};
\gpcolor{rgb color={0.863,0.149,0.498}}
\gpsetlinetype{gp lt border}
\gpsetdashtype{gp dt solid}
\gpsetlinewidth{3.00}
\draw[gp path](1.640,6.101)--(2.249,6.101);
\gpcolor{rgb color={0.392,0.561,1.000}}
\draw[gp path](4.685,6.101)--(5.294,6.101);
\gpcolor{rgb color={0.471,0.369,0.941}}
\draw[gp path](1.640,5.799)--(2.249,5.799);
\gpcolor{rgb color={0.996,0.380,0.000}}
\draw[gp path](4.685,5.799)--(5.294,5.799);
\gpdefrectangularnode{gp plot 1}{\pgfpoint{0.422cm}{5.461cm}}{\pgfpoint{8.034cm}{6.349cm}}
\gpcolor{color=gp lt color axes}
\gpsetlinetype{gp lt axes}
\gpsetdashtype{gp dt axes}
\gpsetlinewidth{0.50}
\draw[gp path] (0.676,3.365)--(3.551,3.365);
\gpcolor{color=gp lt color border}
\gpsetlinetype{gp lt border}
\gpsetdashtype{gp dt solid}
\gpsetlinewidth{1.00}
\draw[gp path] (0.676,3.365)--(0.766,3.365);
\node[gp node right] at (0.529,3.365) {$0$};
\gpcolor{color=gp lt color axes}
\gpsetlinetype{gp lt axes}
\gpsetdashtype{gp dt axes}
\gpsetlinewidth{0.50}
\draw[gp path] (0.676,3.759)--(3.551,3.759);
\gpcolor{color=gp lt color border}
\gpsetlinetype{gp lt border}
\gpsetdashtype{gp dt solid}
\gpsetlinewidth{1.00}
\draw[gp path] (0.676,3.759)--(0.766,3.759);
\node[gp node right] at (0.529,3.759) {$30$};
\gpcolor{color=gp lt color axes}
\gpsetlinetype{gp lt axes}
\gpsetdashtype{gp dt axes}
\gpsetlinewidth{0.50}
\draw[gp path] (0.676,4.152)--(3.551,4.152);
\gpcolor{color=gp lt color border}
\gpsetlinetype{gp lt border}
\gpsetdashtype{gp dt solid}
\gpsetlinewidth{1.00}
\draw[gp path] (0.676,4.152)--(0.766,4.152);
\node[gp node right] at (0.529,4.152) {$60$};
\gpcolor{color=gp lt color axes}
\gpsetlinetype{gp lt axes}
\gpsetdashtype{gp dt axes}
\gpsetlinewidth{0.50}
\draw[gp path] (0.676,4.546)--(3.551,4.546);
\gpcolor{color=gp lt color border}
\gpsetlinetype{gp lt border}
\gpsetdashtype{gp dt solid}
\gpsetlinewidth{1.00}
\draw[gp path] (0.676,4.546)--(0.766,4.546);
\node[gp node right] at (0.529,4.546) {$90$};
\gpcolor{color=gp lt color axes}
\gpsetlinetype{gp lt axes}
\gpsetdashtype{gp dt axes}
\gpsetlinewidth{0.50}
\draw[gp path] (0.676,4.939)--(3.551,4.939);
\gpcolor{color=gp lt color border}
\gpsetlinetype{gp lt border}
\gpsetdashtype{gp dt solid}
\gpsetlinewidth{1.00}
\draw[gp path] (0.676,4.939)--(0.766,4.939);
\node[gp node right] at (0.529,4.939) {$120$};
\gpcolor{color=gp lt color axes}
\gpsetlinetype{gp lt axes}
\gpsetdashtype{gp dt axes}
\gpsetlinewidth{0.50}
\draw[gp path] (0.676,5.333)--(3.551,5.333);
\gpcolor{color=gp lt color border}
\gpsetlinetype{gp lt border}
\gpsetdashtype{gp dt solid}
\gpsetlinewidth{1.00}
\draw[gp path] (0.676,5.333)--(0.766,5.333);
\node[gp node right] at (0.529,5.333) {$150$};
\draw[gp path] (0.676,3.365)--(0.676,3.455);
\node[gp node center] at (0.676,3.119) {$0$};
\draw[gp path] (1.155,3.365)--(1.155,3.455);
\node[gp node center] at (1.155,3.119) {$1$};
\draw[gp path] (1.634,3.365)--(1.634,3.455);
\node[gp node center] at (1.634,3.119) {$2$};
\draw[gp path] (2.114,3.365)--(2.114,3.455);
\node[gp node center] at (2.114,3.119) {$3$};
\draw[gp path] (2.593,3.365)--(2.593,3.455);
\node[gp node center] at (2.593,3.119) {$4$};
\draw[gp path] (3.072,3.365)--(3.072,3.455);
\node[gp node center] at (3.072,3.119) {$5$};
\draw[gp path] (3.551,3.365)--(3.551,3.455);
\node[gp node center] at (3.551,3.119) {$6$};
\draw[gp path] (3.551,3.365)--(3.461,3.365);
\draw[gp path] (3.551,3.759)--(3.461,3.759);
\draw[gp path] (3.551,4.152)--(3.461,4.152);
\draw[gp path] (3.551,4.546)--(3.461,4.546);
\draw[gp path] (3.551,4.939)--(3.461,4.939);
\draw[gp path] (3.551,5.333)--(3.461,5.333);
\draw[gp path] (0.676,5.333)--(0.676,3.365)--(3.551,3.365)--(3.551,5.333)--cycle;
\gpcolor{rgb color={0.471,0.369,0.941}}
\gpsetlinewidth{3.00}
\draw[gp path] (1.155,4.302)--(1.634,3.759)--(2.114,3.648)--(2.593,3.643)--(3.072,3.399);
\gp3point{gp mark 5}{}{(1.155,4.302)}
\gp3point{gp mark 5}{}{(1.634,3.759)}
\gp3point{gp mark 5}{}{(2.114,3.648)}
\gp3point{gp mark 5}{}{(2.593,3.643)}
\gp3point{gp mark 5}{}{(3.072,3.399)}
\gpcolor{rgb color={0.863,0.149,0.498}}
\draw[gp path] (1.155,5.069)--(1.634,5.050)--(2.114,5.010)--(2.593,4.983)--(3.072,5.007);
\gp3point{gp mark 9}{}{(1.155,5.069)}
\gp3point{gp mark 9}{}{(1.634,5.050)}
\gp3point{gp mark 9}{}{(2.114,5.010)}
\gp3point{gp mark 9}{}{(2.593,4.983)}
\gp3point{gp mark 9}{}{(3.072,5.007)}
\gpcolor{rgb color={0.392,0.561,1.000}}
\draw[gp path] (1.155,4.470)--(1.634,4.363)--(2.114,4.295)--(2.593,4.312)--(3.072,4.265);
\gp3point{gp mark 7}{}{(1.155,4.470)}
\gp3point{gp mark 7}{}{(1.634,4.363)}
\gp3point{gp mark 7}{}{(2.114,4.295)}
\gp3point{gp mark 7}{}{(2.593,4.312)}
\gp3point{gp mark 7}{}{(3.072,4.265)}
\gpcolor{rgb color={0.996,0.380,0.000}}
\draw[gp path] (1.155,3.523)--(1.634,3.528)--(2.114,3.517)--(2.593,3.495)--(3.072,3.483);
\gp3point{gp mark 11}{}{(1.155,3.523)}
\gp3point{gp mark 11}{}{(1.634,3.528)}
\gp3point{gp mark 11}{}{(2.114,3.517)}
\gp3point{gp mark 11}{}{(2.593,3.495)}
\gp3point{gp mark 11}{}{(3.072,3.483)}
\gpcolor{color=gp lt color border}
\gpsetlinewidth{1.00}
\draw[gp path] (0.676,5.333)--(0.676,3.365)--(3.551,3.365)--(3.551,5.333)--cycle;
\node[gp node center,rotate=-270.0] at (-0.218,4.349) {Vic Tput [KReq/s]};
\node[gp node center] at (2.113,2.750) {Adv Cores};
\gpdefrectangularnode{gp plot 2}{\pgfpoint{0.676cm}{3.365cm}}{\pgfpoint{3.551cm}{5.333cm}}
\gpcolor{color=gp lt color axes}
\gpsetlinetype{gp lt axes}
\gpsetdashtype{gp dt axes}
\gpsetlinewidth{0.50}
\draw[gp path] (4.736,3.365)--(7.611,3.365);
\gpcolor{color=gp lt color border}
\gpsetlinetype{gp lt border}
\gpsetdashtype{gp dt solid}
\gpsetlinewidth{1.00}
\draw[gp path] (4.736,3.365)--(4.826,3.365);
\node[gp node right] at (4.589,3.365) {$0$};
\gpcolor{color=gp lt color axes}
\gpsetlinetype{gp lt axes}
\gpsetdashtype{gp dt axes}
\gpsetlinewidth{0.50}
\draw[gp path] (4.736,3.759)--(7.611,3.759);
\gpcolor{color=gp lt color border}
\gpsetlinetype{gp lt border}
\gpsetdashtype{gp dt solid}
\gpsetlinewidth{1.00}
\draw[gp path] (4.736,3.759)--(4.826,3.759);
\node[gp node right] at (4.589,3.759) {$2$};
\gpcolor{color=gp lt color axes}
\gpsetlinetype{gp lt axes}
\gpsetdashtype{gp dt axes}
\gpsetlinewidth{0.50}
\draw[gp path] (4.736,4.152)--(7.611,4.152);
\gpcolor{color=gp lt color border}
\gpsetlinetype{gp lt border}
\gpsetdashtype{gp dt solid}
\gpsetlinewidth{1.00}
\draw[gp path] (4.736,4.152)--(4.826,4.152);
\node[gp node right] at (4.589,4.152) {$4$};
\gpcolor{color=gp lt color axes}
\gpsetlinetype{gp lt axes}
\gpsetdashtype{gp dt axes}
\gpsetlinewidth{0.50}
\draw[gp path] (4.736,4.546)--(7.611,4.546);
\gpcolor{color=gp lt color border}
\gpsetlinetype{gp lt border}
\gpsetdashtype{gp dt solid}
\gpsetlinewidth{1.00}
\draw[gp path] (4.736,4.546)--(4.826,4.546);
\node[gp node right] at (4.589,4.546) {$6$};
\gpcolor{color=gp lt color axes}
\gpsetlinetype{gp lt axes}
\gpsetdashtype{gp dt axes}
\gpsetlinewidth{0.50}
\draw[gp path] (4.736,4.939)--(7.611,4.939);
\gpcolor{color=gp lt color border}
\gpsetlinetype{gp lt border}
\gpsetdashtype{gp dt solid}
\gpsetlinewidth{1.00}
\draw[gp path] (4.736,4.939)--(4.826,4.939);
\node[gp node right] at (4.589,4.939) {$8$};
\gpcolor{color=gp lt color axes}
\gpsetlinetype{gp lt axes}
\gpsetdashtype{gp dt axes}
\gpsetlinewidth{0.50}
\draw[gp path] (4.736,5.333)--(7.611,5.333);
\gpcolor{color=gp lt color border}
\gpsetlinetype{gp lt border}
\gpsetdashtype{gp dt solid}
\gpsetlinewidth{1.00}
\draw[gp path] (4.736,5.333)--(4.826,5.333);
\node[gp node right] at (4.589,5.333) {$10$};
\draw[gp path] (4.736,3.365)--(4.736,3.455);
\node[gp node center] at (4.736,3.119) {$0$};
\draw[gp path] (5.215,3.365)--(5.215,3.455);
\node[gp node center] at (5.215,3.119) {$1$};
\draw[gp path] (5.694,3.365)--(5.694,3.455);
\node[gp node center] at (5.694,3.119) {$2$};
\draw[gp path] (6.174,3.365)--(6.174,3.455);
\node[gp node center] at (6.174,3.119) {$3$};
\draw[gp path] (6.653,3.365)--(6.653,3.455);
\node[gp node center] at (6.653,3.119) {$4$};
\draw[gp path] (7.132,3.365)--(7.132,3.455);
\node[gp node center] at (7.132,3.119) {$5$};
\draw[gp path] (7.611,3.365)--(7.611,3.455);
\node[gp node center] at (7.611,3.119) {$6$};
\draw[gp path] (7.611,3.365)--(7.521,3.365);
\draw[gp path] (7.611,3.759)--(7.521,3.759);
\draw[gp path] (7.611,4.152)--(7.521,4.152);
\draw[gp path] (7.611,4.546)--(7.521,4.546);
\draw[gp path] (7.611,4.939)--(7.521,4.939);
\draw[gp path] (7.611,5.333)--(7.521,5.333);
\draw[gp path] (4.736,5.333)--(4.736,3.365)--(7.611,3.365)--(7.611,5.333)--cycle;
\gpcolor{rgb color={0.471,0.369,0.941}}
\gpsetlinewidth{3.00}
\draw[gp path] (5.215,3.824)--(5.694,4.268)--(6.174,4.697)--(6.653,5.187)--(7.132,5.200);
\gp3point{gp mark 5}{}{(5.215,3.824)}
\gp3point{gp mark 5}{}{(5.694,4.268)}
\gp3point{gp mark 5}{}{(6.174,4.697)}
\gp3point{gp mark 5}{}{(6.653,5.187)}
\gp3point{gp mark 5}{}{(7.132,5.200)}
\gpcolor{rgb color={0.863,0.149,0.498}}
\draw[gp path] (5.215,3.905)--(5.694,4.373)--(6.174,4.753)--(6.653,4.747)--(7.132,4.737);
\gp3point{gp mark 9}{}{(5.215,3.905)}
\gp3point{gp mark 9}{}{(5.694,4.373)}
\gp3point{gp mark 9}{}{(6.174,4.753)}
\gp3point{gp mark 9}{}{(6.653,4.747)}
\gp3point{gp mark 9}{}{(7.132,4.737)}
\gpcolor{rgb color={0.392,0.561,1.000}}
\draw[gp path] (5.215,3.884)--(5.694,4.293)--(6.174,4.765)--(6.653,5.166)--(7.132,5.125);
\gp3point{gp mark 7}{}{(5.215,3.884)}
\gp3point{gp mark 7}{}{(5.694,4.293)}
\gp3point{gp mark 7}{}{(6.174,4.765)}
\gp3point{gp mark 7}{}{(6.653,5.166)}
\gp3point{gp mark 7}{}{(7.132,5.125)}
\gpcolor{rgb color={0.996,0.380,0.000}}
\draw[gp path] (5.215,3.396)--(5.694,3.419)--(6.174,3.443)--(6.653,3.474)--(7.132,3.501);
\gp3point{gp mark 11}{}{(5.215,3.396)}
\gp3point{gp mark 11}{}{(5.694,3.419)}
\gp3point{gp mark 11}{}{(6.174,3.443)}
\gp3point{gp mark 11}{}{(6.653,3.474)}
\gp3point{gp mark 11}{}{(7.132,3.501)}
\gpcolor{color=gp lt color border}
\gpsetlinewidth{1.00}
\draw[gp path] (4.736,5.333)--(4.736,3.365)--(7.611,3.365)--(7.611,5.333)--cycle;
\node[gp node center,rotate=-270.0] at (3.988,4.349) {Adv Tput [MReq/s]};
\node[gp node center] at (6.173,2.750) {Adv Cores};
\gpdefrectangularnode{gp plot 3}{\pgfpoint{4.736cm}{3.365cm}}{\pgfpoint{7.611cm}{5.333cm}}
\gpcolor{color=gp lt color axes}
\gpsetlinetype{gp lt axes}
\gpsetdashtype{gp dt axes}
\gpsetlinewidth{0.50}
\draw[gp path] (0.676,0.635)--(3.551,0.635);
\gpcolor{color=gp lt color border}
\gpsetlinetype{gp lt border}
\gpsetdashtype{gp dt solid}
\gpsetlinewidth{1.00}
\draw[gp path] (0.676,0.635)--(0.766,0.635);
\node[gp node right] at (0.529,0.635) {$0$};
\gpcolor{color=gp lt color axes}
\gpsetlinetype{gp lt axes}
\gpsetdashtype{gp dt axes}
\gpsetlinewidth{0.50}
\draw[gp path] (0.676,1.095)--(3.551,1.095);
\gpcolor{color=gp lt color border}
\gpsetlinetype{gp lt border}
\gpsetdashtype{gp dt solid}
\gpsetlinewidth{1.00}
\draw[gp path] (0.676,1.095)--(0.766,1.095);
\node[gp node right] at (0.529,1.095) {$20$};
\gpcolor{color=gp lt color axes}
\gpsetlinetype{gp lt axes}
\gpsetdashtype{gp dt axes}
\gpsetlinewidth{0.50}
\draw[gp path] (0.676,1.556)--(3.551,1.556);
\gpcolor{color=gp lt color border}
\gpsetlinetype{gp lt border}
\gpsetdashtype{gp dt solid}
\gpsetlinewidth{1.00}
\draw[gp path] (0.676,1.556)--(0.766,1.556);
\node[gp node right] at (0.529,1.556) {$40$};
\gpcolor{color=gp lt color axes}
\gpsetlinetype{gp lt axes}
\gpsetdashtype{gp dt axes}
\gpsetlinewidth{0.50}
\draw[gp path] (0.676,2.016)--(3.551,2.016);
\gpcolor{color=gp lt color border}
\gpsetlinetype{gp lt border}
\gpsetdashtype{gp dt solid}
\gpsetlinewidth{1.00}
\draw[gp path] (0.676,2.016)--(0.766,2.016);
\node[gp node right] at (0.529,2.016) {$60$};
\gpcolor{color=gp lt color axes}
\gpsetlinetype{gp lt axes}
\gpsetdashtype{gp dt axes}
\gpsetlinewidth{0.50}
\draw[gp path] (0.676,2.476)--(3.551,2.476);
\gpcolor{color=gp lt color border}
\gpsetlinetype{gp lt border}
\gpsetdashtype{gp dt solid}
\gpsetlinewidth{1.00}
\draw[gp path] (0.676,2.476)--(0.766,2.476);
\node[gp node right] at (0.529,2.476) {$80$};
\draw[gp path] (1.155,0.635)--(1.155,0.725);
\node[gp node center] at (1.155,0.389) {128};
\draw[gp path] (1.634,0.635)--(1.634,0.725);
\draw[gp path] (2.114,0.635)--(2.114,0.725);
\node[gp node center] at (2.114,0.389) {512};
\draw[gp path] (2.593,0.635)--(2.593,0.725);
\draw[gp path] (3.072,0.635)--(3.072,0.725);
\node[gp node center] at (3.072,0.389) {2048};
\draw[gp path] (3.551,0.635)--(3.461,0.635);
\draw[gp path] (3.551,1.095)--(3.461,1.095);
\draw[gp path] (3.551,1.556)--(3.461,1.556);
\draw[gp path] (3.551,2.016)--(3.461,2.016);
\draw[gp path] (3.551,2.476)--(3.461,2.476);
\draw[gp path] (0.676,2.476)--(0.676,0.635)--(3.551,0.635)--(3.551,2.476)--cycle;
\gpcolor{rgb color={0.471,0.369,0.941}}
\gpsetlinewidth{3.00}
\draw[gp path] (1.155,1.309)--(1.634,1.237)--(2.114,0.861)--(2.593,0.730)--(3.072,0.746);
\gp3point{gp mark 5}{}{(1.155,1.309)}
\gp3point{gp mark 5}{}{(1.634,1.237)}
\gp3point{gp mark 5}{}{(2.114,0.861)}
\gp3point{gp mark 5}{}{(2.593,0.730)}
\gp3point{gp mark 5}{}{(3.072,0.746)}
\gpcolor{rgb color={0.863,0.149,0.498}}
\draw[gp path] (1.155,2.154)--(1.634,2.116)--(2.114,2.103)--(2.593,2.089)--(3.072,2.073);
\gp3point{gp mark 9}{}{(1.155,2.154)}
\gp3point{gp mark 9}{}{(1.634,2.116)}
\gp3point{gp mark 9}{}{(2.114,2.103)}
\gp3point{gp mark 9}{}{(2.593,2.089)}
\gp3point{gp mark 9}{}{(3.072,2.073)}
\gpcolor{rgb color={0.392,0.561,1.000}}
\draw[gp path] (1.155,1.622)--(1.634,1.636)--(2.114,1.607)--(2.593,1.517)--(3.072,1.543);
\gp3point{gp mark 7}{}{(1.155,1.622)}
\gp3point{gp mark 7}{}{(1.634,1.636)}
\gp3point{gp mark 7}{}{(2.114,1.607)}
\gp3point{gp mark 7}{}{(2.593,1.517)}
\gp3point{gp mark 7}{}{(3.072,1.543)}
\gpcolor{rgb color={0.996,0.380,0.000}}
\draw[gp path] (1.155,0.764)--(1.634,0.765)--(2.114,0.769)--(2.593,0.756)--(3.072,0.770);
\gp3point{gp mark 11}{}{(1.155,0.764)}
\gp3point{gp mark 11}{}{(1.634,0.765)}
\gp3point{gp mark 11}{}{(2.114,0.769)}
\gp3point{gp mark 11}{}{(2.593,0.756)}
\gp3point{gp mark 11}{}{(3.072,0.770)}
\gpcolor{color=gp lt color border}
\gpsetlinewidth{1.00}
\draw[gp path] (0.676,2.476)--(0.676,0.635)--(3.551,0.635)--(3.551,2.476)--cycle;
\node[gp node center,rotate=-270.0] at (-0.072,1.555) {Vic Tput [Mbit/s]};
\node[gp node center] at (2.113,0.020) {Adv Message Size [B]};
\gpdefrectangularnode{gp plot 4}{\pgfpoint{0.676cm}{0.635cm}}{\pgfpoint{3.551cm}{2.476cm}}
\gpcolor{color=gp lt color axes}
\gpsetlinetype{gp lt axes}
\gpsetdashtype{gp dt axes}
\gpsetlinewidth{0.50}
\draw[gp path] (4.736,0.635)--(7.611,0.635);
\gpcolor{color=gp lt color border}
\gpsetlinetype{gp lt border}
\gpsetdashtype{gp dt solid}
\gpsetlinewidth{1.00}
\draw[gp path] (4.736,0.635)--(4.826,0.635);
\node[gp node right] at (4.589,0.635) {$0$};
\gpcolor{color=gp lt color axes}
\gpsetlinetype{gp lt axes}
\gpsetdashtype{gp dt axes}
\gpsetlinewidth{0.50}
\draw[gp path] (4.736,1.003)--(7.611,1.003);
\gpcolor{color=gp lt color border}
\gpsetlinetype{gp lt border}
\gpsetdashtype{gp dt solid}
\gpsetlinewidth{1.00}
\draw[gp path] (4.736,1.003)--(4.826,1.003);
\node[gp node right] at (4.589,1.003) {$2$};
\gpcolor{color=gp lt color axes}
\gpsetlinetype{gp lt axes}
\gpsetdashtype{gp dt axes}
\gpsetlinewidth{0.50}
\draw[gp path] (4.736,1.371)--(7.611,1.371);
\gpcolor{color=gp lt color border}
\gpsetlinetype{gp lt border}
\gpsetdashtype{gp dt solid}
\gpsetlinewidth{1.00}
\draw[gp path] (4.736,1.371)--(4.826,1.371);
\node[gp node right] at (4.589,1.371) {$4$};
\gpcolor{color=gp lt color axes}
\gpsetlinetype{gp lt axes}
\gpsetdashtype{gp dt axes}
\gpsetlinewidth{0.50}
\draw[gp path] (4.736,1.740)--(7.611,1.740);
\gpcolor{color=gp lt color border}
\gpsetlinetype{gp lt border}
\gpsetdashtype{gp dt solid}
\gpsetlinewidth{1.00}
\draw[gp path] (4.736,1.740)--(4.826,1.740);
\node[gp node right] at (4.589,1.740) {$6$};
\gpcolor{color=gp lt color axes}
\gpsetlinetype{gp lt axes}
\gpsetdashtype{gp dt axes}
\gpsetlinewidth{0.50}
\draw[gp path] (4.736,2.108)--(7.611,2.108);
\gpcolor{color=gp lt color border}
\gpsetlinetype{gp lt border}
\gpsetdashtype{gp dt solid}
\gpsetlinewidth{1.00}
\draw[gp path] (4.736,2.108)--(4.826,2.108);
\node[gp node right] at (4.589,2.108) {$8$};
\gpcolor{color=gp lt color axes}
\gpsetlinetype{gp lt axes}
\gpsetdashtype{gp dt axes}
\gpsetlinewidth{0.50}
\draw[gp path] (4.736,2.476)--(7.611,2.476);
\gpcolor{color=gp lt color border}
\gpsetlinetype{gp lt border}
\gpsetdashtype{gp dt solid}
\gpsetlinewidth{1.00}
\draw[gp path] (4.736,2.476)--(4.826,2.476);
\node[gp node right] at (4.589,2.476) {$10$};
\draw[gp path] (5.215,0.635)--(5.215,0.725);
\node[gp node center] at (5.215,0.389) {128};
\draw[gp path] (5.694,0.635)--(5.694,0.725);
\draw[gp path] (6.174,0.635)--(6.174,0.725);
\node[gp node center] at (6.174,0.389) {512};
\draw[gp path] (6.653,0.635)--(6.653,0.725);
\draw[gp path] (7.132,0.635)--(7.132,0.725);
\node[gp node center] at (7.132,0.389) {2048};
\draw[gp path] (7.611,0.635)--(7.521,0.635);
\draw[gp path] (7.611,1.003)--(7.521,1.003);
\draw[gp path] (7.611,1.371)--(7.521,1.371);
\draw[gp path] (7.611,1.740)--(7.521,1.740);
\draw[gp path] (7.611,2.108)--(7.521,2.108);
\draw[gp path] (7.611,2.476)--(7.521,2.476);
\draw[gp path] (4.736,2.476)--(4.736,0.635)--(7.611,0.635)--(7.611,2.476)--cycle;
\gpcolor{rgb color={0.471,0.369,0.941}}
\gpsetlinewidth{3.00}
\draw[gp path] (5.215,1.041)--(5.694,1.341)--(6.174,1.812)--(6.653,2.190)--(7.132,2.133);
\gp3point{gp mark 5}{}{(5.215,1.041)}
\gp3point{gp mark 5}{}{(5.694,1.341)}
\gp3point{gp mark 5}{}{(6.174,1.812)}
\gp3point{gp mark 5}{}{(6.653,2.190)}
\gp3point{gp mark 5}{}{(7.132,2.133)}
\gpcolor{rgb color={0.863,0.149,0.498}}
\draw[gp path] (5.215,1.097)--(5.694,1.294)--(6.174,1.449)--(6.653,1.645)--(7.132,1.704);
\gp3point{gp mark 9}{}{(5.215,1.097)}
\gp3point{gp mark 9}{}{(5.694,1.294)}
\gp3point{gp mark 9}{}{(6.174,1.449)}
\gp3point{gp mark 9}{}{(6.653,1.645)}
\gp3point{gp mark 9}{}{(7.132,1.704)}
\gpcolor{rgb color={0.392,0.561,1.000}}
\draw[gp path] (5.215,1.069)--(5.694,1.320)--(6.174,1.651)--(6.653,2.070)--(7.132,2.012);
\gp3point{gp mark 7}{}{(5.215,1.069)}
\gp3point{gp mark 7}{}{(5.694,1.320)}
\gp3point{gp mark 7}{}{(6.174,1.651)}
\gp3point{gp mark 7}{}{(6.653,2.070)}
\gp3point{gp mark 7}{}{(7.132,2.012)}
\gpcolor{rgb color={0.996,0.380,0.000}}
\draw[gp path] (5.215,0.664)--(5.694,0.688)--(6.174,0.742)--(6.653,0.837)--(7.132,0.972);
\gp3point{gp mark 11}{}{(5.215,0.664)}
\gp3point{gp mark 11}{}{(5.694,0.688)}
\gp3point{gp mark 11}{}{(6.174,0.742)}
\gp3point{gp mark 11}{}{(6.653,0.837)}
\gp3point{gp mark 11}{}{(7.132,0.972)}
\gpcolor{color=gp lt color border}
\gpsetlinewidth{1.00}
\draw[gp path] (4.736,2.476)--(4.736,0.635)--(7.611,0.635)--(7.611,2.476)--cycle;
\node[gp node center,rotate=-270.0] at (3.988,1.555) {Adv Tput [Gbit/s]};
\node[gp node center] at (6.173,0.020) {Adv Message Size [B]};
\gpdefrectangularnode{gp plot 5}{\pgfpoint{4.736cm}{0.635cm}}{\pgfpoint{7.611cm}{2.476cm}}
\end{tikzpicture}
  \caption{Victim and adversarial throughput under increasing adversarial load
    via additional core allocations or larger message sizes.
    \sys's fine-grained core sharing uses spare cycles to serve 
    adversary traffic while keeping victim tail latency stable, 
    matching shared TAS throughput while improving over siloed designs.}%
  \label{fig:iso-throughput}%
\end{figure}

\subsection{\sys Preserves Efficient Sharing}
In \autoref{fig:memcached-efficiency} we show that time protection
does not compromise the efficiency of shared stacks for bursty workloads.
We run three memcached servers that use \sys, shared TAS, and
siloed TAS for the network datapath and then measure 
datapath per-core efficiency after assigning one or three fast path cores
to the \sys and shared TAS baselines, 
normalised by the throughput of siloed TAS with a total of three fast path cores.
Memtier generates sufficient load to saturate three fast path cores.
We model the workload with 2 ms on-off 
burst durations with mean inter-burst intervals of 5 ms, 
drawn from a normal distribution with 1 ms standard deviation;
we compare it to a non-bursty workload (0 ms) that continuously sends requests
at maximum rate. For bursty workloads, 
\sys's per-core throughput with one fast path core is 204\% higher than siloed TAS
and matches shared TAS. The shared design in \sys reduces stranding and 
improves per-core efficiency because the datapath provisions for the 
peak \textit{aggregate} throughput of all
tenants, rather than the peak throughput of individual tenants, 
allowing a pool of cores to service one tenant when others are idle.
For constant rates, \sys's total throughput is just 6.8\%
lower than shared TAS for one datapath core and matches the throughput
of both siloed and shared TAS with three datapath cores, 
showing that \sys preserves throughput close to shared TAS even for
non-bursty traffic.

\begin{figure}%
  \centering%
\begin{tikzpicture}[gnuplot]
\tikzset{every node/.append style={font={\fontsize{8.0pt}{9.6pt}\selectfont}}}
\path (0.000,0.000) rectangle (8.458,3.048);
\gpfill{rgb color={0.392,0.561,1.000}} (1.194,2.946)--(1.583,2.946)--(1.583,3.037)--(1.194,3.037)--cycle;
\gpcolor{color=gp lt color border}
\gpsetlinetype{gp lt border}
\gpsetdashtype{gp dt solid}
\gpsetlinewidth{1.00}
\draw[gp path] (1.194,2.946)--(1.194,3.037)--(1.583,3.037)--(1.583,2.946)--cycle;
\gpfill{rgb color={0.863,0.149,0.498}} (4.617,2.946)--(5.006,2.946)--(5.006,3.037)--(4.617,3.037)--cycle;
\draw[gp path] (4.617,2.946)--(4.617,3.037)--(5.006,3.037)--(5.006,2.946)--cycle;
\gpfill{rgb color={0.471,0.369,0.941}} (1.194,2.681)--(1.583,2.681)--(1.583,2.772)--(1.194,2.772)--cycle;
\draw[gp path] (1.194,2.681)--(1.194,2.772)--(1.583,2.772)--(1.583,2.681)--cycle;
\gpfill{rgb color={0.996,0.380,0.000}} (4.617,2.681)--(5.006,2.681)--(5.006,2.772)--(4.617,2.772)--cycle;
\draw[gp path] (4.617,2.681)--(4.617,2.772)--(5.006,2.772)--(5.006,2.681)--cycle;
\node[gp node left] at (1.738,2.992) {\sysplot-1-core};
\node[gp node left] at (5.162,2.992) {shared-tas-1-core};
\node[gp node left] at (1.738,2.727) {\sysplot-3-cores};
\node[gp node left] at (5.162,2.727) {shared-tas-3-cores};
\gpdefrectangularnode{gp plot 1}{\pgfpoint{0.338cm}{2.590cm}}{\pgfpoint{8.118cm}{3.047cm}}
\gpcolor{color=gp lt color axes}
\gpsetlinetype{gp lt axes}
\gpsetdashtype{gp dt axes}
\gpsetlinewidth{0.50}
\draw[gp path] (0.676,0.365)--(3.551,0.365);
\gpcolor{color=gp lt color border}
\gpsetlinetype{gp lt border}
\gpsetdashtype{gp dt solid}
\gpsetlinewidth{1.00}
\draw[gp path] (0.676,0.365)--(0.766,0.365);
\node[gp node right] at (0.529,0.365) {0};
\gpcolor{color=gp lt color axes}
\gpsetlinetype{gp lt axes}
\gpsetdashtype{gp dt axes}
\gpsetlinewidth{0.50}
\draw[gp path] (0.676,0.712)--(3.551,0.712);
\gpcolor{color=gp lt color border}
\gpsetlinetype{gp lt border}
\gpsetdashtype{gp dt solid}
\gpsetlinewidth{1.00}
\draw[gp path] (0.676,0.712)--(0.766,0.712);
\node[gp node right] at (0.529,0.712) {0.5};
\gpcolor{color=gp lt color axes}
\gpsetlinetype{gp lt axes}
\gpsetdashtype{gp dt axes}
\gpsetlinewidth{0.50}
\draw[gp path] (0.676,1.060)--(3.551,1.060);
\gpcolor{color=gp lt color border}
\gpsetlinetype{gp lt border}
\gpsetdashtype{gp dt solid}
\gpsetlinewidth{1.00}
\draw[gp path] (0.676,1.060)--(0.766,1.060);
\node[gp node right] at (0.529,1.060) {1};
\gpcolor{color=gp lt color axes}
\gpsetlinetype{gp lt axes}
\gpsetdashtype{gp dt axes}
\gpsetlinewidth{0.50}
\draw[gp path] (0.676,1.407)--(3.551,1.407);
\gpcolor{color=gp lt color border}
\gpsetlinetype{gp lt border}
\gpsetdashtype{gp dt solid}
\gpsetlinewidth{1.00}
\draw[gp path] (0.676,1.407)--(0.766,1.407);
\node[gp node right] at (0.529,1.407) {1.5};
\gpcolor{color=gp lt color axes}
\gpsetlinetype{gp lt axes}
\gpsetdashtype{gp dt axes}
\gpsetlinewidth{0.50}
\draw[gp path] (0.676,1.755)--(3.551,1.755);
\gpcolor{color=gp lt color border}
\gpsetlinetype{gp lt border}
\gpsetdashtype{gp dt solid}
\gpsetlinewidth{1.00}
\draw[gp path] (0.676,1.755)--(0.766,1.755);
\node[gp node right] at (0.529,1.755) {2};
\gpcolor{color=gp lt color axes}
\gpsetlinetype{gp lt axes}
\gpsetdashtype{gp dt axes}
\gpsetlinewidth{0.50}
\draw[gp path] (0.676,2.102)--(3.551,2.102);
\gpcolor{color=gp lt color border}
\gpsetlinetype{gp lt border}
\gpsetdashtype{gp dt solid}
\gpsetlinewidth{1.00}
\draw[gp path] (0.676,2.102)--(0.766,2.102);
\node[gp node right] at (0.529,2.102) {2.5};
\draw[gp path] (1.634,0.365)--(1.634,0.455);
\node[gp node center] at (1.634,0.119) {total};
\draw[gp path] (2.593,0.365)--(2.593,0.455);
\node[gp node center] at (2.593,0.119) {per-core};
\draw[gp path] (3.551,0.365)--(3.461,0.365);
\draw[gp path] (3.551,0.712)--(3.461,0.712);
\draw[gp path] (3.551,1.060)--(3.461,1.060);
\draw[gp path] (3.551,1.407)--(3.461,1.407);
\draw[gp path] (3.551,1.755)--(3.461,1.755);
\draw[gp path] (3.551,2.102)--(3.461,2.102);
\draw[gp path] (0.676,2.102)--(0.676,0.365)--(3.551,0.365)--(3.551,2.102)--cycle;
\gpfill{rgb color={0.392,0.561,1.000}} (1.251,0.365)--(1.444,0.365)--(1.444,0.861)--(1.251,0.861)--cycle;
\draw[gp path] (1.251,0.365)--(1.251,0.860)--(1.443,0.860)--(1.443,0.365)--cycle;
\gpfill{rgb color={0.392,0.561,1.000}} (2.209,0.365)--(2.402,0.365)--(2.402,1.852)--(2.209,1.852)--cycle;
\draw[gp path] (2.209,0.365)--(2.209,1.851)--(2.401,1.851)--(2.401,0.365)--cycle;
\gpfill{rgb color={0.863,0.149,0.498}} (1.443,0.365)--(1.635,0.365)--(1.635,0.897)--(1.443,0.897)--cycle;
\draw[gp path] (1.443,0.365)--(1.443,0.896)--(1.634,0.896)--(1.634,0.365)--cycle;
\gpfill{rgb color={0.863,0.149,0.498}} (2.401,0.365)--(2.594,0.365)--(2.594,1.960)--(2.401,1.960)--cycle;
\draw[gp path] (2.401,0.365)--(2.401,1.959)--(2.593,1.959)--(2.593,0.365)--cycle;
\gpfill{rgb color={0.471,0.369,0.941}} (1.634,0.365)--(1.827,0.365)--(1.827,1.045)--(1.634,1.045)--cycle;
\draw[gp path] (1.634,0.365)--(1.634,1.044)--(1.826,1.044)--(1.826,0.365)--cycle;
\gpfill{rgb color={0.471,0.369,0.941}} (2.593,0.365)--(2.785,0.365)--(2.785,1.045)--(2.593,1.045)--cycle;
\draw[gp path] (2.593,0.365)--(2.593,1.044)--(2.784,1.044)--(2.784,0.365)--cycle;
\gpfill{rgb color={0.996,0.380,0.000}} (1.826,0.365)--(2.019,0.365)--(2.019,1.048)--(1.826,1.048)--cycle;
\draw[gp path] (1.826,0.365)--(1.826,1.047)--(2.018,1.047)--(2.018,0.365)--cycle;
\gpfill{rgb color={0.996,0.380,0.000}} (2.784,0.365)--(2.977,0.365)--(2.977,1.048)--(2.784,1.048)--cycle;
\draw[gp path] (2.784,0.365)--(2.784,1.047)--(2.976,1.047)--(2.976,0.365)--cycle;
\draw[gp path] (0.676,2.102)--(0.676,0.365)--(3.551,0.365)--(3.551,2.102)--cycle;
\node[gp node center,rotate=-270.0] at (-0.218,1.233) {Normalised Tput};
\node[gp node center] at (2.113,2.348) {(a) 0 ms};
\gpdefrectangularnode{gp plot 2}{\pgfpoint{0.676cm}{0.365cm}}{\pgfpoint{3.551cm}{2.102cm}}
\gpcolor{color=gp lt color axes}
\gpsetlinetype{gp lt axes}
\gpsetdashtype{gp dt axes}
\gpsetlinewidth{0.50}
\draw[gp path] (4.736,0.365)--(7.611,0.365);
\gpcolor{color=gp lt color border}
\gpsetlinetype{gp lt border}
\gpsetdashtype{gp dt solid}
\gpsetlinewidth{1.00}
\draw[gp path] (4.736,0.365)--(4.826,0.365);
\node[gp node right] at (4.589,0.365) {0};
\gpcolor{color=gp lt color axes}
\gpsetlinetype{gp lt axes}
\gpsetdashtype{gp dt axes}
\gpsetlinewidth{0.50}
\draw[gp path] (4.736,0.613)--(7.611,0.613);
\gpcolor{color=gp lt color border}
\gpsetlinetype{gp lt border}
\gpsetdashtype{gp dt solid}
\gpsetlinewidth{1.00}
\draw[gp path] (4.736,0.613)--(4.826,0.613);
\node[gp node right] at (4.589,0.613) {0.5};
\gpcolor{color=gp lt color axes}
\gpsetlinetype{gp lt axes}
\gpsetdashtype{gp dt axes}
\gpsetlinewidth{0.50}
\draw[gp path] (4.736,0.861)--(7.611,0.861);
\gpcolor{color=gp lt color border}
\gpsetlinetype{gp lt border}
\gpsetdashtype{gp dt solid}
\gpsetlinewidth{1.00}
\draw[gp path] (4.736,0.861)--(4.826,0.861);
\node[gp node right] at (4.589,0.861) {1};
\gpcolor{color=gp lt color axes}
\gpsetlinetype{gp lt axes}
\gpsetdashtype{gp dt axes}
\gpsetlinewidth{0.50}
\draw[gp path] (4.736,1.109)--(7.611,1.109);
\gpcolor{color=gp lt color border}
\gpsetlinetype{gp lt border}
\gpsetdashtype{gp dt solid}
\gpsetlinewidth{1.00}
\draw[gp path] (4.736,1.109)--(4.826,1.109);
\node[gp node right] at (4.589,1.109) {1.5};
\gpcolor{color=gp lt color axes}
\gpsetlinetype{gp lt axes}
\gpsetdashtype{gp dt axes}
\gpsetlinewidth{0.50}
\draw[gp path] (4.736,1.358)--(7.611,1.358);
\gpcolor{color=gp lt color border}
\gpsetlinetype{gp lt border}
\gpsetdashtype{gp dt solid}
\gpsetlinewidth{1.00}
\draw[gp path] (4.736,1.358)--(4.826,1.358);
\node[gp node right] at (4.589,1.358) {2};
\gpcolor{color=gp lt color axes}
\gpsetlinetype{gp lt axes}
\gpsetdashtype{gp dt axes}
\gpsetlinewidth{0.50}
\draw[gp path] (4.736,1.606)--(7.611,1.606);
\gpcolor{color=gp lt color border}
\gpsetlinetype{gp lt border}
\gpsetdashtype{gp dt solid}
\gpsetlinewidth{1.00}
\draw[gp path] (4.736,1.606)--(4.826,1.606);
\node[gp node right] at (4.589,1.606) {2.5};
\gpcolor{color=gp lt color axes}
\gpsetlinetype{gp lt axes}
\gpsetdashtype{gp dt axes}
\gpsetlinewidth{0.50}
\draw[gp path] (4.736,1.854)--(7.611,1.854);
\gpcolor{color=gp lt color border}
\gpsetlinetype{gp lt border}
\gpsetdashtype{gp dt solid}
\gpsetlinewidth{1.00}
\draw[gp path] (4.736,1.854)--(4.826,1.854);
\node[gp node right] at (4.589,1.854) {3};
\gpcolor{color=gp lt color axes}
\gpsetlinetype{gp lt axes}
\gpsetdashtype{gp dt axes}
\gpsetlinewidth{0.50}
\draw[gp path] (4.736,2.102)--(7.611,2.102);
\gpcolor{color=gp lt color border}
\gpsetlinetype{gp lt border}
\gpsetdashtype{gp dt solid}
\gpsetlinewidth{1.00}
\draw[gp path] (4.736,2.102)--(4.826,2.102);
\node[gp node right] at (4.589,2.102) {3.5};
\draw[gp path] (5.694,0.365)--(5.694,0.455);
\node[gp node center] at (5.694,0.119) {total};
\draw[gp path] (6.653,0.365)--(6.653,0.455);
\node[gp node center] at (6.653,0.119) {per-core};
\draw[gp path] (7.611,0.365)--(7.521,0.365);
\draw[gp path] (7.611,0.613)--(7.521,0.613);
\draw[gp path] (7.611,0.861)--(7.521,0.861);
\draw[gp path] (7.611,1.109)--(7.521,1.109);
\draw[gp path] (7.611,1.358)--(7.521,1.358);
\draw[gp path] (7.611,1.606)--(7.521,1.606);
\draw[gp path] (7.611,1.854)--(7.521,1.854);
\draw[gp path] (7.611,2.102)--(7.521,2.102);
\draw[gp path] (4.736,2.102)--(4.736,0.365)--(7.611,0.365)--(7.611,2.102)--cycle;
\gpfill{rgb color={0.392,0.561,1.000}} (5.311,0.365)--(5.504,0.365)--(5.504,0.869)--(5.311,0.869)--cycle;
\draw[gp path] (5.311,0.365)--(5.311,0.868)--(5.503,0.868)--(5.503,0.365)--cycle;
\gpfill{rgb color={0.392,0.561,1.000}} (6.269,0.365)--(6.462,0.365)--(6.462,1.874)--(6.269,1.874)--cycle;
\draw[gp path] (6.269,0.365)--(6.269,1.873)--(6.461,1.873)--(6.461,0.365)--cycle;
\gpfill{rgb color={0.863,0.149,0.498}} (5.503,0.365)--(5.695,0.365)--(5.695,0.861)--(5.503,0.861)--cycle;
\draw[gp path] (5.503,0.365)--(5.503,0.860)--(5.694,0.860)--(5.694,0.365)--cycle;
\gpfill{rgb color={0.863,0.149,0.498}} (6.461,0.365)--(6.654,0.365)--(6.654,1.851)--(6.461,1.851)--cycle;
\draw[gp path] (6.461,0.365)--(6.461,1.850)--(6.653,1.850)--(6.653,0.365)--cycle;
\gpfill{rgb color={0.471,0.369,0.941}} (5.694,0.365)--(5.887,0.365)--(5.887,0.876)--(5.694,0.876)--cycle;
\draw[gp path] (5.694,0.365)--(5.694,0.875)--(5.886,0.875)--(5.886,0.365)--cycle;
\gpfill{rgb color={0.471,0.369,0.941}} (6.653,0.365)--(6.845,0.365)--(6.845,0.876)--(6.653,0.876)--cycle;
\draw[gp path] (6.653,0.365)--(6.653,0.875)--(6.844,0.875)--(6.844,0.365)--cycle;
\gpfill{rgb color={0.996,0.380,0.000}} (5.886,0.365)--(6.079,0.365)--(6.079,0.860)--(5.886,0.860)--cycle;
\draw[gp path] (5.886,0.365)--(5.886,0.859)--(6.078,0.859)--(6.078,0.365)--cycle;
\gpfill{rgb color={0.996,0.380,0.000}} (6.844,0.365)--(7.037,0.365)--(7.037,0.860)--(6.844,0.860)--cycle;
\draw[gp path] (6.844,0.365)--(6.844,0.859)--(7.036,0.859)--(7.036,0.365)--cycle;
\draw[gp path] (4.736,2.102)--(4.736,0.365)--(7.611,0.365)--(7.611,2.102)--cycle;
\node[gp node center,rotate=-270.0] at (3.841,1.233) {Normalised Tput};
\node[gp node center] at (6.173,2.348) {(b) 5 ms};
\gpdefrectangularnode{gp plot 3}{\pgfpoint{4.736cm}{0.365cm}}{\pgfpoint{7.611cm}{2.102cm}}
\end{tikzpicture}
  \caption{TAS and \sys throughput for 1 and 3 fast path cores, 
    normalised to siloed TAS throughput with three cores, 
    under constant transmit rates (0 ms) and inter-burst intervals of 5 ms.
    \sys facilitates sharing and reduces stranding for improved per-core efficiency
    compared to siloed datapaths.}%
  \label{fig:memcached-efficiency}%
\end{figure}

\subsection{\sys Incurs Low Overhead}
\textbf{\textit{Latency.}}
To evaluate the effect of time protection on datapath latency, 
we use a minimal RPC deployment with one client and one server, 
a configuration that minimises queueing and makes any 
added latency directly visible.
On each machine, the application connects to a single local TAS or 
\sys instance, and both systems are configured with one fast path core. 
The client runs a latency-sensitive benchmark with a single connection 
and a single outstanding request, and we record the end-to-end latency 
CDF in \autoref{fig:rpc-latency}. 
\sys achieves a median latency of 6$\mu$s, compared to 7$\mu$s with TAS. 
At the tail, \sys reaches a 99th-percentile latency of 6$\mu$s, versus 
8$\mu$s for TAS. \sys slightly outperforms TAS because time protection 
also introduces a modest pacing effect, 
which reduces transient congestion.
Across the distribution, 
\sys closely tracks TAS, showing that our changes to implement 
time protection in the TAS datapath preserve low RPC latency.

\begin{figure}%
  \centering%
\begin{tikzpicture}[gnuplot]
\tikzset{every node/.append style={font={\fontsize{8.0pt}{9.6pt}\selectfont}}}
\path (0.000,0.000) rectangle (8.458,3.048);
\gpcolor{color=gp lt color axes}
\gpsetlinetype{gp lt axes}
\gpsetdashtype{gp dt axes}
\gpsetlinewidth{0.50}
\draw[gp path] (1.054,0.787)--(8.016,0.787);
\gpcolor{color=gp lt color border}
\gpsetlinetype{gp lt border}
\gpsetdashtype{gp dt solid}
\gpsetlinewidth{1.00}
\draw[gp path] (1.054,0.787)--(1.144,0.787);
\node[gp node right] at (0.907,0.787) {$0$};
\gpcolor{color=gp lt color axes}
\gpsetlinetype{gp lt axes}
\gpsetdashtype{gp dt axes}
\gpsetlinewidth{0.50}
\draw[gp path] (1.054,1.190)--(8.016,1.190);
\gpcolor{color=gp lt color border}
\gpsetlinetype{gp lt border}
\gpsetdashtype{gp dt solid}
\gpsetlinewidth{1.00}
\draw[gp path] (1.054,1.190)--(1.144,1.190);
\node[gp node right] at (0.907,1.190) {$0.2$};
\gpcolor{color=gp lt color axes}
\gpsetlinetype{gp lt axes}
\gpsetdashtype{gp dt axes}
\gpsetlinewidth{0.50}
\draw[gp path] (1.054,1.593)--(8.016,1.593);
\gpcolor{color=gp lt color border}
\gpsetlinetype{gp lt border}
\gpsetdashtype{gp dt solid}
\gpsetlinewidth{1.00}
\draw[gp path] (1.054,1.593)--(1.144,1.593);
\node[gp node right] at (0.907,1.593) {$0.4$};
\gpcolor{color=gp lt color axes}
\gpsetlinetype{gp lt axes}
\gpsetdashtype{gp dt axes}
\gpsetlinewidth{0.50}
\draw[gp path] (1.054,1.995)--(3.269,1.995);
\draw[gp path] (5.801,1.995)--(8.016,1.995);
\gpcolor{color=gp lt color border}
\gpsetlinetype{gp lt border}
\gpsetdashtype{gp dt solid}
\gpsetlinewidth{1.00}
\draw[gp path] (1.054,1.995)--(1.144,1.995);
\node[gp node right] at (0.907,1.995) {$0.6$};
\gpcolor{color=gp lt color axes}
\gpsetlinetype{gp lt axes}
\gpsetdashtype{gp dt axes}
\gpsetlinewidth{0.50}
\draw[gp path] (1.054,2.398)--(3.269,2.398);
\draw[gp path] (5.801,2.398)--(8.016,2.398);
\gpcolor{color=gp lt color border}
\gpsetlinetype{gp lt border}
\gpsetdashtype{gp dt solid}
\gpsetlinewidth{1.00}
\draw[gp path] (1.054,2.398)--(1.144,2.398);
\node[gp node right] at (0.907,2.398) {$0.8$};
\gpcolor{color=gp lt color axes}
\gpsetlinetype{gp lt axes}
\gpsetdashtype{gp dt axes}
\gpsetlinewidth{0.50}
\draw[gp path] (1.054,2.801)--(8.016,2.801);
\gpcolor{color=gp lt color border}
\gpsetlinetype{gp lt border}
\gpsetdashtype{gp dt solid}
\gpsetlinewidth{1.00}
\draw[gp path] (1.054,2.801)--(1.144,2.801);
\node[gp node right] at (0.907,2.801) {$1$};
\draw[gp path] (1.054,0.787)--(1.054,0.877);
\node[gp node center] at (1.054,0.541) {$0$};
\draw[gp path] (2.446,0.787)--(2.446,0.877);
\node[gp node center] at (2.446,0.541) {$20$};
\draw[gp path] (3.839,0.787)--(3.839,0.877);
\node[gp node center] at (3.839,0.541) {$40$};
\draw[gp path] (5.231,0.787)--(5.231,0.877);
\node[gp node center] at (5.231,0.541) {$60$};
\draw[gp path] (6.624,0.787)--(6.624,0.877);
\node[gp node center] at (6.624,0.541) {$80$};
\draw[gp path] (8.016,0.787)--(8.016,0.877);
\node[gp node center] at (8.016,0.541) {$100$};
\draw[gp path] (1.054,2.801)--(1.054,0.787)--(8.016,0.787)--(8.016,2.801)--cycle;
\node[gp node right] at (4.739,2.498) {shared-tas};
\gpcolor{rgb color={0.471,0.369,0.941}}
\gpsetdashtype{gp dt 2}
\gpsetlinewidth{3.00}
\draw[gp path] (4.886,2.498)--(5.654,2.498);
\draw[gp path] (1.054,0.787)--(1.472,1.495)--(1.541,2.756)--(1.611,2.796)--(1.681,2.798)%
  --(1.750,2.799)--(1.820,2.800)--(1.889,2.800)--(1.959,2.800)--(2.029,2.800)--(2.098,2.801)%
  --(2.168,2.801)--(2.238,2.801)--(2.307,2.801)--(2.377,2.801)--(2.446,2.801)--(2.516,2.801)%
  --(2.586,2.801)--(2.655,2.801)--(2.725,2.801)--(2.795,2.801)--(2.864,2.801)--(3.003,2.801)%
  --(3.073,2.801)--(3.351,2.801)--(3.421,2.801)--(3.491,2.801)--(3.560,2.801)--(3.630,2.801)%
  --(3.908,2.801)--(3.978,2.801)--(4.048,2.801)--(4.117,2.801)--(4.257,2.801)--(4.326,2.801)%
  --(4.535,2.801)--(4.605,2.801)--(5.162,2.801)--(6.206,2.801)--(6.902,2.801)--(6.972,2.801)%
  --(7.111,2.801)--(7.250,2.801)--(7.320,2.801)--(8.016,2.801);
\gpcolor{color=gp lt color border}
\node[gp node right] at (4.739,2.252) {\sysplot};
\gpcolor{rgb color={0.392,0.561,1.000}}
\gpsetdashtype{gp dt 1}
\draw[gp path] (4.886,2.252)--(5.654,2.252);
\draw[gp path] (1.054,0.787)--(1.402,1.275)--(1.472,2.793)--(1.541,2.797)--(1.611,2.799)%
  --(1.681,2.800)--(1.750,2.801)--(1.820,2.801)--(1.889,2.801)--(1.959,2.801)--(2.029,2.801)%
  --(2.098,2.801)--(2.168,2.801)--(2.238,2.801)--(2.307,2.801)--(2.377,2.801)--(2.446,2.801)%
  --(2.516,2.801)--(2.586,2.801)--(2.655,2.801)--(2.725,2.801)--(2.795,2.801)--(2.864,2.801)%
  --(3.003,2.801)--(3.143,2.801)--(3.560,2.801)--(3.630,2.801)--(3.700,2.801)--(3.769,2.801)%
  --(3.839,2.801)--(3.908,2.801)--(3.978,2.801)--(4.048,2.801)--(4.257,2.801)--(4.396,2.801)%
  --(4.465,2.801)--(4.535,2.801)--(4.605,2.801)--(4.674,2.801)--(4.744,2.801)--(5.092,2.801)%
  --(5.370,2.801)--(7.738,2.801)--(7.877,2.801)--(8.016,2.801);
\gpcolor{color=gp lt color border}
\node[gp node right] at (4.739,2.006) {linux};
\gpcolor{rgb color={0.996,0.380,0.000}}
\gpsetdashtype{gp dt 3}
\draw[gp path] (4.886,2.006)--(5.654,2.006);
\draw[gp path] (1.054,0.787)--(5.231,0.787)--(5.301,0.787)--(5.370,0.787)--(5.440,0.787)%
  --(5.510,0.787)--(5.579,0.788)--(5.649,0.788)--(5.719,0.788)--(5.788,0.789)--(5.858,0.790)%
  --(5.927,0.791)--(5.997,0.794)--(6.067,0.803)--(6.136,0.812)--(6.206,0.838)--(6.276,0.916)%
  --(6.345,1.136)--(6.415,1.556)--(6.484,1.899)--(6.554,2.051)--(6.624,2.125)--(6.693,2.175)%
  --(6.763,2.228)--(6.832,2.338)--(6.902,2.504)--(6.972,2.640)--(7.041,2.709)--(7.111,2.740)%
  --(7.181,2.756)--(7.250,2.767)--(7.320,2.775)--(7.389,2.780)--(7.459,2.786)--(7.529,2.790)%
  --(7.598,2.793)--(7.668,2.795)--(7.738,2.797)--(7.807,2.797)--(7.877,2.798)--(7.946,2.798)%
  --(8.016,2.799);
\gpcolor{color=gp lt color border}
\gpsetdashtype{gp dt solid}
\gpsetlinewidth{1.00}
\draw[gp path] (1.054,2.801)--(1.054,0.787)--(8.016,0.787)--(8.016,2.801)--cycle;
\node[gp node center,rotate=-270.0] at (0.232,1.794) {CDF};
\node[gp node center] at (4.535,0.172) {Latency [us]};
\gpdefrectangularnode{gp plot 1}{\pgfpoint{1.054cm}{0.787cm}}{\pgfpoint{8.016cm}{2.801cm}}
\end{tikzpicture}
  \caption{RTT Latency CDF for RPC client and server.
    \sys closely matches TAS across the latency distribution, 
    indicating that time protection preserves low RPC latency in the TAS datapath.}%
  \label{fig:rpc-latency}%
\end{figure}

\textbf{\textit{Throughput.}}
To evaluate the impact of time protection on datapath throughput, 
we run three RPC clients and three RPC echo servers across two machines. 
On each machine, all applications connect to a single local TAS or 
\sys instance; each application is assigned three cores,
for a total of nine cores, and both datapaths 
are configured with three fast path cores. We vary the RPC message 
size and record the aggregate client throughput at each size 
under TAS, \sys and Linux. The results in \autoref{fig:rpc-throughput-scaling}
show that \sys closely matches 
TAS for small message sizes and incurs a small 5\% throughput reduction for
4096 B messages. This happens because the budget mechanism slightly
throttles senders, but this difference can be further reduced by 
tuning budget parameters.
Linux achieves higher throughput for large
messages, consistent with TAS's design point on small-message workloads, which
\sys inherits.

\textbf{\textit{Scalability.}}
To evaluate the synchronisation overhead of time protection and its 
ability to scale across cores, we run three RPC clients and three RPC 
echo servers across two machines. On each machine, all applications connect 
to a single local TAS or \sys instance, and each client and server 
is assigned three application cores; each client opens 3000 connections
for a total of 9000 connections and sends 64 B messages. 
We vary the number of fast path cores assigned to TAS and \sys and measure 
aggregate throughput at each configuration. We display the results in
\autoref{fig:rpc-throughput-scaling}.
\sys scales with increasing 
fast path cores at nearly the same rate as TAS, indicating that time 
protection introduces minimal synchronisation overhead while 
preserving multicore scalability.

\begin{figure}%
  \centering%
\begin{tikzpicture}[gnuplot]
\tikzset{every node/.append style={font={\fontsize{8.0pt}{9.6pt}\selectfont}}}
\path (0.000,0.000) rectangle (8.458,3.048);
\gpcolor{color=gp lt color border}
\gpcolor{rgb color={0.471,0.369,0.941}}
\gpsetlinewidth{1.00}
\gpsetpointsize{4.00}
\gp3point{gp mark 5}{}{(1.640,2.860)}
\gpcolor{color=gp lt color border}
\gpcolor{rgb color={0.392,0.561,1.000}}
\gp3point{gp mark 7}{}{(3.847,2.860)}
\gpcolor{color=gp lt color border}
\gpcolor{rgb color={0.996,0.380,0.000}}
\gp3point{gp mark 11}{}{(5.827,2.860)}
\gpcolor{color=gp lt color border}
\node[gp node left] at (1.944,2.860) {shared-tas};
\node[gp node left] at (4.228,2.860) {\sysplot};
\node[gp node left] at (6.512,2.860) {linux};
\gpcolor{rgb color={0.471,0.369,0.941}}
\gpsetlinetype{gp lt border}
\gpsetdashtype{gp dt solid}
\gpsetlinewidth{3.00}
\draw[gp path](1.336,2.860)--(1.945,2.860);
\gpcolor{rgb color={0.392,0.561,1.000}}
\draw[gp path](3.543,2.860)--(4.152,2.860);
\gpcolor{rgb color={0.996,0.380,0.000}}
\draw[gp path](5.523,2.860)--(6.132,2.860);
\gpdefrectangularnode{gp plot 1}{\pgfpoint{0.422cm}{2.621cm}}{\pgfpoint{8.034cm}{3.047cm}}
\gpcolor{color=gp lt color axes}
\gpsetlinetype{gp lt axes}
\gpsetdashtype{gp dt axes}
\gpsetlinewidth{0.50}
\draw[gp path] (0.676,0.365)--(3.551,0.365);
\gpcolor{color=gp lt color border}
\gpsetlinetype{gp lt border}
\gpsetdashtype{gp dt solid}
\gpsetlinewidth{1.00}
\draw[gp path] (0.676,0.365)--(0.766,0.365);
\node[gp node right] at (0.529,0.365) {$0$};
\gpcolor{color=gp lt color axes}
\gpsetlinetype{gp lt axes}
\gpsetdashtype{gp dt axes}
\gpsetlinewidth{0.50}
\draw[gp path] (0.676,0.868)--(3.551,0.868);
\gpcolor{color=gp lt color border}
\gpsetlinetype{gp lt border}
\gpsetdashtype{gp dt solid}
\gpsetlinewidth{1.00}
\draw[gp path] (0.676,0.868)--(0.766,0.868);
\node[gp node right] at (0.529,0.868) {$5$};
\gpcolor{color=gp lt color axes}
\gpsetlinetype{gp lt axes}
\gpsetdashtype{gp dt axes}
\gpsetlinewidth{0.50}
\draw[gp path] (0.676,1.371)--(3.551,1.371);
\gpcolor{color=gp lt color border}
\gpsetlinetype{gp lt border}
\gpsetdashtype{gp dt solid}
\gpsetlinewidth{1.00}
\draw[gp path] (0.676,1.371)--(0.766,1.371);
\node[gp node right] at (0.529,1.371) {$10$};
\gpcolor{color=gp lt color axes}
\gpsetlinetype{gp lt axes}
\gpsetdashtype{gp dt axes}
\gpsetlinewidth{0.50}
\draw[gp path] (0.676,1.873)--(3.551,1.873);
\gpcolor{color=gp lt color border}
\gpsetlinetype{gp lt border}
\gpsetdashtype{gp dt solid}
\gpsetlinewidth{1.00}
\draw[gp path] (0.676,1.873)--(0.766,1.873);
\node[gp node right] at (0.529,1.873) {$15$};
\gpcolor{color=gp lt color axes}
\gpsetlinetype{gp lt axes}
\gpsetdashtype{gp dt axes}
\gpsetlinewidth{0.50}
\draw[gp path] (0.676,2.376)--(3.551,2.376);
\gpcolor{color=gp lt color border}
\gpsetlinetype{gp lt border}
\gpsetdashtype{gp dt solid}
\gpsetlinewidth{1.00}
\draw[gp path] (0.676,2.376)--(0.766,2.376);
\node[gp node right] at (0.529,2.376) {$20$};
\draw[gp path] (1.035,0.365)--(1.035,0.455);
\node[gp node center] at (1.035,0.119) {64};
\draw[gp path] (1.395,0.365)--(1.395,0.455);
\draw[gp path] (1.754,0.365)--(1.754,0.455);
\node[gp node center] at (1.754,0.119) {256};
\draw[gp path] (2.114,0.365)--(2.114,0.455);
\draw[gp path] (2.473,0.365)--(2.473,0.455);
\node[gp node center] at (2.473,0.119) {1024};
\draw[gp path] (2.832,0.365)--(2.832,0.455);
\draw[gp path] (3.192,0.365)--(3.192,0.455);
\node[gp node center] at (3.192,0.119) {4096};
\draw[gp path] (3.551,0.365)--(3.461,0.365);
\draw[gp path] (3.551,0.868)--(3.461,0.868);
\draw[gp path] (3.551,1.371)--(3.461,1.371);
\draw[gp path] (3.551,1.873)--(3.461,1.873);
\draw[gp path] (3.551,2.376)--(3.461,2.376);
\draw[gp path] (0.676,2.376)--(0.676,0.365)--(3.551,0.365)--(3.551,2.376)--cycle;
\gpcolor{rgb color={0.471,0.369,0.941}}
\gpsetlinewidth{3.00}
\draw[gp path] (1.035,0.868)--(1.395,1.180)--(1.754,1.267)--(2.114,1.746)--(2.473,1.925)%
  --(2.832,1.927)--(3.192,1.913);
\gp3point{gp mark 5}{}{(1.035,0.868)}
\gp3point{gp mark 5}{}{(1.395,1.180)}
\gp3point{gp mark 5}{}{(1.754,1.267)}
\gp3point{gp mark 5}{}{(2.114,1.746)}
\gp3point{gp mark 5}{}{(2.473,1.925)}
\gp3point{gp mark 5}{}{(2.832,1.927)}
\gp3point{gp mark 5}{}{(3.192,1.913)}
\gpcolor{rgb color={0.996,0.380,0.000}}
\draw[gp path] (1.035,0.428)--(1.395,0.492)--(1.754,0.609)--(2.114,0.826)--(2.473,1.183)%
  --(2.832,1.700)--(3.192,2.230);
\gp3point{gp mark 11}{}{(1.035,0.428)}
\gp3point{gp mark 11}{}{(1.395,0.492)}
\gp3point{gp mark 11}{}{(1.754,0.609)}
\gp3point{gp mark 11}{}{(2.114,0.826)}
\gp3point{gp mark 11}{}{(2.473,1.183)}
\gp3point{gp mark 11}{}{(2.832,1.700)}
\gp3point{gp mark 11}{}{(3.192,2.230)}
\gpcolor{rgb color={0.392,0.561,1.000}}
\draw[gp path] (1.035,0.856)--(1.395,1.201)--(1.754,1.329)--(2.114,1.703)--(2.473,1.809)%
  --(2.832,1.829)--(3.192,1.831);
\gp3point{gp mark 7}{}{(1.035,0.856)}
\gp3point{gp mark 7}{}{(1.395,1.201)}
\gp3point{gp mark 7}{}{(1.754,1.329)}
\gp3point{gp mark 7}{}{(2.114,1.703)}
\gp3point{gp mark 7}{}{(2.473,1.809)}
\gp3point{gp mark 7}{}{(2.832,1.829)}
\gp3point{gp mark 7}{}{(3.192,1.831)}
\gpcolor{color=gp lt color border}
\gpsetlinewidth{1.00}
\draw[gp path] (0.676,2.376)--(0.676,0.365)--(3.551,0.365)--(3.551,2.376)--cycle;
\node[gp node center,rotate=-270.0] at (-0.072,1.370) {Tput [Gbit/s]};
\node[gp node center] at (2.113,-0.249) {Message Size [B]};
\gpdefrectangularnode{gp plot 2}{\pgfpoint{0.676cm}{0.365cm}}{\pgfpoint{3.551cm}{2.376cm}}
\gpcolor{color=gp lt color axes}
\gpsetlinetype{gp lt axes}
\gpsetdashtype{gp dt axes}
\gpsetlinewidth{0.50}
\draw[gp path] (4.736,0.365)--(7.611,0.365);
\gpcolor{color=gp lt color border}
\gpsetlinetype{gp lt border}
\gpsetdashtype{gp dt solid}
\gpsetlinewidth{1.00}
\draw[gp path] (4.736,0.365)--(4.826,0.365);
\node[gp node right] at (4.589,0.365) {$0$};
\gpcolor{color=gp lt color axes}
\gpsetlinetype{gp lt axes}
\gpsetdashtype{gp dt axes}
\gpsetlinewidth{0.50}
\draw[gp path] (4.736,0.868)--(7.611,0.868);
\gpcolor{color=gp lt color border}
\gpsetlinetype{gp lt border}
\gpsetdashtype{gp dt solid}
\gpsetlinewidth{1.00}
\draw[gp path] (4.736,0.868)--(4.826,0.868);
\node[gp node right] at (4.589,0.868) {$5$};
\gpcolor{color=gp lt color axes}
\gpsetlinetype{gp lt axes}
\gpsetdashtype{gp dt axes}
\gpsetlinewidth{0.50}
\draw[gp path] (4.736,1.371)--(7.611,1.371);
\gpcolor{color=gp lt color border}
\gpsetlinetype{gp lt border}
\gpsetdashtype{gp dt solid}
\gpsetlinewidth{1.00}
\draw[gp path] (4.736,1.371)--(4.826,1.371);
\node[gp node right] at (4.589,1.371) {$10$};
\gpcolor{color=gp lt color axes}
\gpsetlinetype{gp lt axes}
\gpsetdashtype{gp dt axes}
\gpsetlinewidth{0.50}
\draw[gp path] (4.736,1.873)--(7.611,1.873);
\gpcolor{color=gp lt color border}
\gpsetlinetype{gp lt border}
\gpsetdashtype{gp dt solid}
\gpsetlinewidth{1.00}
\draw[gp path] (4.736,1.873)--(4.826,1.873);
\node[gp node right] at (4.589,1.873) {$15$};
\gpcolor{color=gp lt color axes}
\gpsetlinetype{gp lt axes}
\gpsetdashtype{gp dt axes}
\gpsetlinewidth{0.50}
\draw[gp path] (4.736,2.376)--(7.611,2.376);
\gpcolor{color=gp lt color border}
\gpsetlinetype{gp lt border}
\gpsetdashtype{gp dt solid}
\gpsetlinewidth{1.00}
\draw[gp path] (4.736,2.376)--(4.826,2.376);
\node[gp node right] at (4.589,2.376) {$20$};
\draw[gp path] (4.736,0.365)--(4.736,0.455);
\node[gp node center] at (4.736,0.119) {$1$};
\draw[gp path] (5.095,0.365)--(5.095,0.455);
\node[gp node center] at (5.095,0.119) {$2$};
\draw[gp path] (5.455,0.365)--(5.455,0.455);
\node[gp node center] at (5.455,0.119) {$3$};
\draw[gp path] (5.814,0.365)--(5.814,0.455);
\node[gp node center] at (5.814,0.119) {$4$};
\draw[gp path] (6.174,0.365)--(6.174,0.455);
\node[gp node center] at (6.174,0.119) {$5$};
\draw[gp path] (6.533,0.365)--(6.533,0.455);
\node[gp node center] at (6.533,0.119) {$6$};
\draw[gp path] (6.892,0.365)--(6.892,0.455);
\node[gp node center] at (6.892,0.119) {$7$};
\draw[gp path] (7.252,0.365)--(7.252,0.455);
\node[gp node center] at (7.252,0.119) {$8$};
\draw[gp path] (7.611,0.365)--(7.611,0.455);
\node[gp node center] at (7.611,0.119) {$9$};
\draw[gp path] (7.611,0.365)--(7.521,0.365);
\draw[gp path] (7.611,0.868)--(7.521,0.868);
\draw[gp path] (7.611,1.371)--(7.521,1.371);
\draw[gp path] (7.611,1.873)--(7.521,1.873);
\draw[gp path] (7.611,2.376)--(7.521,2.376);
\draw[gp path] (4.736,2.376)--(4.736,0.365)--(7.611,0.365)--(7.611,2.376)--cycle;
\gpcolor{rgb color={0.471,0.369,0.941}}
\gpsetlinewidth{3.00}
\draw[gp path] (4.736,0.861)--(5.095,1.102)--(5.455,1.350)--(5.814,1.538)--(6.174,1.711)%
  --(6.533,1.874)--(6.892,1.987)--(7.252,2.182);
\gp3point{gp mark 5}{}{(4.736,0.861)}
\gp3point{gp mark 5}{}{(5.095,1.102)}
\gp3point{gp mark 5}{}{(5.455,1.350)}
\gp3point{gp mark 5}{}{(5.814,1.538)}
\gp3point{gp mark 5}{}{(6.174,1.711)}
\gp3point{gp mark 5}{}{(6.533,1.874)}
\gp3point{gp mark 5}{}{(6.892,1.987)}
\gp3point{gp mark 5}{}{(7.252,2.182)}
\gpcolor{rgb color={0.392,0.561,1.000}}
\draw[gp path] (4.736,0.900)--(5.095,1.102)--(5.455,1.325)--(5.814,1.572)--(6.174,1.736)%
  --(6.533,1.893)--(6.892,2.014)--(7.252,2.237);
\gp3point{gp mark 7}{}{(4.736,0.900)}
\gp3point{gp mark 7}{}{(5.095,1.102)}
\gp3point{gp mark 7}{}{(5.455,1.325)}
\gp3point{gp mark 7}{}{(5.814,1.572)}
\gp3point{gp mark 7}{}{(6.174,1.736)}
\gp3point{gp mark 7}{}{(6.533,1.893)}
\gp3point{gp mark 7}{}{(6.892,2.014)}
\gp3point{gp mark 7}{}{(7.252,2.237)}
\gpcolor{color=gp lt color border}
\gpsetlinewidth{1.00}
\draw[gp path] (4.736,2.376)--(4.736,0.365)--(7.611,0.365)--(7.611,2.376)--cycle;
\node[gp node center,rotate=-270.0] at (3.988,1.370) {Tput [MReq/s]};
\node[gp node center] at (6.173,-0.249) {Fast Path Cores};
\gpdefrectangularnode{gp plot 3}{\pgfpoint{4.736cm}{0.365cm}}{\pgfpoint{7.611cm}{2.376cm}}
\end{tikzpicture}
  \caption{RPC client and server throughput under increasing 
    message sizes or number of fast path cores.}%
  \label{fig:rpc-throughput-scaling}%
\end{figure}

\subsection{Budget Parameter Sensitivity}
\autoref{fig:sensitivity} shows how the boost, budget cap, and update period
affect victim performance under adversarial interference.
For this experiment, the adversary creates a load imbalance by using 9 cores to
open a total of 900 connections, while the victim opens one connection on one
core.
If the boost parameter is too high, the adversary is not throttled
sufficiently because its budget is fully replenished to the capped amount every
round.
If it is too low, the victim suffers a small tail latency increase due to
throttling.
When varying the budget cap, the adversary bursts every 250 ms for 250 ms.
Large caps allow the adversary to accumulate credits while momentarily idle and
later affect the victim, whereas overly small caps over-regulate both tenants.
Finally, shorter update periods help maintain low $\mu$s-scale tail latencies.

\begin{figure}%
  \centering%
\begin{tikzpicture}[gnuplot]
\tikzset{every node/.append style={font={\fontsize{8.0pt}{9.6pt}\selectfont}}}
\path (0.000,0.000) rectangle (7.366,3.810);
\gpcolor{color=gp lt color axes}
\gpsetlinetype{gp lt axes}
\gpsetdashtype{gp dt axes}
\gpsetlinewidth{0.50}
\draw[gp path] (0.441,1.962)--(2.700,1.962);
\gpcolor{color=gp lt color border}
\gpsetlinetype{gp lt border}
\gpsetdashtype{gp dt solid}
\gpsetlinewidth{1.00}
\draw[gp path] (0.441,1.962)--(0.531,1.962);
\node[gp node right] at (0.441,1.962) {0};
\gpcolor{color=gp lt color axes}
\gpsetlinetype{gp lt axes}
\gpsetdashtype{gp dt axes}
\gpsetlinewidth{0.50}
\draw[gp path] (0.441,2.159)--(2.700,2.159);
\gpcolor{color=gp lt color border}
\gpsetlinetype{gp lt border}
\gpsetdashtype{gp dt solid}
\gpsetlinewidth{1.00}
\draw[gp path] (0.441,2.159)--(0.531,2.159);
\gpcolor{color=gp lt color axes}
\gpsetlinetype{gp lt axes}
\gpsetdashtype{gp dt axes}
\gpsetlinewidth{0.50}
\draw[gp path] (0.441,2.355)--(2.700,2.355);
\gpcolor{color=gp lt color border}
\gpsetlinetype{gp lt border}
\gpsetdashtype{gp dt solid}
\gpsetlinewidth{1.00}
\draw[gp path] (0.441,2.355)--(0.531,2.355);
\node[gp node right] at (0.441,2.355) {400};
\gpcolor{color=gp lt color axes}
\gpsetlinetype{gp lt axes}
\gpsetdashtype{gp dt axes}
\gpsetlinewidth{0.50}
\draw[gp path] (0.441,2.552)--(2.700,2.552);
\gpcolor{color=gp lt color border}
\gpsetlinetype{gp lt border}
\gpsetdashtype{gp dt solid}
\gpsetlinewidth{1.00}
\draw[gp path] (0.441,2.552)--(0.531,2.552);
\gpcolor{color=gp lt color axes}
\gpsetlinetype{gp lt axes}
\gpsetdashtype{gp dt axes}
\gpsetlinewidth{0.50}
\draw[gp path] (0.441,2.749)--(2.700,2.749);
\gpcolor{color=gp lt color border}
\gpsetlinetype{gp lt border}
\gpsetdashtype{gp dt solid}
\gpsetlinewidth{1.00}
\draw[gp path] (0.441,2.749)--(0.531,2.749);
\node[gp node right] at (0.441,2.749) {800};
\gpcolor{color=gp lt color axes}
\gpsetlinetype{gp lt axes}
\gpsetdashtype{gp dt axes}
\gpsetlinewidth{0.50}
\draw[gp path] (0.441,2.945)--(2.700,2.945);
\gpcolor{color=gp lt color border}
\gpsetlinetype{gp lt border}
\gpsetdashtype{gp dt solid}
\gpsetlinewidth{1.00}
\draw[gp path] (0.441,2.945)--(0.531,2.945);
\gpcolor{color=gp lt color axes}
\gpsetlinetype{gp lt axes}
\gpsetdashtype{gp dt axes}
\gpsetlinewidth{0.50}
\draw[gp path] (0.441,3.142)--(2.700,3.142);
\gpcolor{color=gp lt color border}
\gpsetlinetype{gp lt border}
\gpsetdashtype{gp dt solid}
\gpsetlinewidth{1.00}
\draw[gp path] (0.441,3.142)--(0.531,3.142);
\node[gp node right] at (0.441,3.142) {1200};
\gpcolor{color=gp lt color axes}
\gpsetlinetype{gp lt axes}
\gpsetdashtype{gp dt axes}
\gpsetlinewidth{0.50}
\draw[gp path] (0.441,3.339)--(2.700,3.339);
\gpcolor{color=gp lt color border}
\gpsetlinetype{gp lt border}
\gpsetdashtype{gp dt solid}
\gpsetlinewidth{1.00}
\draw[gp path] (0.441,3.339)--(0.531,3.339);
\gpcolor{color=gp lt color axes}
\gpsetlinetype{gp lt axes}
\gpsetdashtype{gp dt axes}
\gpsetlinewidth{0.50}
\draw[gp path] (0.441,3.535)--(2.700,3.535);
\gpcolor{color=gp lt color border}
\gpsetlinetype{gp lt border}
\gpsetdashtype{gp dt solid}
\gpsetlinewidth{1.00}
\draw[gp path] (0.441,3.535)--(0.531,3.535);
\node[gp node right] at (0.441,3.535) {1600};
\draw[gp path] (0.818,1.962)--(0.818,2.052);
\draw[gp path] (1.194,1.962)--(1.194,2.052);
\draw[gp path] (1.571,1.962)--(1.571,2.052);
\draw[gp path] (1.947,1.962)--(1.947,2.052);
\draw[gp path] (2.324,1.962)--(2.324,2.052);
\draw[gp path] (2.700,1.962)--(2.610,1.962);
\draw[gp path] (0.441,1.962)--(0.531,1.962);
\draw[gp path] (2.700,2.159)--(2.610,2.159);
\draw[gp path] (0.441,2.159)--(0.531,2.159);
\draw[gp path] (2.700,2.355)--(2.610,2.355);
\draw[gp path] (0.441,2.355)--(0.531,2.355);
\draw[gp path] (2.700,2.552)--(2.610,2.552);
\draw[gp path] (0.441,2.552)--(0.531,2.552);
\draw[gp path] (2.700,2.749)--(2.610,2.749);
\draw[gp path] (0.441,2.749)--(0.531,2.749);
\draw[gp path] (2.700,2.945)--(2.610,2.945);
\draw[gp path] (0.441,2.945)--(0.531,2.945);
\draw[gp path] (2.700,3.142)--(2.610,3.142);
\draw[gp path] (0.441,3.142)--(0.531,3.142);
\draw[gp path] (2.700,3.339)--(2.610,3.339);
\draw[gp path] (0.441,3.339)--(0.531,3.339);
\draw[gp path] (2.700,3.535)--(2.610,3.535);
\draw[gp path] (0.441,3.535)--(0.531,3.535);
\draw[gp path] (0.441,3.732)--(0.441,1.962)--(2.700,1.962)--(2.700,3.732)--cycle;
\gpcolor{rgb color={0.392,0.561,1.000}}
\gpsetlinewidth{3.00}
\draw[gp path] (0.818,2.375)--(1.194,2.793)--(1.571,2.787)--(1.947,2.462)--(2.324,2.388);
\gpsetpointsize{4.00}
\gp3point{gp mark 7}{}{(0.818,2.375)}
\gp3point{gp mark 7}{}{(1.194,2.793)}
\gp3point{gp mark 7}{}{(1.571,2.787)}
\gp3point{gp mark 7}{}{(1.947,2.462)}
\gp3point{gp mark 7}{}{(2.324,2.388)}
\gpcolor{rgb color={0.471,0.369,0.941}}
\draw[gp path] (0.818,2.223)--(1.194,2.560)--(1.571,2.896)--(1.947,3.088)--(2.324,3.088);
\gp3point{gp mark 5}{}{(0.818,2.223)}
\gp3point{gp mark 5}{}{(1.194,2.560)}
\gp3point{gp mark 5}{}{(1.571,2.896)}
\gp3point{gp mark 5}{}{(1.947,3.088)}
\gp3point{gp mark 5}{}{(2.324,3.088)}
\gpcolor{color=gp lt color border}
\gpsetlinewidth{1.00}
\draw[gp path] (0.441,3.732)--(0.441,1.962)--(2.700,1.962)--(2.700,3.732)--cycle;
\node[gp node center,rotate=-270.0] at (-0.380,2.847) {Tput [kRps]};
\gpdefrectangularnode{gp plot 1}{\pgfpoint{0.441cm}{1.962cm}}{\pgfpoint{2.700cm}{3.732cm}}
\gpcolor{color=gp lt color axes}
\gpsetlinetype{gp lt axes}
\gpsetdashtype{gp dt axes}
\gpsetlinewidth{0.50}
\draw[gp path] (2.700,1.962)--(4.959,1.962);
\gpcolor{color=gp lt color border}
\gpsetlinetype{gp lt border}
\gpsetdashtype{gp dt solid}
\gpsetlinewidth{1.00}
\draw[gp path] (2.700,1.962)--(2.790,1.962);
\gpcolor{color=gp lt color axes}
\gpsetlinetype{gp lt axes}
\gpsetdashtype{gp dt axes}
\gpsetlinewidth{0.50}
\draw[gp path] (2.700,2.159)--(4.959,2.159);
\gpcolor{color=gp lt color border}
\gpsetlinetype{gp lt border}
\gpsetdashtype{gp dt solid}
\gpsetlinewidth{1.00}
\draw[gp path] (2.700,2.159)--(2.790,2.159);
\gpcolor{color=gp lt color axes}
\gpsetlinetype{gp lt axes}
\gpsetdashtype{gp dt axes}
\gpsetlinewidth{0.50}
\draw[gp path] (2.700,2.355)--(4.959,2.355);
\gpcolor{color=gp lt color border}
\gpsetlinetype{gp lt border}
\gpsetdashtype{gp dt solid}
\gpsetlinewidth{1.00}
\draw[gp path] (2.700,2.355)--(2.790,2.355);
\gpcolor{color=gp lt color axes}
\gpsetlinetype{gp lt axes}
\gpsetdashtype{gp dt axes}
\gpsetlinewidth{0.50}
\draw[gp path] (2.700,2.552)--(4.959,2.552);
\gpcolor{color=gp lt color border}
\gpsetlinetype{gp lt border}
\gpsetdashtype{gp dt solid}
\gpsetlinewidth{1.00}
\draw[gp path] (2.700,2.552)--(2.790,2.552);
\gpcolor{color=gp lt color axes}
\gpsetlinetype{gp lt axes}
\gpsetdashtype{gp dt axes}
\gpsetlinewidth{0.50}
\draw[gp path] (2.700,2.749)--(4.959,2.749);
\gpcolor{color=gp lt color border}
\gpsetlinetype{gp lt border}
\gpsetdashtype{gp dt solid}
\gpsetlinewidth{1.00}
\draw[gp path] (2.700,2.749)--(2.790,2.749);
\gpcolor{color=gp lt color axes}
\gpsetlinetype{gp lt axes}
\gpsetdashtype{gp dt axes}
\gpsetlinewidth{0.50}
\draw[gp path] (2.700,2.945)--(4.959,2.945);
\gpcolor{color=gp lt color border}
\gpsetlinetype{gp lt border}
\gpsetdashtype{gp dt solid}
\gpsetlinewidth{1.00}
\draw[gp path] (2.700,2.945)--(2.790,2.945);
\gpcolor{color=gp lt color axes}
\gpsetlinetype{gp lt axes}
\gpsetdashtype{gp dt axes}
\gpsetlinewidth{0.50}
\draw[gp path] (2.700,3.142)--(4.959,3.142);
\gpcolor{color=gp lt color border}
\gpsetlinetype{gp lt border}
\gpsetdashtype{gp dt solid}
\gpsetlinewidth{1.00}
\draw[gp path] (2.700,3.142)--(2.790,3.142);
\gpcolor{color=gp lt color axes}
\gpsetlinetype{gp lt axes}
\gpsetdashtype{gp dt axes}
\gpsetlinewidth{0.50}
\draw[gp path] (2.700,3.339)--(4.959,3.339);
\gpcolor{color=gp lt color border}
\gpsetlinetype{gp lt border}
\gpsetdashtype{gp dt solid}
\gpsetlinewidth{1.00}
\draw[gp path] (2.700,3.339)--(2.790,3.339);
\gpcolor{color=gp lt color axes}
\gpsetlinetype{gp lt axes}
\gpsetdashtype{gp dt axes}
\gpsetlinewidth{0.50}
\draw[gp path] (2.700,3.535)--(4.959,3.535);
\gpcolor{color=gp lt color border}
\gpsetlinetype{gp lt border}
\gpsetdashtype{gp dt solid}
\gpsetlinewidth{1.00}
\draw[gp path] (2.700,3.535)--(2.790,3.535);
\draw[gp path] (2.905,1.962)--(2.905,2.052);
\draw[gp path] (3.521,1.962)--(3.521,2.052);
\draw[gp path] (4.138,1.962)--(4.138,2.052);
\draw[gp path] (4.754,1.962)--(4.754,2.052);
\draw[gp path] (4.959,1.962)--(4.869,1.962);
\draw[gp path] (2.700,1.962)--(2.790,1.962);
\draw[gp path] (4.959,2.159)--(4.869,2.159);
\draw[gp path] (2.700,2.159)--(2.790,2.159);
\draw[gp path] (4.959,2.355)--(4.869,2.355);
\draw[gp path] (2.700,2.355)--(2.790,2.355);
\draw[gp path] (4.959,2.552)--(4.869,2.552);
\draw[gp path] (2.700,2.552)--(2.790,2.552);
\draw[gp path] (4.959,2.749)--(4.869,2.749);
\draw[gp path] (2.700,2.749)--(2.790,2.749);
\draw[gp path] (4.959,2.945)--(4.869,2.945);
\draw[gp path] (2.700,2.945)--(2.790,2.945);
\draw[gp path] (4.959,3.142)--(4.869,3.142);
\draw[gp path] (2.700,3.142)--(2.790,3.142);
\draw[gp path] (4.959,3.339)--(4.869,3.339);
\draw[gp path] (2.700,3.339)--(2.790,3.339);
\draw[gp path] (4.959,3.535)--(4.869,3.535);
\draw[gp path] (2.700,3.535)--(2.790,3.535);
\draw[gp path] (2.700,3.732)--(2.700,1.962)--(4.959,1.962)--(4.959,3.732)--cycle;
\gpcolor{rgb color={0.392,0.561,1.000}}
\gpsetlinewidth{3.00}
\draw[gp path] (2.905,2.746)--(3.521,2.733)--(4.138,2.485)--(4.754,2.155);
\gp3point{gp mark 7}{}{(2.905,2.746)}
\gp3point{gp mark 7}{}{(3.521,2.733)}
\gp3point{gp mark 7}{}{(4.138,2.485)}
\gp3point{gp mark 7}{}{(4.754,2.155)}
\gpcolor{rgb color={0.471,0.369,0.941}}
\draw[gp path] (2.905,2.565)--(3.521,3.038)--(4.138,3.413)--(4.754,3.621);
\gp3point{gp mark 5}{}{(2.905,2.565)}
\gp3point{gp mark 5}{}{(3.521,3.038)}
\gp3point{gp mark 5}{}{(4.138,3.413)}
\gp3point{gp mark 5}{}{(4.754,3.621)}
\gpcolor{color=gp lt color border}
\gpsetlinewidth{1.00}
\draw[gp path] (2.700,3.732)--(2.700,1.962)--(4.959,1.962)--(4.959,3.732)--cycle;
\gpdefrectangularnode{gp plot 2}{\pgfpoint{2.700cm}{1.962cm}}{\pgfpoint{4.959cm}{3.732cm}}
\gpcolor{color=gp lt color axes}
\gpsetlinetype{gp lt axes}
\gpsetdashtype{gp dt axes}
\gpsetlinewidth{0.50}
\draw[gp path] (4.959,1.962)--(7.217,1.962);
\gpcolor{color=gp lt color border}
\gpsetlinetype{gp lt border}
\gpsetdashtype{gp dt solid}
\gpsetlinewidth{1.00}
\draw[gp path] (4.959,1.962)--(5.049,1.962);
\gpcolor{color=gp lt color axes}
\gpsetlinetype{gp lt axes}
\gpsetdashtype{gp dt axes}
\gpsetlinewidth{0.50}
\draw[gp path] (4.959,2.159)--(7.217,2.159);
\gpcolor{color=gp lt color border}
\gpsetlinetype{gp lt border}
\gpsetdashtype{gp dt solid}
\gpsetlinewidth{1.00}
\draw[gp path] (4.959,2.159)--(5.049,2.159);
\gpcolor{color=gp lt color axes}
\gpsetlinetype{gp lt axes}
\gpsetdashtype{gp dt axes}
\gpsetlinewidth{0.50}
\draw[gp path] (4.959,2.355)--(7.217,2.355);
\gpcolor{color=gp lt color border}
\gpsetlinetype{gp lt border}
\gpsetdashtype{gp dt solid}
\gpsetlinewidth{1.00}
\draw[gp path] (4.959,2.355)--(5.049,2.355);
\gpcolor{color=gp lt color axes}
\gpsetlinetype{gp lt axes}
\gpsetdashtype{gp dt axes}
\gpsetlinewidth{0.50}
\draw[gp path] (4.959,2.552)--(7.217,2.552);
\gpcolor{color=gp lt color border}
\gpsetlinetype{gp lt border}
\gpsetdashtype{gp dt solid}
\gpsetlinewidth{1.00}
\draw[gp path] (4.959,2.552)--(5.049,2.552);
\gpcolor{color=gp lt color axes}
\gpsetlinetype{gp lt axes}
\gpsetdashtype{gp dt axes}
\gpsetlinewidth{0.50}
\draw[gp path] (4.959,2.749)--(7.217,2.749);
\gpcolor{color=gp lt color border}
\gpsetlinetype{gp lt border}
\gpsetdashtype{gp dt solid}
\gpsetlinewidth{1.00}
\draw[gp path] (4.959,2.749)--(5.049,2.749);
\gpcolor{color=gp lt color axes}
\gpsetlinetype{gp lt axes}
\gpsetdashtype{gp dt axes}
\gpsetlinewidth{0.50}
\draw[gp path] (4.959,2.945)--(7.217,2.945);
\gpcolor{color=gp lt color border}
\gpsetlinetype{gp lt border}
\gpsetdashtype{gp dt solid}
\gpsetlinewidth{1.00}
\draw[gp path] (4.959,2.945)--(5.049,2.945);
\gpcolor{color=gp lt color axes}
\gpsetlinetype{gp lt axes}
\gpsetdashtype{gp dt axes}
\gpsetlinewidth{0.50}
\draw[gp path] (4.959,3.142)--(5.079,3.142);
\draw[gp path] (7.096,3.142)--(7.217,3.142);
\gpcolor{color=gp lt color border}
\gpsetlinetype{gp lt border}
\gpsetdashtype{gp dt solid}
\gpsetlinewidth{1.00}
\draw[gp path] (4.959,3.142)--(5.049,3.142);
\gpcolor{color=gp lt color axes}
\gpsetlinetype{gp lt axes}
\gpsetdashtype{gp dt axes}
\gpsetlinewidth{0.50}
\draw[gp path] (4.959,3.339)--(5.079,3.339);
\draw[gp path] (7.096,3.339)--(7.217,3.339);
\gpcolor{color=gp lt color border}
\gpsetlinetype{gp lt border}
\gpsetdashtype{gp dt solid}
\gpsetlinewidth{1.00}
\draw[gp path] (4.959,3.339)--(5.049,3.339);
\gpcolor{color=gp lt color axes}
\gpsetlinetype{gp lt axes}
\gpsetdashtype{gp dt axes}
\gpsetlinewidth{0.50}
\draw[gp path] (4.959,3.535)--(5.079,3.535);
\draw[gp path] (7.096,3.535)--(7.217,3.535);
\gpcolor{color=gp lt color border}
\gpsetlinetype{gp lt border}
\gpsetdashtype{gp dt solid}
\gpsetlinewidth{1.00}
\draw[gp path] (4.959,3.535)--(5.049,3.535);
\draw[gp path] (5.335,1.962)--(5.335,2.052);
\draw[gp path] (5.712,1.962)--(5.712,2.052);
\draw[gp path] (6.088,1.962)--(6.088,2.052);
\draw[gp path] (6.464,1.962)--(6.464,2.052);
\draw[gp path] (6.841,1.962)--(6.841,2.052);
\draw[gp path] (7.217,1.962)--(7.127,1.962);
\draw[gp path] (4.959,1.962)--(5.049,1.962);
\draw[gp path] (7.217,2.159)--(7.127,2.159);
\draw[gp path] (4.959,2.159)--(5.049,2.159);
\draw[gp path] (7.217,2.355)--(7.127,2.355);
\draw[gp path] (4.959,2.355)--(5.049,2.355);
\draw[gp path] (7.217,2.552)--(7.127,2.552);
\draw[gp path] (4.959,2.552)--(5.049,2.552);
\draw[gp path] (7.217,2.749)--(7.127,2.749);
\draw[gp path] (4.959,2.749)--(5.049,2.749);
\draw[gp path] (7.217,2.945)--(7.127,2.945);
\draw[gp path] (4.959,2.945)--(5.049,2.945);
\draw[gp path] (7.217,3.142)--(7.127,3.142);
\draw[gp path] (4.959,3.142)--(5.049,3.142);
\draw[gp path] (7.217,3.339)--(7.127,3.339);
\draw[gp path] (4.959,3.339)--(5.049,3.339);
\draw[gp path] (7.217,3.535)--(7.127,3.535);
\draw[gp path] (4.959,3.535)--(5.049,3.535);
\draw[gp path] (4.959,3.732)--(4.959,1.962)--(7.217,1.962)--(7.217,3.732)--cycle;
\node[gp node right] at (6.402,3.429) {victim};
\gpcolor{rgb color={0.392,0.561,1.000}}
\gpsetlinewidth{3.00}
\draw[gp path] (6.549,3.429)--(6.949,3.429);
\draw[gp path] (5.335,2.830)--(5.712,2.370)--(6.088,2.242)--(6.464,2.175)--(6.841,2.135);
\gp3point{gp mark 7}{}{(5.335,2.830)}
\gp3point{gp mark 7}{}{(5.712,2.370)}
\gp3point{gp mark 7}{}{(6.088,2.242)}
\gp3point{gp mark 7}{}{(6.464,2.175)}
\gp3point{gp mark 7}{}{(6.841,2.135)}
\gp3point{gp mark 7}{}{(6.749,3.429)}
\gpcolor{color=gp lt color border}
\node[gp node right] at (6.402,3.183) {adversary};
\gpcolor{rgb color={0.471,0.369,0.941}}
\draw[gp path] (6.549,3.183)--(6.949,3.183);
\draw[gp path] (5.335,2.863)--(5.712,2.369)--(6.088,2.243)--(6.464,2.171)--(6.841,2.134);
\gp3point{gp mark 5}{}{(5.335,2.863)}
\gp3point{gp mark 5}{}{(5.712,2.369)}
\gp3point{gp mark 5}{}{(6.088,2.243)}
\gp3point{gp mark 5}{}{(6.464,2.171)}
\gp3point{gp mark 5}{}{(6.841,2.134)}
\gp3point{gp mark 5}{}{(6.749,3.183)}
\gpcolor{color=gp lt color border}
\gpsetlinewidth{1.00}
\draw[gp path] (4.959,3.732)--(4.959,1.962)--(7.217,1.962)--(7.217,3.732)--cycle;
\gpdefrectangularnode{gp plot 3}{\pgfpoint{4.959cm}{1.962cm}}{\pgfpoint{7.217cm}{3.732cm}}
\gpcolor{color=gp lt color axes}
\gpsetlinetype{gp lt axes}
\gpsetdashtype{gp dt axes}
\gpsetlinewidth{0.50}
\draw[gp path] (0.441,0.190)--(2.700,0.190);
\gpcolor{color=gp lt color border}
\gpsetlinetype{gp lt border}
\gpsetdashtype{gp dt solid}
\gpsetlinewidth{1.00}
\draw[gp path] (0.441,0.190)--(0.531,0.190);
\node[gp node right] at (0.441,0.190) {0};
\gpcolor{color=gp lt color axes}
\gpsetlinetype{gp lt axes}
\gpsetdashtype{gp dt axes}
\gpsetlinewidth{0.50}
\draw[gp path] (0.441,0.387)--(2.700,0.387);
\gpcolor{color=gp lt color border}
\gpsetlinetype{gp lt border}
\gpsetdashtype{gp dt solid}
\gpsetlinewidth{1.00}
\draw[gp path] (0.441,0.387)--(0.531,0.387);
\gpcolor{color=gp lt color axes}
\gpsetlinetype{gp lt axes}
\gpsetdashtype{gp dt axes}
\gpsetlinewidth{0.50}
\draw[gp path] (0.441,0.584)--(2.700,0.584);
\gpcolor{color=gp lt color border}
\gpsetlinetype{gp lt border}
\gpsetdashtype{gp dt solid}
\gpsetlinewidth{1.00}
\draw[gp path] (0.441,0.584)--(0.531,0.584);
\node[gp node right] at (0.441,0.584) {200};
\gpcolor{color=gp lt color axes}
\gpsetlinetype{gp lt axes}
\gpsetdashtype{gp dt axes}
\gpsetlinewidth{0.50}
\draw[gp path] (0.441,0.780)--(2.700,0.780);
\gpcolor{color=gp lt color border}
\gpsetlinetype{gp lt border}
\gpsetdashtype{gp dt solid}
\gpsetlinewidth{1.00}
\draw[gp path] (0.441,0.780)--(0.531,0.780);
\gpcolor{color=gp lt color axes}
\gpsetlinetype{gp lt axes}
\gpsetdashtype{gp dt axes}
\gpsetlinewidth{0.50}
\draw[gp path] (0.441,0.977)--(2.700,0.977);
\gpcolor{color=gp lt color border}
\gpsetlinetype{gp lt border}
\gpsetdashtype{gp dt solid}
\gpsetlinewidth{1.00}
\draw[gp path] (0.441,0.977)--(0.531,0.977);
\node[gp node right] at (0.441,0.977) {400};
\gpcolor{color=gp lt color axes}
\gpsetlinetype{gp lt axes}
\gpsetdashtype{gp dt axes}
\gpsetlinewidth{0.50}
\draw[gp path] (0.441,1.174)--(2.700,1.174);
\gpcolor{color=gp lt color border}
\gpsetlinetype{gp lt border}
\gpsetdashtype{gp dt solid}
\gpsetlinewidth{1.00}
\draw[gp path] (0.441,1.174)--(0.531,1.174);
\gpcolor{color=gp lt color axes}
\gpsetlinetype{gp lt axes}
\gpsetdashtype{gp dt axes}
\gpsetlinewidth{0.50}
\draw[gp path] (0.441,1.371)--(2.700,1.371);
\gpcolor{color=gp lt color border}
\gpsetlinetype{gp lt border}
\gpsetdashtype{gp dt solid}
\gpsetlinewidth{1.00}
\draw[gp path] (0.441,1.371)--(0.531,1.371);
\node[gp node right] at (0.441,1.371) {600};
\gpcolor{color=gp lt color axes}
\gpsetlinetype{gp lt axes}
\gpsetdashtype{gp dt axes}
\gpsetlinewidth{0.50}
\draw[gp path] (0.441,1.567)--(2.700,1.567);
\gpcolor{color=gp lt color border}
\gpsetlinetype{gp lt border}
\gpsetdashtype{gp dt solid}
\gpsetlinewidth{1.00}
\draw[gp path] (0.441,1.567)--(0.531,1.567);
\gpcolor{color=gp lt color axes}
\gpsetlinetype{gp lt axes}
\gpsetdashtype{gp dt axes}
\gpsetlinewidth{0.50}
\draw[gp path] (0.441,1.764)--(2.700,1.764);
\gpcolor{color=gp lt color border}
\gpsetlinetype{gp lt border}
\gpsetdashtype{gp dt solid}
\gpsetlinewidth{1.00}
\draw[gp path] (0.441,1.764)--(0.531,1.764);
\node[gp node right] at (0.441,1.764) {800};
\draw[gp path] (0.818,0.190)--(0.818,0.280);
\node[gp node center] at (0.818,-0.056) {50};
\draw[gp path] (1.194,0.190)--(1.194,0.280);
\draw[gp path] (1.571,0.190)--(1.571,0.280);
\node[gp node center] at (1.571,-0.056) {150};
\draw[gp path] (1.947,0.190)--(1.947,0.280);
\draw[gp path] (2.324,0.190)--(2.324,0.280);
\node[gp node center] at (2.324,-0.056) {250};
\draw[gp path] (2.700,0.190)--(2.610,0.190);
\draw[gp path] (2.700,0.387)--(2.610,0.387);
\draw[gp path] (2.700,0.584)--(2.610,0.584);
\draw[gp path] (2.700,0.780)--(2.610,0.780);
\draw[gp path] (2.700,0.977)--(2.610,0.977);
\draw[gp path] (2.700,1.174)--(2.610,1.174);
\draw[gp path] (2.700,1.371)--(2.610,1.371);
\draw[gp path] (2.700,1.567)--(2.610,1.567);
\draw[gp path] (2.700,1.764)--(2.610,1.764);
\draw[gp path] (2.700,1.961)--(2.610,1.961);
\draw[gp path] (0.441,1.961)--(0.441,0.190)--(2.700,0.190)--(2.700,1.961)--cycle;
\gpcolor{rgb color={0.392,0.561,1.000}}
\gpsetlinewidth{3.00}
\draw[gp path] (0.818,0.555)--(1.194,0.345)--(1.571,0.387)--(1.947,1.744)--(2.324,1.832);
\gp3point{gp mark 7}{}{(0.818,0.555)}
\gp3point{gp mark 7}{}{(1.194,0.345)}
\gp3point{gp mark 7}{}{(1.571,0.387)}
\gp3point{gp mark 7}{}{(1.947,1.744)}
\gp3point{gp mark 7}{}{(2.324,1.832)}
\gpcolor{color=gp lt color border}
\gpsetlinewidth{1.00}
\draw[gp path] (0.441,1.961)--(0.441,0.190)--(2.700,0.190)--(2.700,1.961)--cycle;
\node[gp node center,rotate=-270.0] at (-0.380,1.075) {99p Lat [µs]};
\node[gp node center] at (1.570,-0.424) {Cap [kCycles]};
\gpdefrectangularnode{gp plot 4}{\pgfpoint{0.441cm}{0.190cm}}{\pgfpoint{2.700cm}{1.961cm}}
\gpcolor{color=gp lt color axes}
\gpsetlinetype{gp lt axes}
\gpsetdashtype{gp dt axes}
\gpsetlinewidth{0.50}
\draw[gp path] (2.700,0.190)--(4.959,0.190);
\gpcolor{color=gp lt color border}
\gpsetlinetype{gp lt border}
\gpsetdashtype{gp dt solid}
\gpsetlinewidth{1.00}
\draw[gp path] (2.700,0.190)--(2.790,0.190);
\gpcolor{color=gp lt color axes}
\gpsetlinetype{gp lt axes}
\gpsetdashtype{gp dt axes}
\gpsetlinewidth{0.50}
\draw[gp path] (2.700,0.387)--(4.959,0.387);
\gpcolor{color=gp lt color border}
\gpsetlinetype{gp lt border}
\gpsetdashtype{gp dt solid}
\gpsetlinewidth{1.00}
\draw[gp path] (2.700,0.387)--(2.790,0.387);
\gpcolor{color=gp lt color axes}
\gpsetlinetype{gp lt axes}
\gpsetdashtype{gp dt axes}
\gpsetlinewidth{0.50}
\draw[gp path] (2.700,0.584)--(4.959,0.584);
\gpcolor{color=gp lt color border}
\gpsetlinetype{gp lt border}
\gpsetdashtype{gp dt solid}
\gpsetlinewidth{1.00}
\draw[gp path] (2.700,0.584)--(2.790,0.584);
\gpcolor{color=gp lt color axes}
\gpsetlinetype{gp lt axes}
\gpsetdashtype{gp dt axes}
\gpsetlinewidth{0.50}
\draw[gp path] (2.700,0.780)--(4.959,0.780);
\gpcolor{color=gp lt color border}
\gpsetlinetype{gp lt border}
\gpsetdashtype{gp dt solid}
\gpsetlinewidth{1.00}
\draw[gp path] (2.700,0.780)--(2.790,0.780);
\gpcolor{color=gp lt color axes}
\gpsetlinetype{gp lt axes}
\gpsetdashtype{gp dt axes}
\gpsetlinewidth{0.50}
\draw[gp path] (2.700,0.977)--(4.959,0.977);
\gpcolor{color=gp lt color border}
\gpsetlinetype{gp lt border}
\gpsetdashtype{gp dt solid}
\gpsetlinewidth{1.00}
\draw[gp path] (2.700,0.977)--(2.790,0.977);
\gpcolor{color=gp lt color axes}
\gpsetlinetype{gp lt axes}
\gpsetdashtype{gp dt axes}
\gpsetlinewidth{0.50}
\draw[gp path] (2.700,1.174)--(4.959,1.174);
\gpcolor{color=gp lt color border}
\gpsetlinetype{gp lt border}
\gpsetdashtype{gp dt solid}
\gpsetlinewidth{1.00}
\draw[gp path] (2.700,1.174)--(2.790,1.174);
\gpcolor{color=gp lt color axes}
\gpsetlinetype{gp lt axes}
\gpsetdashtype{gp dt axes}
\gpsetlinewidth{0.50}
\draw[gp path] (2.700,1.371)--(4.959,1.371);
\gpcolor{color=gp lt color border}
\gpsetlinetype{gp lt border}
\gpsetdashtype{gp dt solid}
\gpsetlinewidth{1.00}
\draw[gp path] (2.700,1.371)--(2.790,1.371);
\gpcolor{color=gp lt color axes}
\gpsetlinetype{gp lt axes}
\gpsetdashtype{gp dt axes}
\gpsetlinewidth{0.50}
\draw[gp path] (2.700,1.567)--(4.959,1.567);
\gpcolor{color=gp lt color border}
\gpsetlinetype{gp lt border}
\gpsetdashtype{gp dt solid}
\gpsetlinewidth{1.00}
\draw[gp path] (2.700,1.567)--(2.790,1.567);
\gpcolor{color=gp lt color axes}
\gpsetlinetype{gp lt axes}
\gpsetdashtype{gp dt axes}
\gpsetlinewidth{0.50}
\draw[gp path] (2.700,1.764)--(4.959,1.764);
\gpcolor{color=gp lt color border}
\gpsetlinetype{gp lt border}
\gpsetdashtype{gp dt solid}
\gpsetlinewidth{1.00}
\draw[gp path] (2.700,1.764)--(2.790,1.764);
\gpcolor{color=gp lt color axes}
\gpsetlinetype{gp lt axes}
\gpsetdashtype{gp dt axes}
\gpsetlinewidth{0.50}
\draw[gp path] (2.700,1.961)--(4.959,1.961);
\gpcolor{color=gp lt color border}
\gpsetlinetype{gp lt border}
\gpsetdashtype{gp dt solid}
\gpsetlinewidth{1.00}
\draw[gp path] (2.700,1.961)--(2.790,1.961);
\draw[gp path] (2.905,0.190)--(2.905,0.280);
\node[gp node center] at (2.905,-0.056) {0.6};
\draw[gp path] (3.521,0.190)--(3.521,0.280);
\node[gp node center] at (3.521,-0.056) {0.9};
\draw[gp path] (4.138,0.190)--(4.138,0.280);
\node[gp node center] at (4.138,-0.056) {1.2};
\draw[gp path] (4.754,0.190)--(4.754,0.280);
\node[gp node center] at (4.754,-0.056) {1.5};
\draw[gp path] (4.959,0.190)--(4.869,0.190);
\draw[gp path] (4.959,0.387)--(4.869,0.387);
\draw[gp path] (4.959,0.584)--(4.869,0.584);
\draw[gp path] (4.959,0.780)--(4.869,0.780);
\draw[gp path] (4.959,0.977)--(4.869,0.977);
\draw[gp path] (4.959,1.174)--(4.869,1.174);
\draw[gp path] (4.959,1.371)--(4.869,1.371);
\draw[gp path] (4.959,1.567)--(4.869,1.567);
\draw[gp path] (4.959,1.764)--(4.869,1.764);
\draw[gp path] (4.959,1.961)--(4.869,1.961);
\draw[gp path] (2.700,1.961)--(2.700,0.190)--(4.959,0.190)--(4.959,1.961)--cycle;
\gpcolor{rgb color={0.392,0.561,1.000}}
\gpsetlinewidth{3.00}
\draw[gp path] (2.905,0.394)--(3.521,0.300)--(4.138,0.368)--(4.754,1.082);
\gp3point{gp mark 7}{}{(2.905,0.394)}
\gp3point{gp mark 7}{}{(3.521,0.300)}
\gp3point{gp mark 7}{}{(4.138,0.368)}
\gp3point{gp mark 7}{}{(4.754,1.082)}
\gpcolor{color=gp lt color border}
\gpsetlinewidth{1.00}
\draw[gp path] (2.700,1.961)--(2.700,0.190)--(4.959,0.190)--(4.959,1.961)--cycle;
\node[gp node center] at (3.829,-0.424) {Boost};
\gpdefrectangularnode{gp plot 5}{\pgfpoint{2.700cm}{0.190cm}}{\pgfpoint{4.959cm}{1.961cm}}
\gpcolor{color=gp lt color axes}
\gpsetlinetype{gp lt axes}
\gpsetdashtype{gp dt axes}
\gpsetlinewidth{0.50}
\draw[gp path] (4.959,0.190)--(7.217,0.190);
\gpcolor{color=gp lt color border}
\gpsetlinetype{gp lt border}
\gpsetdashtype{gp dt solid}
\gpsetlinewidth{1.00}
\draw[gp path] (4.959,0.190)--(5.049,0.190);
\gpcolor{color=gp lt color axes}
\gpsetlinetype{gp lt axes}
\gpsetdashtype{gp dt axes}
\gpsetlinewidth{0.50}
\draw[gp path] (4.959,0.387)--(7.217,0.387);
\gpcolor{color=gp lt color border}
\gpsetlinetype{gp lt border}
\gpsetdashtype{gp dt solid}
\gpsetlinewidth{1.00}
\draw[gp path] (4.959,0.387)--(5.049,0.387);
\gpcolor{color=gp lt color axes}
\gpsetlinetype{gp lt axes}
\gpsetdashtype{gp dt axes}
\gpsetlinewidth{0.50}
\draw[gp path] (4.959,0.584)--(7.217,0.584);
\gpcolor{color=gp lt color border}
\gpsetlinetype{gp lt border}
\gpsetdashtype{gp dt solid}
\gpsetlinewidth{1.00}
\draw[gp path] (4.959,0.584)--(5.049,0.584);
\gpcolor{color=gp lt color axes}
\gpsetlinetype{gp lt axes}
\gpsetdashtype{gp dt axes}
\gpsetlinewidth{0.50}
\draw[gp path] (4.959,0.780)--(7.217,0.780);
\gpcolor{color=gp lt color border}
\gpsetlinetype{gp lt border}
\gpsetdashtype{gp dt solid}
\gpsetlinewidth{1.00}
\draw[gp path] (4.959,0.780)--(5.049,0.780);
\gpcolor{color=gp lt color axes}
\gpsetlinetype{gp lt axes}
\gpsetdashtype{gp dt axes}
\gpsetlinewidth{0.50}
\draw[gp path] (4.959,0.977)--(7.217,0.977);
\gpcolor{color=gp lt color border}
\gpsetlinetype{gp lt border}
\gpsetdashtype{gp dt solid}
\gpsetlinewidth{1.00}
\draw[gp path] (4.959,0.977)--(5.049,0.977);
\gpcolor{color=gp lt color axes}
\gpsetlinetype{gp lt axes}
\gpsetdashtype{gp dt axes}
\gpsetlinewidth{0.50}
\draw[gp path] (4.959,1.174)--(7.217,1.174);
\gpcolor{color=gp lt color border}
\gpsetlinetype{gp lt border}
\gpsetdashtype{gp dt solid}
\gpsetlinewidth{1.00}
\draw[gp path] (4.959,1.174)--(5.049,1.174);
\gpcolor{color=gp lt color axes}
\gpsetlinetype{gp lt axes}
\gpsetdashtype{gp dt axes}
\gpsetlinewidth{0.50}
\draw[gp path] (4.959,1.371)--(7.217,1.371);
\gpcolor{color=gp lt color border}
\gpsetlinetype{gp lt border}
\gpsetdashtype{gp dt solid}
\gpsetlinewidth{1.00}
\draw[gp path] (4.959,1.371)--(5.049,1.371);
\gpcolor{color=gp lt color axes}
\gpsetlinetype{gp lt axes}
\gpsetdashtype{gp dt axes}
\gpsetlinewidth{0.50}
\draw[gp path] (4.959,1.567)--(7.217,1.567);
\gpcolor{color=gp lt color border}
\gpsetlinetype{gp lt border}
\gpsetdashtype{gp dt solid}
\gpsetlinewidth{1.00}
\draw[gp path] (4.959,1.567)--(5.049,1.567);
\gpcolor{color=gp lt color axes}
\gpsetlinetype{gp lt axes}
\gpsetdashtype{gp dt axes}
\gpsetlinewidth{0.50}
\draw[gp path] (4.959,1.764)--(7.217,1.764);
\gpcolor{color=gp lt color border}
\gpsetlinetype{gp lt border}
\gpsetdashtype{gp dt solid}
\gpsetlinewidth{1.00}
\draw[gp path] (4.959,1.764)--(5.049,1.764);
\gpcolor{color=gp lt color axes}
\gpsetlinetype{gp lt axes}
\gpsetdashtype{gp dt axes}
\gpsetlinewidth{0.50}
\draw[gp path] (4.959,1.961)--(7.217,1.961);
\gpcolor{color=gp lt color border}
\gpsetlinetype{gp lt border}
\gpsetdashtype{gp dt solid}
\gpsetlinewidth{1.00}
\draw[gp path] (4.959,1.961)--(5.049,1.961);
\draw[gp path] (5.335,0.190)--(5.335,0.280);
\node[gp node center] at (5.335,-0.056) {100};
\draw[gp path] (5.712,0.190)--(5.712,0.280);
\draw[gp path] (6.088,0.190)--(6.088,0.280);
\node[gp node center] at (6.088,-0.056) {300};
\draw[gp path] (6.464,0.190)--(6.464,0.280);
\draw[gp path] (6.841,0.190)--(6.841,0.280);
\node[gp node center] at (6.841,-0.056) {500};
\draw[gp path] (7.217,0.190)--(7.127,0.190);
\draw[gp path] (7.217,0.387)--(7.127,0.387);
\draw[gp path] (7.217,0.584)--(7.127,0.584);
\draw[gp path] (7.217,0.780)--(7.127,0.780);
\draw[gp path] (7.217,0.977)--(7.127,0.977);
\draw[gp path] (7.217,1.174)--(7.127,1.174);
\draw[gp path] (7.217,1.371)--(7.127,1.371);
\draw[gp path] (7.217,1.567)--(7.127,1.567);
\draw[gp path] (7.217,1.764)--(7.127,1.764);
\draw[gp path] (7.217,1.961)--(7.127,1.961);
\draw[gp path] (4.959,1.961)--(4.959,0.190)--(7.217,0.190)--(7.217,1.961)--cycle;
\gpcolor{rgb color={0.392,0.561,1.000}}
\gpsetlinewidth{3.00}
\draw[gp path] (5.335,0.284)--(5.712,0.524)--(6.088,0.683)--(6.464,0.842)--(6.841,1.010);
\gp3point{gp mark 7}{}{(5.335,0.284)}
\gp3point{gp mark 7}{}{(5.712,0.524)}
\gp3point{gp mark 7}{}{(6.088,0.683)}
\gp3point{gp mark 7}{}{(6.464,0.842)}
\gp3point{gp mark 7}{}{(6.841,1.010)}
\gpcolor{color=gp lt color border}
\gpsetlinewidth{1.00}
\draw[gp path] (4.959,1.961)--(4.959,0.190)--(7.217,0.190)--(7.217,1.961)--cycle;
\node[gp node center] at (6.088,-0.424) {Period [µs]};
\gpdefrectangularnode{gp plot 6}{\pgfpoint{4.959cm}{0.190cm}}{\pgfpoint{7.217cm}{1.961cm}}
\end{tikzpicture}
  \caption{Victim tenant latency and throughput
      with variable boost, budget caps,
      and update periods, under adversarial interference.}%
  \label{fig:sensitivity}%
\end{figure}

\subsection{TAS \texttt{RX}-Side Feedback for Open-Loop Traffic}
Under open-loop traffic, \autoref{sec:design} leaves receiver-side protection
to post-attribution drops or explicit feedback.
In TAS, we realize this feedback path with the ECN-based extension from
\autoref{sec:impl}.
We vary the ECN marking threshold in the TAS implementation and record tenant
throughput and packet drops.
In \autoref{fig:ecn-drops}, we measure the effectiveness of
this mechanism. We measure throughput and packet 
drops for different ECN marking thresholds expressed 
as a fraction of the budget cap. 
At an ECN threshold of 0, no packets are marked, 
so senders do not slow down and drops increase. When we raise
the threshold, early congestion control feedback reduces drops 
and improves throughput, but high thresholds 
trigger overly aggressive rate reduction.

\begin{figure}%
\centering%
\begin{tikzpicture}[gnuplot]
\tikzset{every node/.append style={font={\fontsize{8.0pt}{9.6pt}\selectfont}}}
\path (0.000,0.000) rectangle (8.458,3.048);
\gpcolor{color=gp lt color axes}
\gpsetlinetype{gp lt axes}
\gpsetdashtype{gp dt axes}
\gpsetlinewidth{0.50}
\draw[gp path] (0.676,0.365)--(3.551,0.365);
\gpcolor{color=gp lt color border}
\gpsetlinetype{gp lt border}
\gpsetdashtype{gp dt solid}
\gpsetlinewidth{1.00}
\draw[gp path] (0.676,0.365)--(0.766,0.365);
\node[gp node right] at (0.529,0.365) {$0$};
\gpcolor{color=gp lt color axes}
\gpsetlinetype{gp lt axes}
\gpsetdashtype{gp dt axes}
\gpsetlinewidth{0.50}
\draw[gp path] (0.676,0.767)--(3.551,0.767);
\gpcolor{color=gp lt color border}
\gpsetlinetype{gp lt border}
\gpsetdashtype{gp dt solid}
\gpsetlinewidth{1.00}
\draw[gp path] (0.676,0.767)--(0.766,0.767);
\node[gp node right] at (0.529,0.767) {$2$};
\gpcolor{color=gp lt color axes}
\gpsetlinetype{gp lt axes}
\gpsetdashtype{gp dt axes}
\gpsetlinewidth{0.50}
\draw[gp path] (0.676,1.169)--(3.551,1.169);
\gpcolor{color=gp lt color border}
\gpsetlinetype{gp lt border}
\gpsetdashtype{gp dt solid}
\gpsetlinewidth{1.00}
\draw[gp path] (0.676,1.169)--(0.766,1.169);
\node[gp node right] at (0.529,1.169) {$4$};
\gpcolor{color=gp lt color axes}
\gpsetlinetype{gp lt axes}
\gpsetdashtype{gp dt axes}
\gpsetlinewidth{0.50}
\draw[gp path] (0.676,1.572)--(3.551,1.572);
\gpcolor{color=gp lt color border}
\gpsetlinetype{gp lt border}
\gpsetdashtype{gp dt solid}
\gpsetlinewidth{1.00}
\draw[gp path] (0.676,1.572)--(0.766,1.572);
\node[gp node right] at (0.529,1.572) {$6$};
\gpcolor{color=gp lt color axes}
\gpsetlinetype{gp lt axes}
\gpsetdashtype{gp dt axes}
\gpsetlinewidth{0.50}
\draw[gp path] (0.676,1.974)--(2.563,1.974);
\draw[gp path] (3.404,1.974)--(3.551,1.974);
\gpcolor{color=gp lt color border}
\gpsetlinetype{gp lt border}
\gpsetdashtype{gp dt solid}
\gpsetlinewidth{1.00}
\draw[gp path] (0.676,1.974)--(0.766,1.974);
\node[gp node right] at (0.529,1.974) {$8$};
\gpcolor{color=gp lt color axes}
\gpsetlinetype{gp lt axes}
\gpsetdashtype{gp dt axes}
\gpsetlinewidth{0.50}
\draw[gp path] (0.676,2.376)--(3.551,2.376);
\gpcolor{color=gp lt color border}
\gpsetlinetype{gp lt border}
\gpsetdashtype{gp dt solid}
\gpsetlinewidth{1.00}
\draw[gp path] (0.676,2.376)--(0.766,2.376);
\node[gp node right] at (0.529,2.376) {$10$};
\draw[gp path] (0.676,0.365)--(0.676,0.455);
\node[gp node center] at (0.676,0.119) {0};
\draw[gp path] (1.118,0.365)--(1.118,0.455);
\draw[gp path] (1.561,0.365)--(1.561,0.455);
\node[gp node center] at (1.561,0.119) {0.2};
\draw[gp path] (2.003,0.365)--(2.003,0.455);
\draw[gp path] (2.445,0.365)--(2.445,0.455);
\node[gp node center] at (2.445,0.119) {0.4};
\draw[gp path] (2.888,0.365)--(2.888,0.455);
\draw[gp path] (3.330,0.365)--(3.330,0.455);
\node[gp node center] at (3.330,0.119) {0.6};
\draw[gp path] (3.551,0.365)--(3.461,0.365);
\draw[gp path] (3.551,0.767)--(3.461,0.767);
\draw[gp path] (3.551,1.169)--(3.461,1.169);
\draw[gp path] (3.551,1.572)--(3.461,1.572);
\draw[gp path] (3.551,1.974)--(3.461,1.974);
\draw[gp path] (3.551,2.376)--(3.461,2.376);
\draw[gp path] (0.676,2.376)--(0.676,0.365)--(3.551,0.365)--(3.551,2.376)--cycle;
\node[gp node right] at (2.710,2.073) {\sysplot};
\gpcolor{rgb color={0.392,0.561,1.000}}
\gpsetlinewidth{3.00}
\draw[gp path] (2.857,2.073)--(3.257,2.073);
\draw[gp path] (0.676,2.196)--(1.118,1.517)--(1.561,1.048)--(2.003,0.740)--(2.445,0.430)%
  --(2.888,0.400)--(3.330,0.399);
\gpsetpointsize{4.00}
\gp3point{gp mark 7}{}{(0.676,2.196)}
\gp3point{gp mark 7}{}{(1.118,1.517)}
\gp3point{gp mark 7}{}{(1.561,1.048)}
\gp3point{gp mark 7}{}{(2.003,0.740)}
\gp3point{gp mark 7}{}{(2.445,0.430)}
\gp3point{gp mark 7}{}{(2.888,0.400)}
\gp3point{gp mark 7}{}{(3.330,0.399)}
\gp3point{gp mark 7}{}{(3.057,2.073)}
\gpcolor{color=gp lt color border}
\gpsetlinewidth{1.00}
\draw[gp path] (0.676,2.376)--(0.676,0.365)--(3.551,0.365)--(3.551,2.376)--cycle;
\node[gp node center,rotate=-270.0] at (-0.072,1.370) {KDrop/s};
\node[gp node center] at (2.113,-0.249) {ECN Threshold};
\gpdefrectangularnode{gp plot 1}{\pgfpoint{0.676cm}{0.365cm}}{\pgfpoint{3.551cm}{2.376cm}}
\gpcolor{color=gp lt color axes}
\gpsetlinetype{gp lt axes}
\gpsetdashtype{gp dt axes}
\gpsetlinewidth{0.50}
\draw[gp path] (4.736,0.365)--(7.611,0.365);
\gpcolor{color=gp lt color border}
\gpsetlinetype{gp lt border}
\gpsetdashtype{gp dt solid}
\gpsetlinewidth{1.00}
\draw[gp path] (4.736,0.365)--(4.826,0.365);
\node[gp node right] at (4.589,0.365) {$0$};
\gpcolor{color=gp lt color axes}
\gpsetlinetype{gp lt axes}
\gpsetdashtype{gp dt axes}
\gpsetlinewidth{0.50}
\draw[gp path] (4.736,0.868)--(7.611,0.868);
\gpcolor{color=gp lt color border}
\gpsetlinetype{gp lt border}
\gpsetdashtype{gp dt solid}
\gpsetlinewidth{1.00}
\draw[gp path] (4.736,0.868)--(4.826,0.868);
\node[gp node right] at (4.589,0.868) {$0.3$};
\gpcolor{color=gp lt color axes}
\gpsetlinetype{gp lt axes}
\gpsetdashtype{gp dt axes}
\gpsetlinewidth{0.50}
\draw[gp path] (4.736,1.371)--(7.611,1.371);
\gpcolor{color=gp lt color border}
\gpsetlinetype{gp lt border}
\gpsetdashtype{gp dt solid}
\gpsetlinewidth{1.00}
\draw[gp path] (4.736,1.371)--(4.826,1.371);
\node[gp node right] at (4.589,1.371) {$0.6$};
\gpcolor{color=gp lt color axes}
\gpsetlinetype{gp lt axes}
\gpsetdashtype{gp dt axes}
\gpsetlinewidth{0.50}
\draw[gp path] (4.736,1.873)--(7.611,1.873);
\gpcolor{color=gp lt color border}
\gpsetlinetype{gp lt border}
\gpsetdashtype{gp dt solid}
\gpsetlinewidth{1.00}
\draw[gp path] (4.736,1.873)--(4.826,1.873);
\node[gp node right] at (4.589,1.873) {$0.9$};
\gpcolor{color=gp lt color axes}
\gpsetlinetype{gp lt axes}
\gpsetdashtype{gp dt axes}
\gpsetlinewidth{0.50}
\draw[gp path] (4.736,2.376)--(7.611,2.376);
\gpcolor{color=gp lt color border}
\gpsetlinetype{gp lt border}
\gpsetdashtype{gp dt solid}
\gpsetlinewidth{1.00}
\draw[gp path] (4.736,2.376)--(4.826,2.376);
\node[gp node right] at (4.589,2.376) {$1.2$};
\draw[gp path] (4.736,0.365)--(4.736,0.455);
\node[gp node center] at (4.736,0.119) {0};
\draw[gp path] (5.178,0.365)--(5.178,0.455);
\draw[gp path] (5.621,0.365)--(5.621,0.455);
\node[gp node center] at (5.621,0.119) {0.2};
\draw[gp path] (6.063,0.365)--(6.063,0.455);
\draw[gp path] (6.505,0.365)--(6.505,0.455);
\node[gp node center] at (6.505,0.119) {0.4};
\draw[gp path] (6.948,0.365)--(6.948,0.455);
\draw[gp path] (7.390,0.365)--(7.390,0.455);
\node[gp node center] at (7.390,0.119) {0.6};
\draw[gp path] (7.611,0.365)--(7.521,0.365);
\draw[gp path] (7.611,0.868)--(7.521,0.868);
\draw[gp path] (7.611,1.371)--(7.521,1.371);
\draw[gp path] (7.611,1.873)--(7.521,1.873);
\draw[gp path] (7.611,2.376)--(7.521,2.376);
\draw[gp path] (4.736,2.376)--(4.736,0.365)--(7.611,0.365)--(7.611,2.376)--cycle;
\gpcolor{rgb color={0.392,0.561,1.000}}
\gpsetlinewidth{3.00}
\draw[gp path] (4.736,2.059)--(5.178,2.010)--(5.621,2.021)--(6.063,2.079)--(6.505,2.091)%
  --(6.948,1.851)--(7.390,1.551);
\gp3point{gp mark 7}{}{(4.736,2.059)}
\gp3point{gp mark 7}{}{(5.178,2.010)}
\gp3point{gp mark 7}{}{(5.621,2.021)}
\gp3point{gp mark 7}{}{(6.063,2.079)}
\gp3point{gp mark 7}{}{(6.505,2.091)}
\gp3point{gp mark 7}{}{(6.948,1.851)}
\gp3point{gp mark 7}{}{(7.390,1.551)}
\gpcolor{color=gp lt color border}
\gpsetlinewidth{1.00}
\draw[gp path] (4.736,2.376)--(4.736,0.365)--(7.611,0.365)--(7.611,2.376)--cycle;
\node[gp node center,rotate=-270.0] at (3.841,1.370) {Tput [MReq/s]};
\node[gp node center] at (6.173,-0.249) {ECN Threshold};
\gpdefrectangularnode{gp plot 2}{\pgfpoint{4.736cm}{0.365cm}}{\pgfpoint{7.611cm}{2.376cm}}
\end{tikzpicture}
\caption{Packet drops and tenant throughput under 
  increasing ECN marking thresholds, expressed as a ratio of the tenant budget cap. 
  Coupling congestion control with budget accounting 
  reduces the number of drops when tenants are out-of-budget.}%
\Description[ECN mark thresholds]{ECN threshold increases and number 
  of drops and throughput decreases}%
\label{fig:ecn-drops}%
\end{figure}
\section{Discussion and Future Work}
\label{sec:discussion}

We discuss the scope of \sys and directions to extend it beyond its current
design.
These directions explore how to better adapt to workload diversity, leverage
hardware support, and integrate with resource management.
Together they highlight the opportunities and challenges in sustaining
tail latency isolation alongside high efficiency.

\textbf{\textit{Applicability beyond TAS.}}
Instantiating \sys in TAS suggests that the design can generalise beyond a
kernel-bypass TCP stack.
The essential requirements are architectural rather than TAS-specific: a
datapath needs a bounded fast path, identifiable intervention points where work
can be charged and gated, and a separate control path that can coordinate
budgets at a coarser timescale with low synchronisation overhead.
At the same time, datapaths with long or highly variable critical-path
operations or pervasive cross-core shared state may require additional
restructuring and may only support coarser enforcement.

\textbf{\textit{Deployment on DPUs and SmartNICs.}}
DPUs~\cite{spec:nvidia_bluefield3,spec:marvell_octeon10,spec:amd_pensando_elba}
and SoC SmartNICs~\cite{spec:corigine_agilio_cx,spec:broadcom_stingray,spec:amd_xilinx_alveo_u25n}
are also a direct target for \sys when they run shared software datapaths on
general-purpose cores.
In this setting, the core design remains unchanged: tenants still contend for
CPU time in batched run-to-completion loops, and the same accounting and
intervention-point mechanisms apply.
Compared to host CPUs, these deployments may provide cleaner isolation from
unrelated host activity, but they also introduce different core-performance and
queueing tradeoffs that may require retuning budget parameters.

\textbf{\textit{Adaptive control and placement.}}
A next step is to tune budget parameters dynamically rather than
fixing them statically, since they encode a tradeoff between latency
isolation and throughput efficiency.
Throughput-oriented workloads may benefit from more permissive settings that
allow larger batches and better amortization, whereas latency-sensitive
workloads may require tighter settings to curb interference.
Such adaptation would require online profiling, application hints, or both,
using signals already visible at intervention points such as buffer occupancy
and transmission rates.
The challenge is to track workload shifts quickly enough without inducing
oscillation or transient SLO violations.

A complementary direction is to expose \sys to existing placement and
load-balancing mechanisms.
These systems separate latency-sensitive and throughput-oriented
traffic when capacity permits; \sys can then bound the interference that
remains once cores are shared.
This would present a fallback when burstiness or 
limited capacity forces co-location.
The challenge is to surface useful signals to these mechanisms and react
quickly enough to workload shifts without causing oscillation or excessive
migration.

\section{Related Work}%
\label{sec:related}

\textbf{\textit{Microsecond network dataplanes.}}
Microsecond scale dataplanes reduce tail latency at the boundary 
between an application and the datapath. 
Caladan~\cite{fried:caladan} reacts to interference by using 
a centralised scheduler plus a kernel module that bypasses Linux 
scheduling to support rapid monitoring and core reallocation. 
Shenango~\cite{ousterhout:shenango} reaches similar timescales with an IOKernel 
that steers packets and reallocates cores across applications 
based on thread and packet queuing delays. ZygOS~\cite{prekas:zygos} makes
scheduling work-conserving by using cross-core work stealing and 
a shuffle layer to avoid head-of-line 
blocking. Shinjuku~\cite{kaffes:shinjuku} 
instead centralises 
dispatch and uses low-overhead preemption to prevent 
long requests from inflating the tail. 
Prior systems highlight the need for careful load balancing 
and resource allocation to control $\mu$s tail latency. 
Their controls are enforced at the application boundary, 
rather than inside a shared tenant multiplexed datapath, and
can afford higher overheads. 
In a shared datapath with cross-tenant batching, 
tail latency isolation must be enforced inside the datapath under
no preemption and minimal synchronisation constraints 
to prevent tail latency interference.

\textbf{\textit{Microsecond storage dataplanes.}}
As in microsecond scale network dataplanes, the key pathology in 
storage is that shared, non-preemptive, kernel work creates head-of-line 
blocking that inflates the tail.
Blk-switch~\cite{hwang:blkswitch} curbs tail latency by steering requests and 
prioritising latency-sensitive I/O, which reduces head-of-line blocking. 
It does not, however, meter CPU time per tenant or repay overshoot when 
tenants couple through batching; our contribution is a 
CPU-time accounting that stays enforceable even under such coupling.

\textbf{\textit{Microsecond scheduling policies.}}
Microsecond scale scheduling policies address queueing delay to prevent
tail latency inflation, yet enforcing tail optimal policies can impose
enough overhead to lower maximum throughput. 
Concord~\cite{iyer:concord}
addresses the throughput-tail tension by approximating optimal policies:
it replaces interrupt-driven preemption with compiler-enforced cooperative
yield points, substitutes a single queue with bounded per worker queues to
reduce coherence stalls, and makes the dispatcher work-conserving by stealing
work under load.
Tiny Quanta~\cite{luo:tiny_quanta} uses blind job scheduling with
compiler-inserted yield points to enable efficient scheduling of
$\mu$s-level tasks.
Persephone~\cite{demoulin:persephone} biases capacity 
toward short requests:
cores are reserved for short requests and short requests steal workers 
from long requests when they are threatened, thus sacrificing work conservation
to preserve the tail. 
These designs solve cross-application head-of-line blocking, 
but they do not solve shared datapath interference, 
where latency comes from unattributed CPU time and cross-tenant batching, 
so changing the per-application queueing or preemption
policy cannot bound how much CPU 
time a noisy tenant can consume within the shared datapath. 
Furthermore, the tasks scheduled at the application layer are coarser-grained
and can thus afford the switching and cache pollution from yielding:
these overheads would be prohibitive with sub-µs tasks.
Nonetheless,
both datapath time protection and scheduling policies at the application
boundary can be combined for end-to-end isolation.

\textbf{\textit{Time-based accounting.}}
Prior systems have also used time as the accounting currency, 
but they apply it at coarser boundaries than our shared datapath. 
Andromeda~\cite{dalton:andromeda} attributes time to each VM and
enforces ingress control by 
scheduling per VM queues based on which VM has recently consumed the least 
cycles. 
PicNIC~\cite{kumar:picnic} uses time primarily to enforce network shaping
and to cap per VM buffering, 
which targets bandwidth and buffer interference rather than shared core CPU interference. 
In contrast, our work enforces short window, 
per tenant CPU time budgets inside shared, 
batched, run-to-completion phases that improve efficiency under
strict latency constraints.
CPU time based resource accounting has also been used to
more accurately account for the resources used by 
VMs~\cite{djomgwe:billing_cpu_time}
and containers~\cite{bacou:wysiwyg}. Nonetheless, previous approaches
do not deal with the challenges imposed by tight latency
budgets and mitigating interference at sub microsecond scales.
\section{Conclusion}
Shared software datapaths need fine-grained sharing for efficiency, but that
sharing creates cross-tenant tail latency interference that packet- and
byte-based controls cannot bound.
This paper showed that tail latency isolation in shared datapaths must be
expressed in CPU time.
\sys realizes this with time protection: it enforces per-tenant CPU-time
budgets at low-overhead intervention points while preserving batching and
handling delayed \texttt{RX} attribution.
Instantiated in TAS, a demanding shared transport datapath, \sys
substantially reduces cross-tenant interference while retaining most of the
efficiency of shared execution.
More broadly, our results show that strong tail latency isolation and
efficient fine-grained sharing are compatible in shared software datapaths.
 \if \ANON 0
\section*{Acknowledgments}
We thank Rajath Shashidhara
for his contributions in the long running discussions over
the course of this project.
 \fi

\bibliographystyle{plain}
\bibliography{paper, bibdb/papers, bibdb/strings, bibdb/defs}

\label{page:last}
\end{document}